\pdfoutput=1

\documentclass[preprint,12pt]{elsarticle}

\usepackage{mwe}

\usepackage{setspace} 
\usepackage{lineno,hyperref}
\modulolinenumbers[5]

\usepackage[a4paper]{geometry}
\usepackage{amssymb}
\usepackage[]{amsmath}
\usepackage{upgreek}
\usepackage{pgfplots}
\pgfplotsset{compat=1.5}
\usepgfplotslibrary{units}
\usetikzlibrary{spy}
\usetikzlibrary{pgfplots.groupplots}

\usepackage{units}
\usepackage[latin1]{inputenc}
\usepackage{lscape}
\usepackage{booktabs}
\usepackage{gensymb}
\usepackage{multirow}

\usepackage{graphicx}
\usepackage{caption}
\usepackage{subcaption}

\usepackage{floatrow} 
\usepackage[flushleft]{threeparttable}
\usepackage{tikz}
\usepackage{placeins}
\usepackage{color}
\usepackage{hyperref}

\usepackage{listings}
\usepackage{courier}
\usepackage{stackengine}

\newcommand\eatpunct[1]{}

\begin{document}

\begin{frontmatter}

\title{Finite volume simulations of particle-laden viscoelastic fluid flows: application to hydraulic fracture processes}

\author[mymainaddress]{C. Fernandes\corref{mycorrespondingauthor}}
\author[mysecondaryaddress2]{S. A. Faroughi}
\author[mymainaddress]{R. Ribeiro}
\author[mymainaddress]{A. Isabel}
\author[mythirdaddress]{and G. H. McKinley}
\cortext[mycorrespondingauthor]{Corresponding author: cbpf@dep.uminho.pt}

\address[mymainaddress]{Institute for Polymers and Composites, Department of Polymer Engineering, University of Minho, Campus de Azur\'em, 4800-058 Guimar\~aes, Portugal}
\address[mysecondaryaddress2]{Geo-Intelligence Laboratory, Ingram School of Engineering, Texas State University, San Marcos, Texas 78666, USA}
\address[mythirdaddress]{Hatsopoulos Microfluids Laboratory, Department of Mechanical Engineering, Massachusetts Institute of Technology, Cambridge, Massachusetts 02139, USA}

\begin{abstract}

Accurately resolving the coupled momentum transfer between the liquid and solid phases of complex fluids is a fundamental problem in multiphase transport processes, such as hydraulic fracture operations. Specifically we need to characterize the dependence of the normalized average fluid-particle force $\langle F \rangle$ on the volume fraction of the dispersed solid phase and on the rheology of the complex fluid matrix, parameterized through the Weissenberg number $Wi$ measuring the relative magnitude of elastic to viscous stresses in the fluid. Here we use direct numerical simulations (DNS) to study the creeping flow ($Re\ll 1$) of viscoelastic fluids through static random arrays of monodisperse spherical particles using a finite volume Navier-Stokes/Cauchy momentum solver. The numerical study consists of $N=150$ different systems, in which the normalized average fluid-particle force $\langle F \rangle$ is obtained as a function of the volume fraction $\phi$ $(0 < \phi \leq 0.2)$ of the dispersed solid phase and the Weissenberg number $Wi$ $(0 \leq Wi \leq 4)$. From these predictions a closure law $\langle F \rangle(Wi,\phi)$ for the drag force is derived for the quasi-linear Oldroyd-B viscoelastic fluid model (with fixed retardation ratio $\beta = 0.5$) which is, on average, within $5.7\%$ of the DNS results. Additionally, a flow solver able to couple Eulerian and Lagrangian phases (in which the particulate phase is modeled by the discrete particle method (DPM)) is developed, which incorporates the viscoelastic nature of the continuum phase and the closed-form drag law. Two case studies were simulated using this solver, in order to assess the accuracy and robustness of the newly-developed approach for handling particle-laden viscoelastic flow configurations with $O(10^5-10^6)$ rigid spheres that are representative of hydraulic fracture operations. Three-dimensional settling processes in a Newtonian fluid and in a quasi-linear Oldroyd-B viscoelastic fluid are both investigated using a rectangular channel and an annular pipe domain. Good agreement is obtained for the particle distribution measured in a Newtonian fluid, when comparing numerical results with experimental data. For the cases in which the continuous fluid phase is viscoelastic we compute the evolution in the velocity fields and predicted particle distributions are presented at different elasticity numbers $0\leq El \leq 30$ (where $El=Wi/Re$) and for different suspension particle volume fractions.

\end{abstract}


\end{frontmatter}

\section{Introduction}

Understanding the force balance that governs the migration of rigid particles suspended in a viscoelastic fluid is fundamental to a wide range of engineering and technology applications. Examples include polymer processing of highly-filled viscoelastic melts and elastomers \cite{liff2007high}, the processing of semi-solid conductive flow battery slurries \cite{olsen2016coupled}, the flow-induced migration of circulating cancer cells in biopolymeric media such as blood \cite{lim2014inertio}, magma eruption dynamics \cite{parmigiani2016bubble}, and hydraulic fracturing operations using solids-filled muds, slurries and foams \cite{barbati2016complex,faroughi2018rheological}. 

In many ways, developing robust and accurate tools to simulate such behaviors may be viewed as an unsolved grand challenge in the dynamics of complex fluids, involving the effects of nonlinear material rheology, fluid inertia, elasticity, flow-unsteadiness plus many-body interactions. In a fluid with Newtonian or non-Newtonian rheology, the presence of a cloud of particles dramatically changes the transmission of stress between both phases (fluid and particles), specifically in terms of the rate at which the constituents of the mixture exchange momentum,  known as hindrance effect \cite{faroughi2015unifying,faroughi2016theoretical} .  When particles are suspended in a viscoelastic fluid (e.g. a polymer solution or polymer melt), the problem becomes even more challenging, because the fluid may shear-thin or shear-thicken as well as exhibit viscoelasticity and yield stress attributes \cite{Shaqfeh2019,Tanner2019}. Therefore, understanding and predicting both the bulk/macro-scale and particle-level response of these complex multiphase suspensions remains an open problem. As a first step, quantifying the momentum exchange between the constituent phases (i.e. a viscoelastic fluid matrix and a suspended phase of rigid spherical particles) remains a challenging and important problem to be solved in non-Newtonian fluid dynamics.

Because of the linear response between stress and deformation rate, the hydrodynamic behavior of rigid spheres in a Newtonian fluid has received considerable attention since the pioneering work of \citet{Stokes1851} (see for example the monographs by \citet{Happel1983,Kim2005} and \citet{Guazzelli2011}). In the limit of infinite dilution, and when inertial effects can be neglected, the drag force, $\textbf{F}_d$, exerted by the fluid on the solid object takes the Stokes-Einstein form $\textbf{F}_d=6\pi a \eta_0 \overline{u}$, where $a$ is the radius of the spherical particle, $\eta_0$ is the fluid shear viscosity and $\overline{u}$ is the superficial fluid velocity, defined as the fluid velocity averaged over the total volume of the system \cite{Hoef2005}. Higher order corrections to the Stokes-Einstein drag acting on a single particle arising from the presence of neighboring particles have been evaluated in terms of an expansion in the particle volume fraction $\phi$. For small packing fractions ($\phi < 0.10$), the first few terms can be worked out analytically \cite{Hasimoto1959}, but for larger packing fractions, the drag force has to be estimated from approximate theoretical methods \cite{Brinkman1947, Kim1985}, or from empirical data via experimental measurements \cite{Carman1937}. Additionally, numerical simulations \cite{Hoef2005, Hill2001, Koch1999} can provide data to derive drag force expressions for the creeping flow of random arrays of spheres surrounded by a Newtonian fluid. 

For the creeping flow of a sphere through an unbounded viscoelastic fluid, measurements of the changes to the force acting on the sphere are typically represented in terms of a dimensionless drag correction factor $X(Wi)$, which is the ratio of the measured drag coefficient compared to the well-known Stokes drag, $X(Wi)\equiv C_D(Wi)/C_D(Wi=0)=C_D(Wi)/(24/Re)$  where $Re$ and $Wi$ are the dimensionless Reynolds and Weissenberg numbers, respectively. For the inertialess flow of a viscoelastic fluid, perturbation solutions predict the departure of the drag from the Stokes result to be quadratic in the Weissenberg number for spheres \cite{leslie1961slow}, and linear in the case of long rod-like particles \cite{Leal1975}. Recently, two reviews comparing experimental data with computations for non-Brownian suspension rheology with non-Newtonian matrices have been published \cite{Shaqfeh2019, Tanner2019}. \citet{Tanner2019} compared and contrasted inelastic fluids with rate-dependent viscosity, materials with a yield stress, as well as viscoelastic fluids - highlighting the need for rheological modeling improvement, possibly with multiple relaxation times. Additionally, he concludes that several aspects of suspension rheology, such as roughness, ionic forces, particle shape, and polydispersity all need to be addressed. Finally, \citet{Tanner2019} also reported experimental results for steady viscometric flows, unsteady shear flows and uniaxial elongational flows. However, good agreement between computation and experiment is scarce, because there are, as yet, few computational studies which allow careful comparison with experimental data, further emphasizing that progress in rheological modelling and improved computational methods are needed. In a recent perspective, \citet{Shaqfeh2019} notes that the foundations for the development of suspension mechanics in viscoelastic fluids, as well as the development of computational methods to accurately simulate with particle-level non-Brownian suspensions, have been established. Nevertheless, numerous unanswered questions remain, including the rheological behavior of these suspensions for different matrix fluid rheologies, particle shapes, deformability, flow histories, etc. All these questions can be addressed, in principle, by employing theoretical/computational frameworks to systematically explore the coupling between the kinematics and momentum distribution of the fluid phase and the resulting evolution of the dispersed particulate phase. \citet{Shaqfeh2021} performed 3D transient simulations of the bulk shear rheology of particle suspensions in Boger fluids for a range of $Wi \leq 6$ and finite strains and calculated the per-particle extra viscosity of the suspension. They categorize the per-particle viscosity calculations as contributions from either the particle-induced fluid stress (PIFS) or stresslet contributions. It was concluded that in the dilute limit, the PIFS increases monotonically with shear strain; however, the stresslet contribution shows a non-monotonic evolution to steady state at large $Wi$. The total combined per-particle viscosity contribution, however, shows a monotonic evolution to steady state. Additionally, \citet{Shaqfeh2021} performed multiple-particle simulations using the IB method to examine the effect of particle-particle hydrodynamic interactions on the per-particle viscosity calculation. It was concluded from transient immersed boundary simulations that the steady values of per-particle viscosity increase with $\phi$, but the per-particle contribution to the primary normal coefficient was independent of $\phi$ (up to 10\% particle volume fraction) at the two values of Weissenberg number investigated ($Wi$ = 3 and 6).

The nonlinear interactions of fluid inertia with viscosity and elasticity cause unexpected phenomena (e.g. negative wakes, shear-induced migration/chaining) in the dynamics of particles suspended in a viscoelastic matrix \cite{Avino2012,Loon2014,Jaensson2016}. These interactions may be expected to change the evolution in the viscoelastic drag correction factor with Weissenberg number. Extensive research efforts over the past 20 years have been directed at the elucidation of the role of fluid rheology and wall effects on the drag of a sphere and the wake developed behind it. Excellent reviews are available in the literature \cite{Caswell2004, walters1992, mckinley2002steady}. There have been a number of computational studies investigating the effect of fluid rheology on the motion of the sphere; however, a common limitation is that results are not convergent for Weissenberg numbers beyond a certain critical limiting value (i.e. typically $Wi_c\approx$ 2 or 3) \cite{Caswell2004}. In the work we extend viscoelastic suspension flow calculations up to $Wi=4$, by employing the log-conformation approach \cite{Fattal2004,Fattal2005,Habla2014,Francisco2017}.

Our previous work \cite{Salah2019} proposed, for the first time, a closure model for the viscoelastic drag coefficient of a single sphere translating in a quasi-linear viscoelastic fluid, which can be well described by the Oldroyd-B constitutive equation. In the present work, we extend the proposed model in order to be able to describe moderate volume fraction viscoelastic suspensions ($\phi\leq 0.2$), which are commonly encountered in a wide range of industrial operations. We focus on non-colloidal suspensions with Newtonian and viscoelastic fluid matrices, and the net effect of other particles in the flow is studied by computing an effective average drag force acting on a particle \cite{TannerHousiadas2014}. 

For Newtonian matrices, \citet{Brinkman1947} presented a modification of Darcy's equation in porous media, in which the viscous force exerted on a dense suspension of rigid particles by the Newtonian fluid is calculated. The main idea is that, from the point of view of a single particle, the other distributed particles effectively act as a porous or Darcy medium. \citet{Durlofsky1987} performed numerical simulations to compare against the predictions of Brinkman's model; however, they only obtained good agreement with the analytical solution of Brinkman for very small volume fractions up to $\phi=5\%$, emphasizing the importance of including the configurational effects of more distant particles. The reason that the Brinkman approach starts to break down at what appears to be a rather low volume fraction is related to the fact that at $\phi=5\%$ a characteristic inter-particle spacing is only slightly larger than four particle radii; so that neighboring particles are in fact quite close together and hydrodynamically interact. For dense random arrays of spheres ($\phi \ge 40\%$) the empirical Carman-Kozeny (CK) relation \cite{Carman1937} is found to describe well the drag force exerted by the fluid flow in the dispersed phase. The idea behind the CK relation is that the suspended medium can be considered as a system of tortuous channels, from which the pressure drop across the porous medium is calculated using the Darcy equation \cite{Darcy1937}. In the present work, we report results for volume fractions of $\phi=~$ 4\%, 8\%, 12\%, 16\% and 20\%, representative of semi-dilute non-colloidal suspension behaviour \cite{Tanner2013}. 

Additionally, we note that fully-resolved particle-laden viscoelastic solvers \cite{Fernandes2019,Shaqfeh2020} are presently only able to directly resolve $O(10^3)$ particles \cite{Hager2014}, which limits their application to large industrial case studies \cite{Steven2020}. To overcome this limitation we implement an Eulerian-Lagrangian viscoelastic solver ($DPMviscoelastic$), which employs the closure drag model that we develop here for moderately dense suspensions with a viscoelastic matrix fluid to quantify the momentum exchange between the two constituent phases (a moderate volume fraction of rigid spherical particles and a non-shear thinning viscoelastic matrix fluid).
  
The present paper describes the simulation method employed to measure the drag force on randomly-dispersed particle arrays immersed in viscoelastic fluids that can be described by the quasi-linear Oldroyd-B constitutive equation, which predicts constant values of the shear viscosity and first normal stress coefficient. The extension of this work to more complex viscoelastic matrix-based fluids, for example models predicting shear-thinning fluid behavior (such as the Giesekus model), is also currently being studied. In this case the magnitude of the stresses is typically smaller and easier to resolve computationally, but the dimensionality of the problem is higher due to the additional nonlinear model parameter(s) required to characterize the shear-thinning. To accomplish the required developments, an open-source library, OpenFOAM \cite{openfoam2019}, was modified to be able to calculate the average drag force acting on random particle arrays. The latter information is then used to formulate a new drag force correlation for the creeping flow of an Oldroyd-B fluid through randomly distributed arrays of spherical particles, with solid volume fractions $0<\phi\leq 0.2$, over a range of Weissenberg numbers $(Wi \leq 4)$. To ensure stability at high $Wi$, the polymer stress contribution is computed using the log-conformation formulation \cite{Fattal2004, Fattal2005, Habla2014, Francisco2017}. Finally, to the best of the authors' knowledge, the current work presents for the first time an Eulerian-Lagrangian solver, in which the fluid continuum phase has viscoelastic rheology and the dynamics of the particulate phase are computed following a discrete particle method (DPM). The momentum transfer between both phases is calculated using the drag force law proposed in the present work (see Section~\ref{sec:viscoelasticDrag}). The newly-developed $DPMviscoelastic$ solver is employed to predict particle settling effects in rectangular channels (which mimics a vertical fissure such as those encountered in gas/oil extractions during hydraulic fracturing operation) and in an annular pipe (a model of pumped transport of a suspension along a drill string during horizontal drilling operations). 

The paper is organized in the following manner: in Section 2 we present the governing equations and numerical method used to compute the solution of the appropriate balance equations for the viscoelastic fluid flows considered in this work. Section 3 provides the details of the simulation methodology used to compute the drag force values exerted by the fluid on the particulate phase. In Section 4 these results are verified for Stokes flow of a particle suspension dispersed in a viscous Newtonian fluid. The results are then used to derive a new closure law for the average fluid-particle drag force acting on random arrays of spheres immersed in an Oldroyd-B fluid. Section 5 is dedicated to description of the development of the $DPMviscoelastic$ solver for up-scaled three-dimensional simulations of particle-laden viscoelastic flows, in which the dispersed phase is modeled by the discrete particle method. Additionally, we illustrate the capability of the newly-developed code to solve challenging physical problems, specifically two canonical proppant transport problems, which are commonly encountered in hydraulic fracturing operations. Finally, in Section 6, we summarize the main conclusions of this work.

\section{Governing equations and numerical method}

\subsection{Governing equations}
\label{subsec:mathematicalmodel}

Following the work of \citet{Salah2019}, as a first step, we consider the problem of moderately dense ($0<\phi\leq 0.2$) suspensions constituted from a continuum viscoelastic matrix fluid and a monodisperse static random array of rigid spheres. For the continuum fluid phase the familiar Oldroyd-B constitutive model was chosen, representing an elastic fluid with a constant shear viscosity, which has been shown by \citet{Dai2020} to fairly well describe the response of highly elastic Boger fluid suspensions in steady shear and uniaxial elongation. By adopting the Oldroyd-B model, we confine the dimensionality of the viscoelastic fluid calculations to only two degrees of freedom (i.e. the relaxation and retardation times, or equivalently). However, the consideration of moderately dense suspensions increases the dimensionality of the problem to four degrees of freedom, due to the addition of two more variables; the particle volume fraction present in the suspension and the number of random particle configurations studied (to obtain statistical significance in the DNS results). Thus, for the current study, we have also fixed the retardation time of the fluid at $\beta=\eta_S/(\eta_S+\eta_P)=\eta_S/\eta_0=1/2$, where $\eta_0$ is the total matrix fluid viscosity, with $\eta_S$ and $\eta_P$ being the solvent and polymeric viscosities, respectively. Within these constraints, the drag correction expression developed in this study will form a foundation for higher-dimensional parameterizations, which should also consider a range of solvent viscosities as well the effect of more complex fluid rheology (e.g. shear-thinning) on random arrays of particles in viscoelastic fluids, by using machine learning algorithms such as convolutional neural networks to capture the non-linear effects of all constitutive parameters on the resulting drag coefficient expressions acting on the particle arrays.

The dimensionless conservation equations governing transient, incompressible and isothermal laminar flow of an Oldroyd-B fluid are given by
\begin{eqnarray}
\nabla \cdot \textbf{\~u} = 0,
\label{eqn:continuitydimensionless}
\end{eqnarray}
\begin{eqnarray}
Re_a\left(\frac{\partial\textbf{\~u}}{\partial \textit{\~t}} + \textbf{\~u}\cdot\nabla\textbf{\~u}\right)-\nabla^2\textbf{\~u} = -\nabla \textit{\~p} -\nabla\cdot\left[(1-\beta) \nabla\textbf{\~u}\right] + \nabla \cdot \tilde{\boldsymbol\tau_P},
\label{eqn:momentumdimensionless}
\end{eqnarray}
\begin{eqnarray}
\tilde{\boldsymbol\tau_P} + Wi \left(\frac{\partial\tilde{\boldsymbol\tau_P}}{\partial \textit{\~t}} + \textbf{\~u} \cdot \nabla \tilde{\boldsymbol\tau_P} - \tilde{\boldsymbol\tau_P} \cdot \nabla \textbf{\~u} - \nabla \textbf{\~u}^T \cdot \tilde{\boldsymbol\tau_P} \right) = (1-\beta) \left(\nabla \textbf{\~u} + \nabla \textbf{\~u}^T\right),
\label{eqn:oldroydBeqdimensionless}
\end{eqnarray}
where the following dimensionless quantities are used
\begin{eqnarray}
\textbf{\~x}=\frac{\textbf{x}}{L},~\textbf{\~u}=\frac{\textbf{u}}{U},
~\textit{\~t}=\frac{U}{L}t,~\textit{\~p}=\frac{L}{\eta_0 U}p,
~\tilde{\boldsymbol\tau}_P=\frac{L}{\eta_0 U}\boldsymbol\tau_P,
\end{eqnarray}
with $L$ and $U$ being the characteristic length and velocity values, respectively, $\textbf{x}$ the position vector, $\textbf{u}$ the velocity vector, $t$ the time, $p$ the pressure and $\boldsymbol \tau_P$ the polymeric contribution to the extra-stress tensor. As mentioned above, the retardation ratio is fixed with the value of $\beta=0.5$ for all the calculations performed in this work.

For the present problem with $L=a$, where $a$ is the radius of a single suspended particle, and $U$ the average fluid velocity at the inlet of the channel, we define the Reynolds and Weissenberg numbers as follows,
\begin{subequations}
\begin{align}
\label{eqn:reynolds}
Re_D = 2Re_a = \frac{2 a\rho U}{\eta_0},\\
\label{eqn:weissenberg}
Wi = \frac{\lambda U}{a},
\end{align}
\end{subequations}
where $\rho$ is the fluid density and $\lambda$ is the relaxation time. Notice that for the case of a Newtonian fluid flow $\lambda=0$ and $\eta_0 = \eta_S$.  

To ensure computational stability over a wide range of fluid elasticities, including suspension flows at high Weissenberg number, we incorporate the log-conformation approach for calculating the polymeric extra-stress tensor. In the present work, we follow the implementation of the log-conformation approach in the OpenFOAM computational library \cite{openfoam2019}, presented in \citet{Habla2014} and \citet{Francisco2017}. Details on the mathematical formulation behind the log-conformation approach can be found in the original works of \citet{Fattal2004,Fattal2005}.

\subsection{Numerical method}

The equations presented in Section~\ref{subsec:mathematicalmodel} are discretized using the finite-volume method (FVM) implemented in the OpenFOAM framework \cite{openfoam2019}. 

Pressure-velocity coupling was accomplished using segregated methods, in which the continuity equation is used to formulate an equation for the pressure, using a semi-discretized form of Eq.~(\ref{eqn:continuitydimensionless}) \cite{Ferziger1995}. The resulting equation set is solved by a segregated approach, using the SIMPLEC (Semi-Implicit Method for Pressure-Linked Equations-Consistent) algorithm \cite{Doomaal1984}, which does not require under-relaxation of pressure and velocity (except for non-orthogonal grids, where the pressure needs to be under-relaxed \cite{Francisco2017}). Additionally, the computational cost per iteration of this algorithm is lower than in the PISO (Pressure-Implicit Split Operator) algorithm \cite{Issa1986}, because the pressure equation is only solved once per cycle. The coupling between stress and velocity fields is established using a special second-order derivative of the velocity field in the explicit diffusive term added by the iBSD (improved both-sides diffusion) technique \cite{Fernandes2017}. The velocity gradient is calculated using a second-order accurate least-squares approach, and the diffusive term in the momentum balance is discretized using second-order accurate linear interpolation. For non-orthogonal meshes the minimum correction approach is used, as explained in \citet{Jasak1996}, in order to retain second-order accuracy. The advective terms in the momentum and constitutive equations are discretized using the high-resolution scheme CUBISTA \cite{Alves2003} following a component-wise and deferred correction approach, enhancing the numerical stability. The time derivatives are discretized with the bounded second-order implicit Crank-Nicolson scheme \cite{Crank1947}. Here, a Poisson-type equation for the pressure field is solved with a conjugate gradient method with a Cholesky preconditioner, and the linear systems of equations for the velocity and stress are solved using BiCGstab with Incomplete Lower-Upper (ILU) preconditioning \cite{Lee2003, Jacobs1980, Ajiz1984}. The absolute tolerance for pressure, velocity and stress fields was set as $10^{-10}$. The simulations are performed including transient terms, but the time marching is used only for relaxation purposes as we will just be looking for the steady-state solution, i.e., when the drag coefficient ceased to vary in the third decimal place.

\section{Simulation methodology}
\label{sec:simmeth}

\begin{figure}[H]
\centering
\includegraphics[scale=0.4]{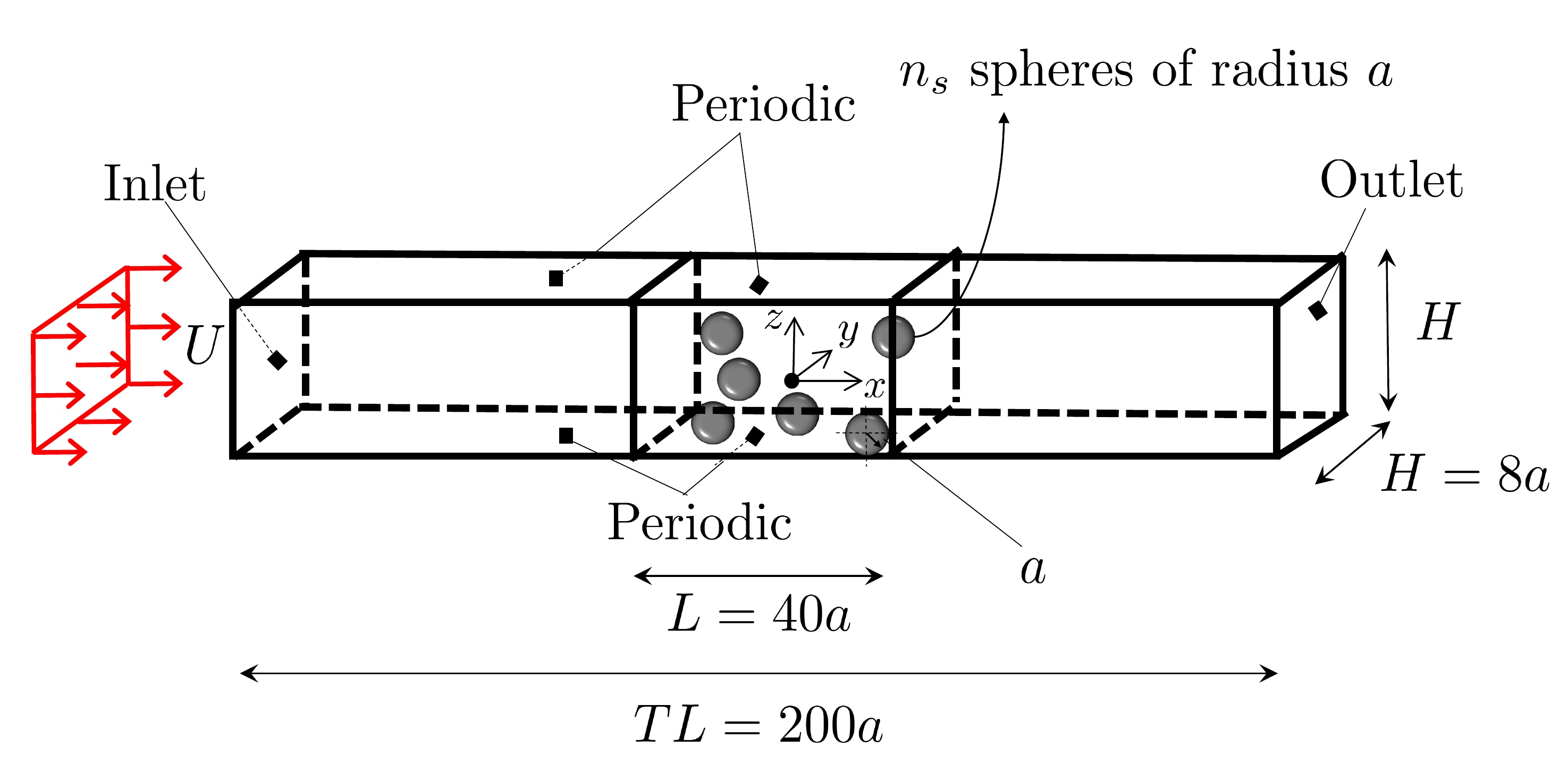}
\caption{Schematic of the channel cross-section used for DNS of random arrays of spheres immersed in Newtonian and quasi-linear Oldroyd-B viscoelastic fluids. The particle volume fractions considered include $\phi = 0.04,~0.08,~0.12,~0.16$ and $0.20$.} 
\label{fig:DNSChannel}
\end{figure}

In order to develop our computational methodology, we address only non-colloidal suspensions with viscoelastic matrices, and focus on both dilute and moderately dense suspensions. Following \citet{TannerHousiadas2014}, we consider the effect of other particles in the flow by assuming that from the point of view of a single particle at any instant, the remaining particles act effectively like a porous medium \cite{Brinkman1947}. Fig.~\ref{fig:DNSChannel} illustrates schematically the computational domain, which is used here to simulate the steady-state flow of unbounded viscoelastic fluids around random arrays of spheres. The porous media considered in the present work have volume fractions $0<\phi\leq 0.2$, with particles that are randomly distributed in a box of square cross-section with dimensions $L\times H \times H$, by means of a constrained approach to prevent particle-particle overlap. The computational domain has a total length of $TL = 200a$ and a square cross-section with height $H=8a$. For Newtonian fluids, the applied boundary conditions are fixed velocity, $\mathbf{u}_{in}=(U,0,0)$, and zero pressure gradient at the inlet. Periodic boundary conditions are applied for the front, back, top and bottom boundaries. At the outlet we enforce zero velocity gradient and zero reference pressure. Finally, on each of the sphere surfaces a no-slip velocity condition is applied. For the viscoelastic calculations the applied boundary conditions for velocity and pressure are the same as those listed for the Newtonian fluid, and for the polymeric extra-stress components, we also apply zero gradient boundary condition at the outlet, periodic boundary conditions at the front, back, top and bottom boundaries, zero stress at the inlet, and linear extrapolation, from the adjacent fluid cell centroid \cite{Francisco2017}, at the sphere surface.

The number of spherical particles is chosen such that the solid volume fraction in the simulation $\phi = \frac{n_s (4/3) \pi a^3}{LH^2}$ is as close as possible to the desired packing fraction, where $n_s$ is the number of spheres enclosed in the square duct (whose volume is $LH^2$) and $a$ is the sphere radius. For the purpose of simulating the proppant transport phenomena that occur during hydraulic fracturing operations, we consider in this work moderately dense suspensions where the particle volume fraction range is $0<\phi\leq 0.2$ \cite{barbati2016complex,Steven2020,Chris2018}. Following \citet{Hill2001} we note that the system to be studied must include a sufficient number of spheres, $n_s$, to minimize artifacts and statistical oscillations coming from the finite size of the computational domain. In practice, $n_s$ is  chosen  to  be  large  enough  to  avoid  periodic  artifacts (typically $24\leq n_s \leq 122$, see Table~\ref{tab:newtonian} in Section~\ref{sec:newtonianStokes}), and statistical uncertainty  is  reduced  by  ensemble  averaging  the  results  from $n_c$ random sphere configurations (in this work $n_c=5$ was found to be sufficiently large to obtain a standard error for the average below $5\%$ of its actual value, as shown in Section~\ref{sec:DNS}, Tables~\ref{tab:newtonian} and \ref{tab:oldroydB}). 

The numerical model employed in this work was comprehensively tested against a similar computational challenge (see \citet{Salah2019} for the bounded and unbounded flow past a single sphere in a fluid described by the Oldroyd-B constitutive model). The meshes employed in this work have the same level of mesh refinement as the most refined mesh (M1) used by \citet{Salah2019}, which resulted from a grid refinement study. 

In all cases, the magnitude of the fluid velocity imposed at the inflow was such that the Reynolds number based on the particle diameter $D$, Eq.~(\ref{eqn:reynolds}), was equal to $Re_D=0.05$, representative of creeping flow conditions. At this point we should note that there exists some ambiguity in the literature \cite{Hoef2005} on the proper definition of the drag force, specifically, if the pressure gradient term should contribute to the drag force or not. It is known that the two definitions differ by a factor $(1-\phi)$, i.e., the relation between the total average force that the fluid exerts on each particle, $\textbf{F}_t$, and the drag force, $\textbf{F}_d$, which results from the friction between the particle and the fluid at the surface of the particle, is $\textbf{F}_t=\textbf{F}_d/(1-\phi)$. Note that in some literature (\citet{Hill2001}), the total force on the particle is defined as the drag force. In this work, the results will be presented in terms of an average dimensionless drag force, $\langle F \rangle$, which is defined as
\begin{eqnarray}
\langle F \rangle = \frac{\langle\textbf{F}_d\rangle\cdot\mathbf{e_\textit{x}}}{6\pi \eta_0 a \overline{u}}=\frac{(1-\phi)\langle\textbf{F}_{t,i}\rangle\cdot\mathbf{e_\textit{x}}}{6\pi \eta_0 a \overline{u}},
\label{eq:dragforce}
\end{eqnarray}
where $\overline{u}=(1-\phi)U$ is the superficial fluid velocity, $\langle\textbf{F}_{t,i}\rangle$ is the average drag force on the random array of spheres computed as $\langle\textbf{F}_{t,i}\rangle=\frac{1}{n_s}\displaystyle\sum_{i=0}^{n_s}\textbf{F}_{t,i}$, where $\textbf{F}_{t,i}$ is the drag force on sphere $i$ from an ensemble of $n_s$ spheres, and $\mathbf{e_\textit{x}}$ is the unit vector in the $x$-direction. The denominator on the right-hand side of Eq.~(\ref{eq:dragforce}) is the Stokes drag force, obtained in the limit of infinite dilution and when inertial effects can be neglected \cite{Stokes1851}. In this work, the uncertainty in the computed average force, referred as the standard error, is calculated from
\begin{eqnarray}
\Delta \langle F\rangle = \sqrt{\frac{\displaystyle\frac{1}{n_c}\displaystyle\sum_{i=1}^{n_c}\left(\langle F\rangle_i-\overline{\langle F\rangle_i}\right)^2}{n_c-1}}.
\label{eq:standarderror}
\end{eqnarray}
Note that the factor $n_c-1$ in the denominator on the right-hand side of Eq.~(\ref{eq:standarderror}) corrects for the fact that there are $n_c-1$ degrees of freedom, since the average is used to calculate the variance in the numerator \cite{Bevington1992}. This is important since the number of random configurations, $n_c$, used to calculate the average of $\langle F \rangle$ is small. 

\section{Results: simulation of flow in random particle arrays}
\label{sec:DNS}

\subsection{Verification: Stokes flow of suspensions with Newtonian fluid matrices}
\label{sec:newtonianStokes}

In this section the creeping flow of random arrays of spheres surrounded by a Newtonian fluid is studied. This way the simulation methodology presented in Section~\ref{sec:simmeth} can be verified against results found in the literature. 

One of the earliest drag force models for describing Stokes flow through an array of spherical particles is the \citet{Carman1937} relation,
\begin{eqnarray}
\langle F \rangle = \frac{10\phi}{(1-\phi)^2}.
\label{eq:carman}
\end{eqnarray}
This relation is only valid for dense arrays $((1-\phi)\ll 1)$, which can be seen by the fact that it does not have the correct limit $\langle F \rangle\to 1$ for $\phi\to 0$. For the limit of dilute systems, \citet{Kim1985} derived a closed-form expression for $\langle F\rangle$:
\begin{eqnarray}
\langle F \rangle = (1-\phi)\left(1 + \frac{3}{\sqrt{2}}\phi^{1/2} + 16.456\phi + \frac{135}{64}\phi~\text{ln}~ \phi + O(\phi^{3/2})\ldots\right).
\label{eq:kim}
\end{eqnarray}
The computational results obtained by \citet{Hill2001}, using Lattice-Boltzmann simulations, were found to be in very good agreement with Kim and Russel's~\cite{Kim1985} drag force expression (Eq.~(\ref{eq:kim})) for dilute arrays of particles $(\phi \le 0.1)$. 

\begin{figure}[H]
\centering
\includegraphics[width=0.8\textwidth]{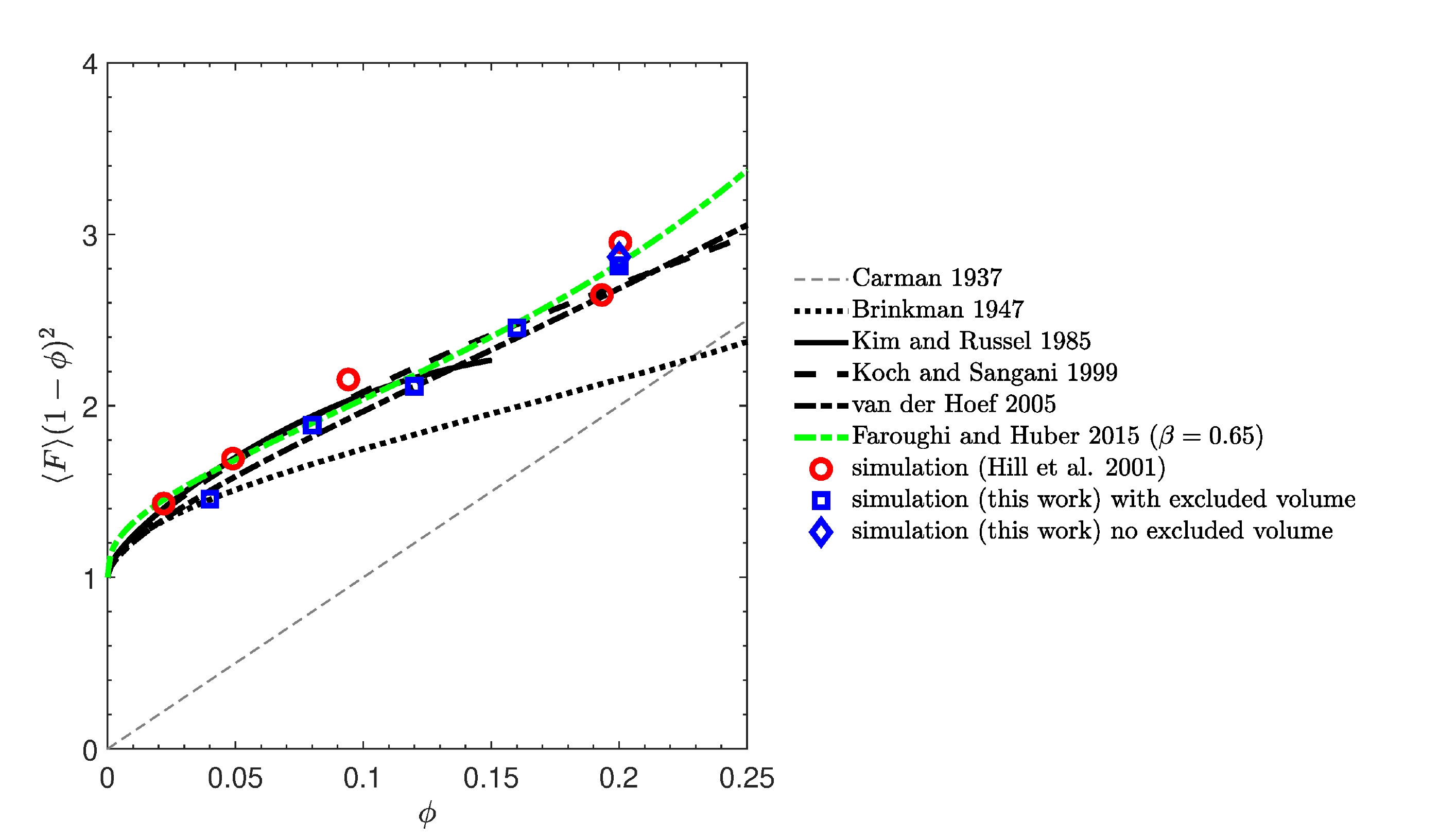}
\caption{The average value of the dimensionless drag force (multiplied by the porosity squared) for creeping flow of a Newtonian fluid past an array of spheres as a function of the packing fraction $\phi$ for $n_c=5$ different configurations. The symbols represent the simulation data, from this work (squares and diamond) and from \citet{Hill2001} (circles). Also shown are the correlations by \citet{Carman1937} (grey line), \citet{Brinkman1947} (dotted line), \citet{Kim1985} (solid line), \citet{Koch1999} (dashed line), \citet{Hoef2005} (dotted-dashed line) and \citet{faroughi2015unifying} (green line).} 
\label{fig:newtonian}
\end{figure}

Subsequently, several expressions have been developed to find an accurate drag force model that is valid over the full solid fraction range. Using a modification of the Darcy equation \cite{Darcy1937}, \citet{Brinkman1947} derived the well-known drag force model,   
\begin{eqnarray}
\langle F \rangle = (1-\phi)\left(1 + \frac{3}{4}\phi\left(1-\sqrt{\frac{8}{\phi}-3}\right)\right)^{-1}.
\label{eq:brinkman}
\end{eqnarray}
\citet{Koch1999} proposed the following expression for the drag force:
\begin{eqnarray}
\langle F\rangle = \begin{cases} 
	\displaystyle\frac{(1-\phi)\left(1+\frac{3}{\sqrt{2}}\phi^{1/2} + 16.456\phi + \frac{135}{64}\phi~\text{ln}~\phi\right)}{1 + 0.681\phi - 8.48\phi^2 + 8.16\phi^3} & \text{for } \phi\leq 0.4\\
	\displaystyle\frac{10\phi}{(1-\phi)^2} & \text{for } \phi\geq 0.4,
	\end{cases}
\label{eq:koch}
\end{eqnarray}
which for low solid volume fraction is equal to Eq.~(\ref{eq:kim}) to $O(\phi~\text{ln}~\phi)$, whereas for large solid volume fractions is the drag force given by the Carman expression Eq.~(\ref{eq:carman}). Finally, \citet{Hoef2005} presented a best fit to simulation data obtained using a Lattice-Boltzmann method (for $\phi \le 0.6$), which takes the following simple form:
\begin{eqnarray}
\langle F \rangle = \frac{10\phi}{(1-\phi)^2}+(1-\phi)^2(1+1.5\sqrt{\phi}),
\label{eq:hoef}
\end{eqnarray}
which is the Carman expression with a correction term for the limiting case of $\phi\to 0$. Recently, \citet{faroughi2015unifying} predicted the correction to the drag coefficient for monosized spherical particles as:
\begin{eqnarray}
\langle F \rangle = \left(\frac{1-\displaystyle\frac{\phi}{\phi_m}}{1-\phi}\right)\times\displaystyle\left(\frac{1-\displaystyle\frac{\phi}{\phi_m}}{1-\phi}\right)^{\displaystyle\frac{-2.5\phi_m}{1-\phi_m}}\times\left[1-\frac{3}{2}\beta\left(\frac{\phi}{\phi_m}\right)^{1/3}+\frac{\beta^3}{2}\frac{\phi}{\phi_m}\right]^{-1},
\label{eq:salah}
\end{eqnarray}
where $\phi_m$ denotes the maximum random close packing fraction, which, for monosized spherical particles, we take to be $\phi_m\approx 0.637$ \cite{Boyer2011}, and $\beta$ is a geometrical proportionality constant, which is related to the shape of the streamtube in the real flow field. The value of $\beta = 0.65$ provides the best fit to numerical simulations shown in Fig.~\ref{fig:newtonian}. The first term in Eq.~(\ref{eq:salah}) accounts for a vertical drag correction due to the reduction in drag even at low Reynolds number which is expected owing to the fact that particles aligned with gravity experience significant acceleration due to viscous forces (Smoluchowski effect). The second term in Eq.~(\ref{eq:salah}) accounts for the change in dynamic viscosity of the suspensions due to the existence of a cloud of particles inside the medium. Finally, the last term in Eq.~(\ref{eq:salah}) accounts for the horizontal drag correction resultant from the hindrance associated with the return flow of ambient fluid in a bounded system.

Fig.~\ref{fig:newtonian} shows our finite-volume simulation results for the dimensionless drag force $\langle F \rangle$ in a random array of spheres, immersed on a Newtonian fluid, at solid volume fractions up to $\phi=0.2$. Additionally, the results of \citet{Hill2001} obtained with lattice-Boltzmann simulations, the \citet{Carman1937} (Eq.~(\ref{eq:carman})), \citet{Kim1985} (Eq.~(\ref{eq:kim})), \citet{Brinkman1947} (Eq.~(\ref{eq:brinkman})), \citet{Koch1999} (Eq.~(\ref{eq:koch})), \citet{Hoef2005} (Eq.~(\ref{eq:hoef})) and \citet{faroughi2015unifying} (Eq.~(\ref{eq:salah})) expressions are also represented in Fig.~\ref{fig:newtonian} for comparison. For all the simulated particle volume fractions, the dimensionless drag force $\langle F \rangle$ obtained from our numerical algorithm is similar (within an average error value of 5\% and a maximum error value of 8.9\%) to both the theories for dilute and semi-concentrated suspensions and with the Lattice-Boltzman numerical results of \citet{Hill2001}. Notice that up to moderate volume fractions, the dimensionless average drag force computed for the case in which the locations of all of the spheres are in the interior of the channel region (i.e. where we take into account an excluded volume region provided by rigid bounding walls) is similar (within 2\%) to the results obtained when we allow the surface of the individual spheres to overlap the periodic channel boundaries, i.e. where we neglect the excluded volume provided by rigid bounding walls (see results in Fig.~\ref{fig:newtonian} and Table~\ref{tab:newtonian} for $\phi=0.2$). For a more detailed discussion regarding the excluded volume provided by impenetrable channel side-walls refer to Appendix A.

The sphere volume fractions, the number of spheres used on each simulation, the number of mesh points, the dimensionless average drag force $\langle F \rangle$ on the spheres and the respective standard errors are listed in Table~\ref{tab:newtonian}. In all cases, the standard errors of $\langle F \rangle$, which measure the statistical accuracy achieved by averaging the results with five random configurations, were below $3.5\%$ of the average.
\begin{table}[H]
	\centering
  \begin{threeparttable}
	\scriptsize
	\caption{Parameters used for the creeping flow of a Newtonian fluid past random arrays of spheres. $\Delta \langle F \rangle$ is the standard error in the average of $\langle F \rangle$.}
	\centering
    \begin{tabular}{lrrccc}
    \toprule
$\phi$ &  $n_s$ & $n_c$ & Number of mesh points & $\langle F \rangle$ &  $\Delta \langle F \rangle/\langle F \rangle$  \\ 
\midrule
0.04 & 24 & 5 & 1341845 & 1.581 & 0.028  \\
0.08 & 49 & 5 & 1293067 & 2.231 & 0.030   \\
0.12 & 73 & 5 & 1249529 & 2.734 & 0.024  \\
0.16 & 98 & 5 & 1203460 & 3.485 & 0.015  \\
0.20 & 122 & 5 & 1158730& 4.402 & 0.035 \\
0.20$^*$ & 122 & 5 & 1199453 & 4.481 & 0.011 \\
\bottomrule
\end{tabular}
\scriptsize $^*$ includes particles located in the excluded volume region near the periodic channel walls
\label{tab:newtonian}
\end{threeparttable}
\end{table}

\begin{figure}[H]
\centering
\captionsetup[subfloatrow]{format = hang, labelfont = up, textfont = up}
\captionsetup{labelfont = up, textfont = up}
\ffigbox{%
\hspace{-4cm}
   \begin{subfloatrow}[1]
       \includegraphics[width=0.7\textwidth]{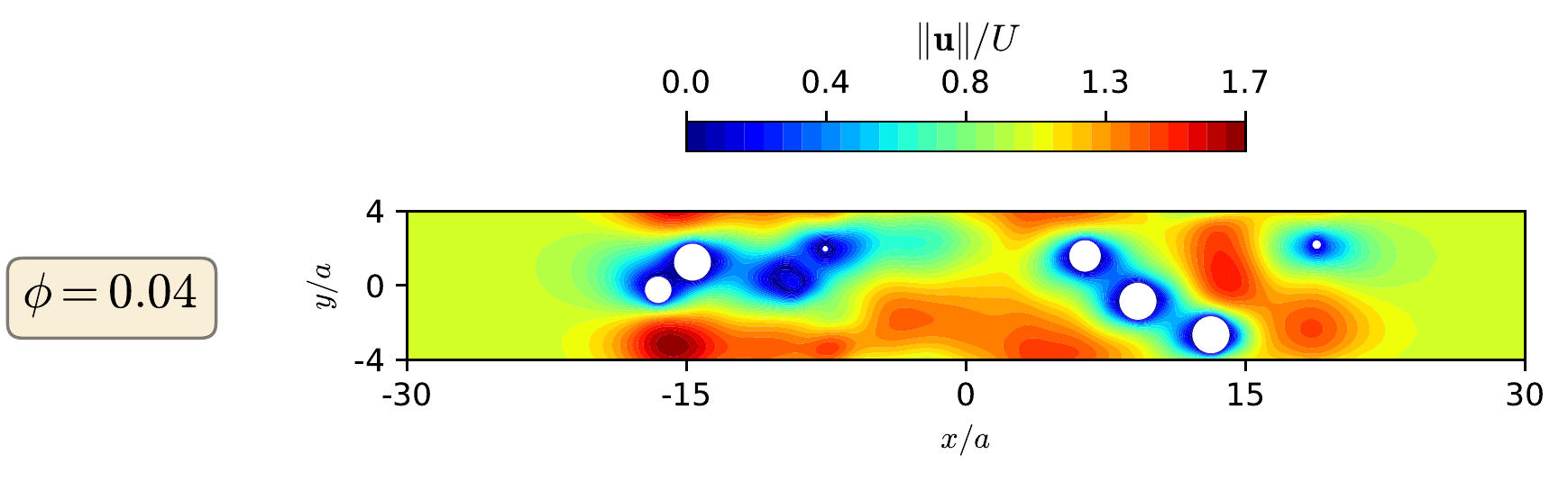}
       {\caption{}}
   \end{subfloatrow}\\
   \hspace{-4cm}
   \begin{subfloatrow}[1]
		\includegraphics[width=0.7\textwidth]{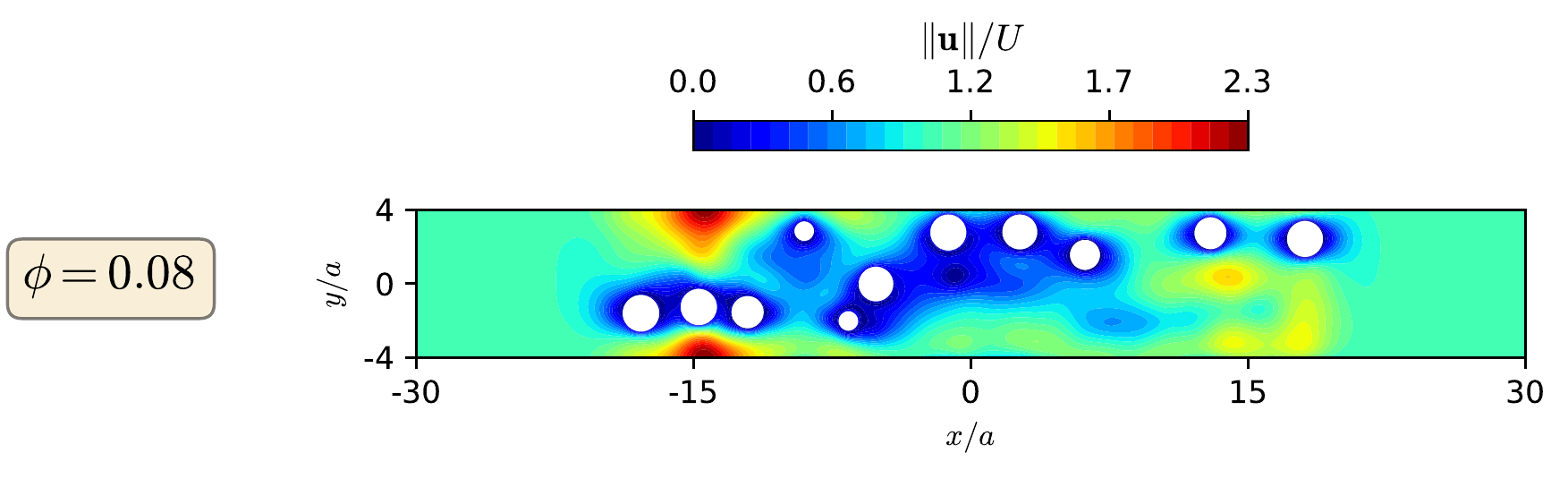}
       {\caption{}}
   \end{subfloatrow}\\ 
   \hspace{-4cm}
      \begin{subfloatrow}[1]
      	\includegraphics[width=0.7\textwidth]{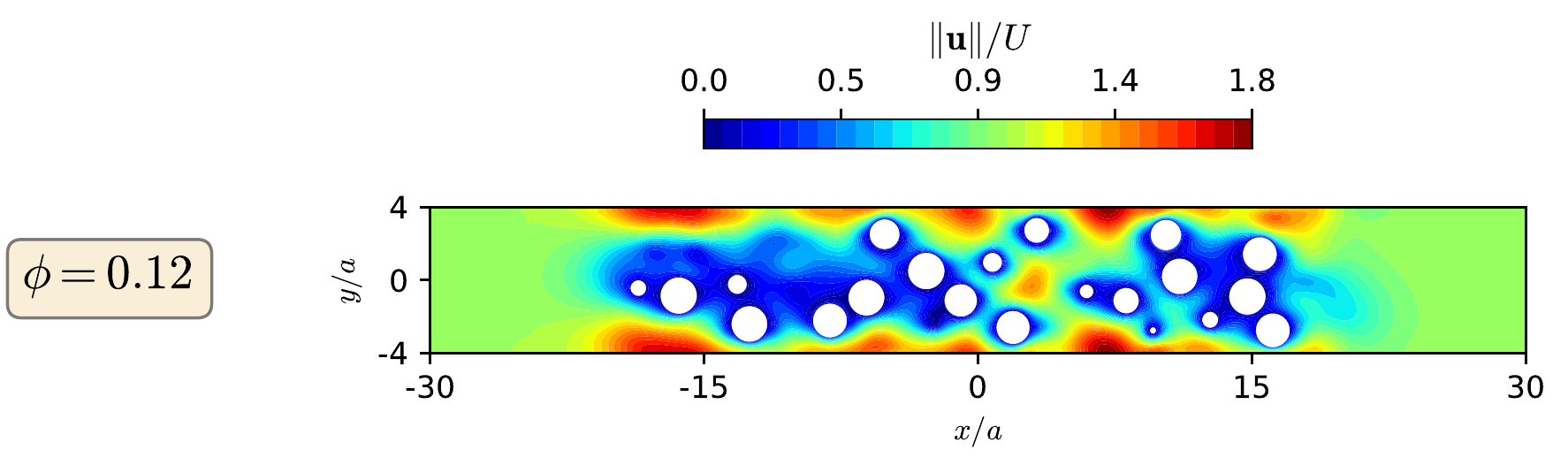}
       {\caption{}}
   \end{subfloatrow}\\
   \hspace{-4cm}
      \begin{subfloatrow}[1]
        \includegraphics[width=0.7\textwidth]{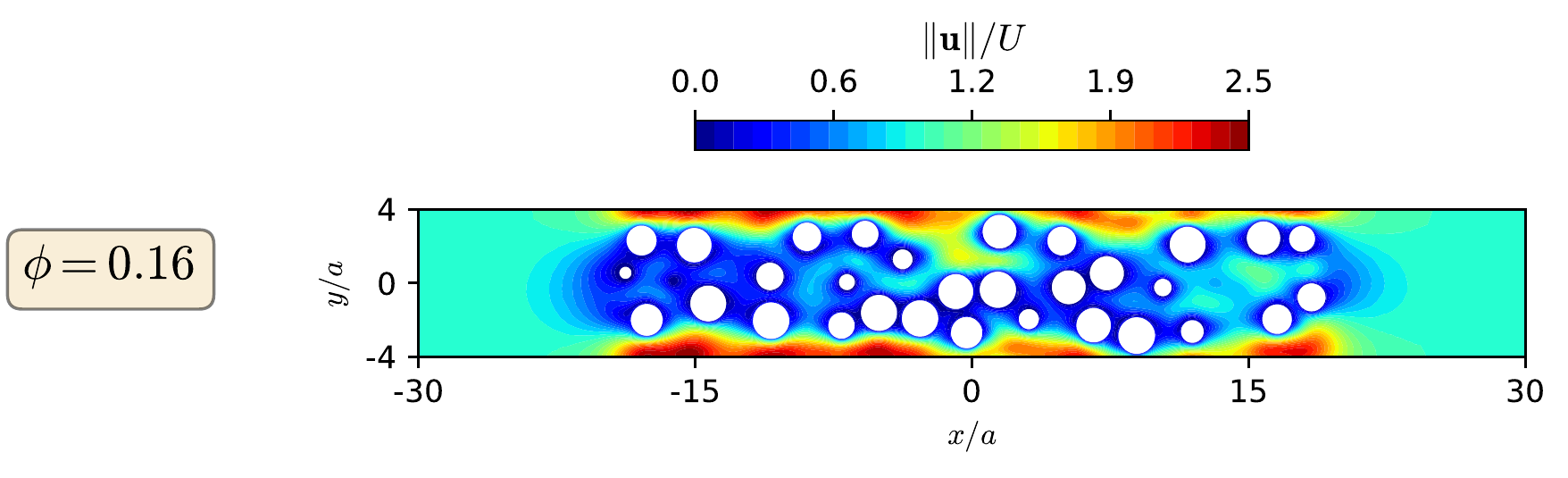}
       {\caption{}}
   \end{subfloatrow}\\
   \hspace{-4cm}
      \begin{subfloatrow}[1]
		 \includegraphics[width=0.7\textwidth]{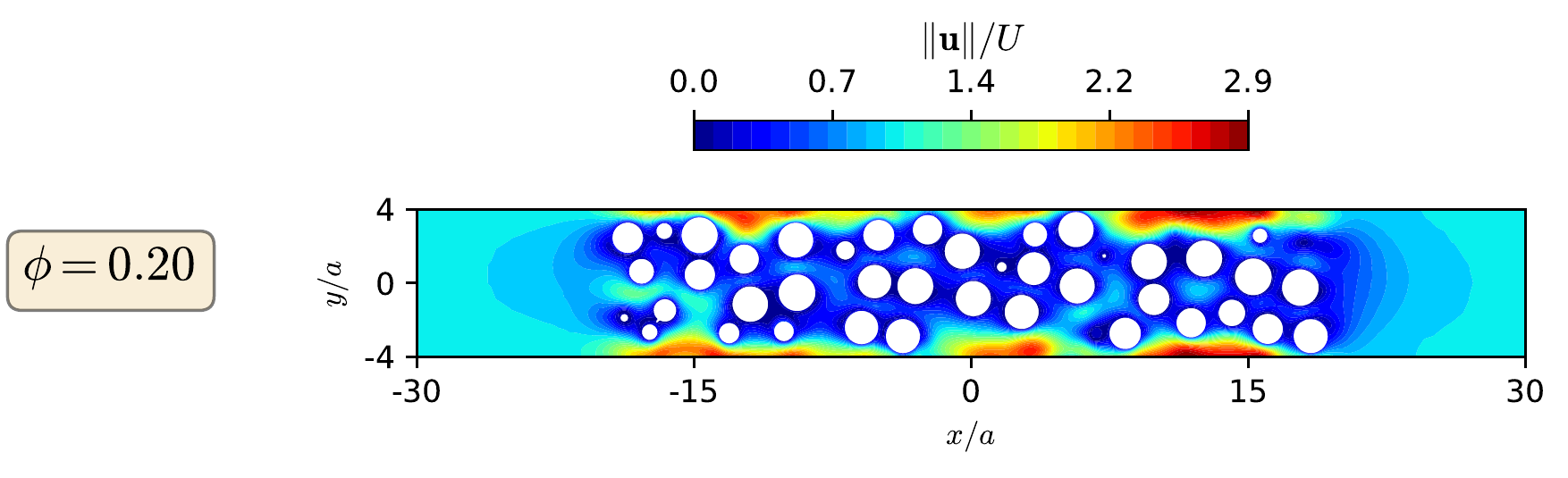}   
       {\caption{}}
   \end{subfloatrow}\\
      \hspace{-4cm}
      \begin{subfloatrow}[1]
          \includegraphics[width=0.7\textwidth]{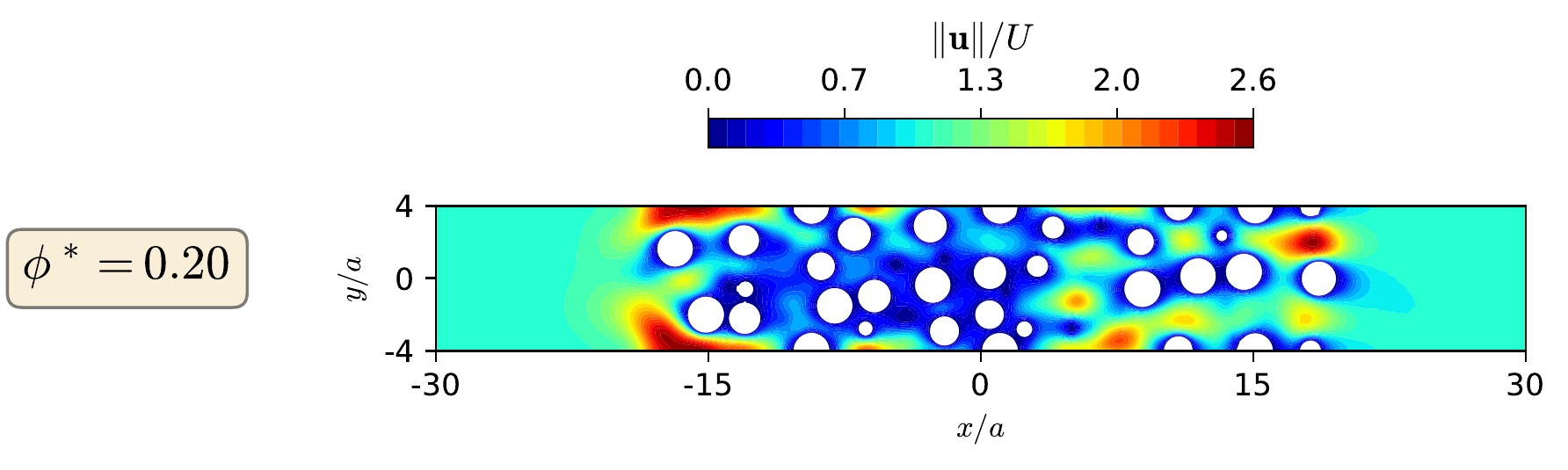}
       {\caption{}}
   \end{subfloatrow}\\
}{\caption{Cross-sections of the steady flow field around one representative random particle array in a channel filled with Newtonian fluid. Contours of the dimensionless velocity field are represented for each of the particle volume fractions used on the simulations, $\phi=0.04, 0.08, 0.12, 0.16$ and $0.2$, in the $x-y$ plane at $z=0$. The configuration denoted $\phi^*=0.20$ corresponds to the more realistic case of penetrable side-walls in which the randomly placed particles can also be located in the excluded volume region of thickness $a$ near each bounding side wall.} 
\label{fig:streamDe0}}
\end{figure}

In Fig.~\ref{fig:streamDe0} we show normalized axial velocity contours obtained from the numerical simulation of random particle arrays in a channel filled with a Newtonian fluid. As can be seen from the velocity contours, as we increase the particle volume fraction more of the fluid is forced to flow through the tortuous paths in the interstitial spaces between the spheres, rather than as a continuous fluid stream that is mildly perturbed by widely separated spheres. This is also visible by the higher magnitude of the fluid velocities near the channel walls where the fluid is squeezed. Notice that for the higher particle volume fraction employed $\phi=0.2$ we have also conducted simulations where particles can be located in the excluded volume region provided by the rigid bounding walls ($\phi^*=0.2$), and the dimensionless average drag force $\langle F \rangle$ remains similar (approximately 2\% higher) to the case with impenetrable side walls where the particles are located only in the central portion of the channel beyond an excluded volume region of thickness $a$ that is adjacent to the bounding side walls region (see Fig.~\ref{fig:newtonian} and Table~\ref{tab:newtonian}). Placing spheres uniformly throughout the entire domain results in a more uniform velocity profile across the channel cross-section.

\subsection{Computational study: drag force on spheres within an Oldroyd-B fluid}
\label{sec:viscoelasticDrag}

We performed finite volume simulations of viscoelastic creeping flows (with the Oldroyd-B constitutive equation) past fixed random configurations of particles at Weissenberg numbers up to $Wi=4$ and for solid volume fractions in the range $0<\phi\leq 0.2$. The numerical resolutions used in the simulations are comparable with those used by \citet{Salah2019}.

The dimensionless average drag force $\langle F(\phi,Wi) \rangle$ on the spheres (see Eq.~(\ref{eq:dragforce})) and the respective standard errors are listed in Table~\ref{tab:oldroydB} for the different kinematic conditions stated above. In all cases, the standard errors of $\langle F(\phi,Wi) \rangle$, which measure the statistical accuracy achieved by averaging the results with five random configurations, were below $4.7\%$ of the average drag force.
\begin{table}[H]
	\centering
  \begin{threeparttable}
	\scriptsize
	\caption{Parameters used for the creeping flow of an Oldroyd-B viscoelastic fluid past random arrays of spheres. $\Delta \langle F \rangle$ is the standard error in the average of $\langle F(\phi,Wi) \rangle$. For each case we consider $n_c=5$ configurations and keep $\eta_S/\eta_0=0.5$ and $Re_D=0.05$.}
	\centering
    \begin{tabular}{cccccccccccc}
    \toprule
$Wi$ & $\phi$ &  $\langle F \rangle$ &  $\Delta \langle F \rangle/\langle F \rangle$ & $Wi$ & $\phi$ &  $\langle F \rangle$ &  $\Delta \langle F \rangle/\langle F \rangle$ & $Wi$ & $\phi$ &  $\langle F \rangle$ &  $\Delta \langle F \rangle/\langle F \rangle$  \\ 
\midrule
\multirow{5}{*}{0.5} & 
0.04 &  1.577 & 0.028 & \multirow{5}{*}{1} & 
0.04 &  1.563 & 0.030 & \multirow{5}{*}{2} & 
0.04 &  1.701 & 0.035 \\

&0.08 &  2.209 & 0.029 & &0.08 &  2.210 & 0.030 & &0.08 &  2.274 & 0.031 \\

&0.12 &  2.685 & 0.024 & &0.12 &  2.826 & 0.031  & &0.12 &  2.896 & 0.025\\

&0.16 &  3.417 & 0.015 & &0.16 &  3.661 & 0.020 & &0.16 &  3.760 & 0.018 \\

&0.20 &  4.229 & 0.022 & &0.20 &  4.476 & 0.022 & &0.20 &  4.709 & 0.020 \\

\midrule

\multirow{5}{*}{3} & 
0.04 &  1.785 & 0.045 & \multirow{5}{*}{4} & 
0.04 & 1.849 & 0.047 &&&& \\

&0.08 &  2.415 & 0.033 & &0.08 & 2.619 & 0.034 &&&&  \\

&0.12 &  3.045 & 0.029 & &0.12 & 3.343 & 0.028 &&&& \\

&0.16 &  3.979 & 0.017 & &0.16 & 4.331 & 0.015 &&&& \\

&0.20 &  4.956 & 0.023 & &0.20 & 5.409 & 0.022 &&&& \\

\bottomrule
\end{tabular}
\label{tab:oldroydB}
\end{threeparttable}
\end{table}

\begin{figure}[H]
\captionsetup[subfigure]{justification=justified,singlelinecheck=false}
    \centering
    {\renewcommand{\arraystretch}{0}
    \begin{tabular}{c@{}c}
    \begin{subfigure}[b]{.5\columnwidth}
        \centering
        \caption{{}}
        \includegraphics[width=\columnwidth]{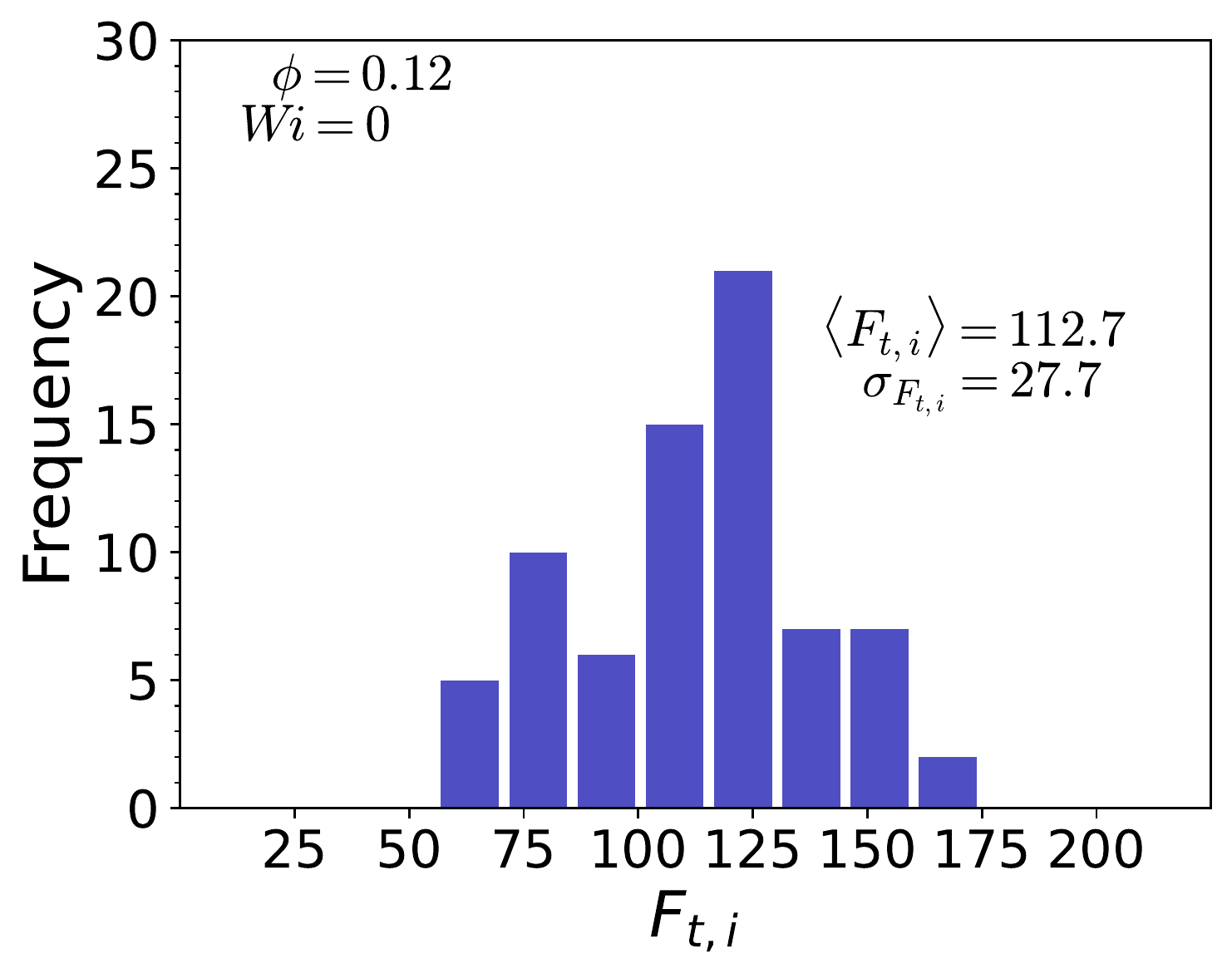}%
        \label{}
    \end{subfigure}&
    \begin{subfigure}[b]{.5\columnwidth}  
        \centering
        \caption{{}}
        \includegraphics[width=\columnwidth]{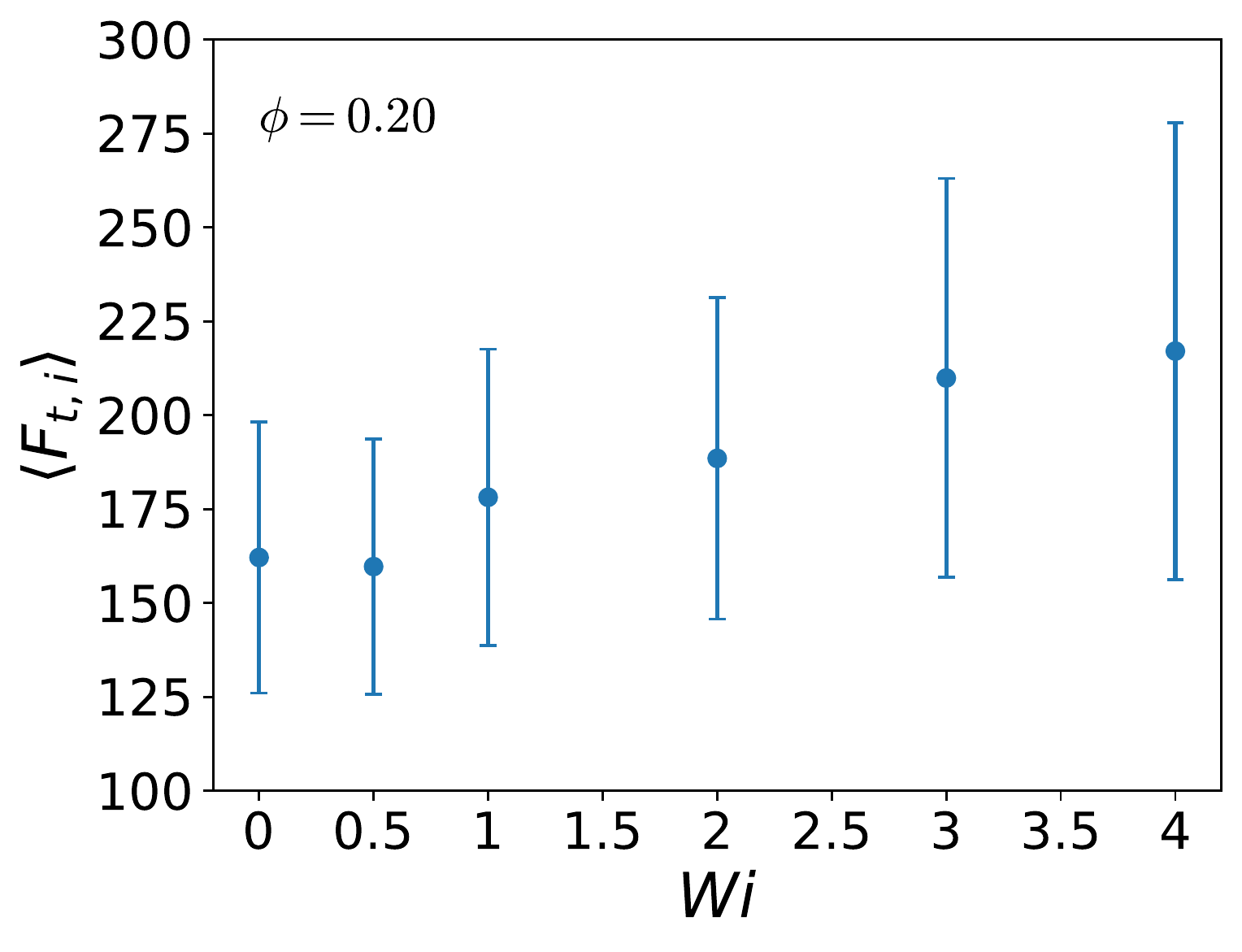}%
        \label{}
    \end{subfigure}\\
    \begin{subfigure}[b]{.5\columnwidth}   
        \centering 
        \caption{{}}
        \includegraphics[width=\columnwidth]{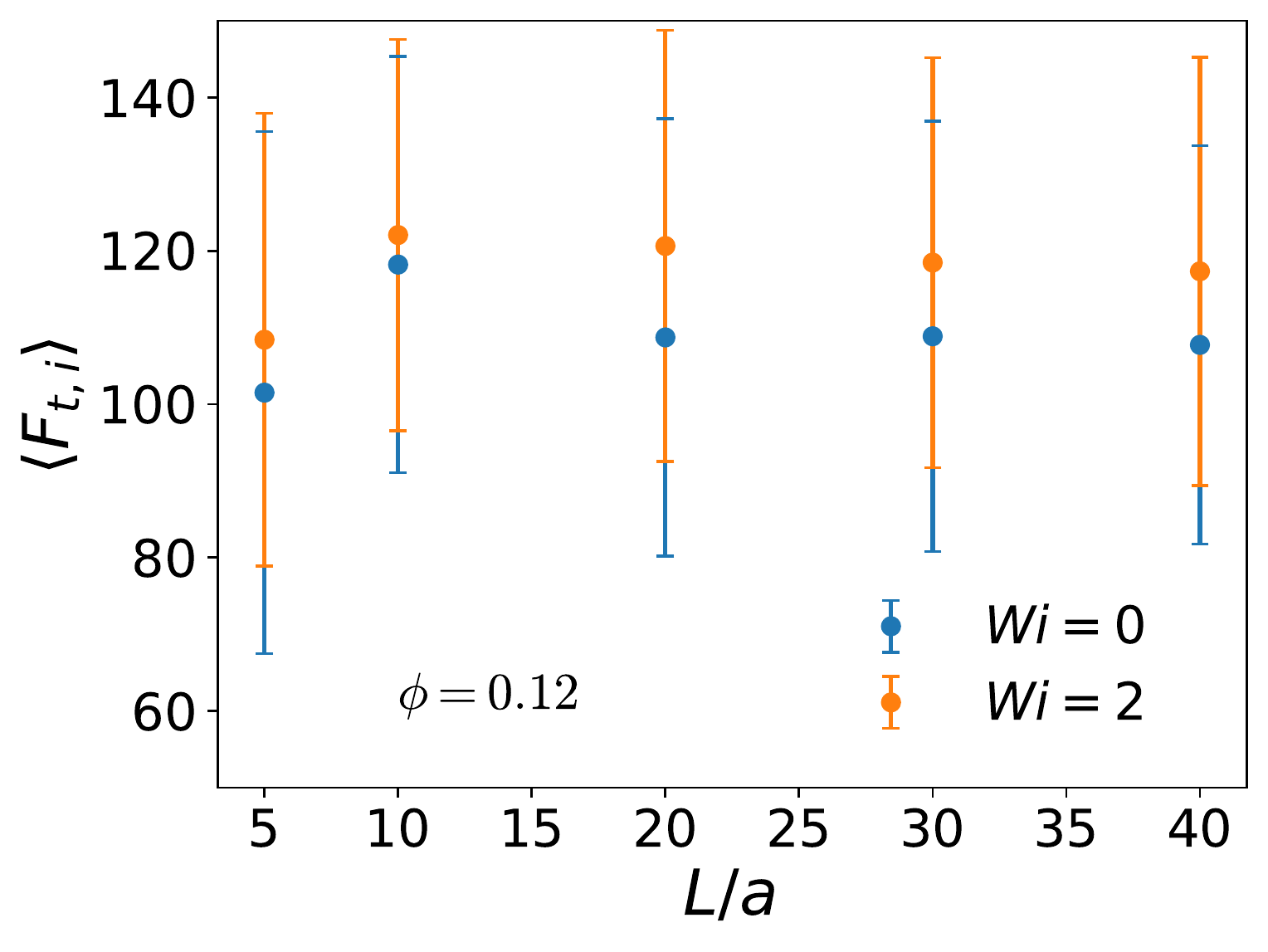}%
        \label{}
    \end{subfigure}&
    \begin{subfigure}[b]{.5\columnwidth}   
        \centering 
        \caption{{}}
        \includegraphics[width=\columnwidth]{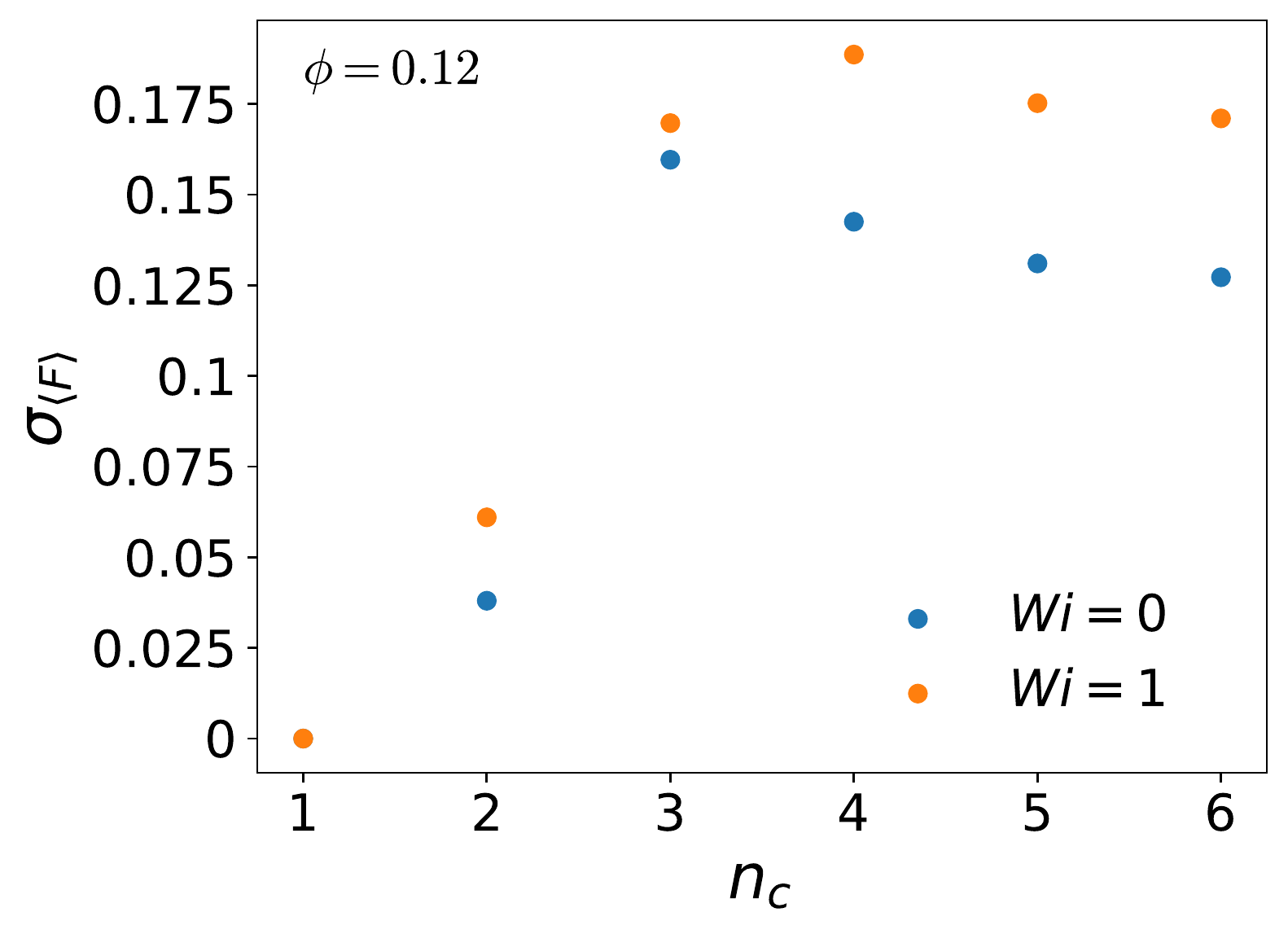}%
        \label{}
    \end{subfigure}
    \end{tabular}}
    \caption[]
    {(a) Frequency distribution (for one configuration, $n_c=1$) of the dimensionless drag force on individual spheres, $F_{t,i}$, for volume fraction $\phi=0.12$ (corresponding to a total number of spheres equal to $n_s=73$) and Newtonian fluid ($Wi=0$), (b) average drag force, $\langle F_{t,i} \rangle$, and standard deviation, $\sigma_{F_{t,i}}$, on the random array of spheres for $\phi=0.20$ at different $Wi$ numbers, (c) average drag force, $\langle F_{t,i} \rangle$, and standard deviation, $\sigma_{F_{t,i}}$, on one single random array of spheres for $\phi=0.12$ at $Wi=0$ and $Wi=2$, using different confinement lengths $L$, and (d) evolution in the standard deviation of the dimensionless average drag force $\sigma_{\langle F \rangle}$ on the random array of spheres for $\phi=0.12$ at $Wi=0$ and $Wi=1$, using a progressively increasing number of configurations $n_c$.} 
    \label{fig:statistics}
\end{figure}

Additionally, we show in Fig.~\ref{fig:statistics} statistical measures of the total drag force exerted by the Newtonian and viscoelastic fluids in the spheres. In the panels of Fig.~\ref{fig:statistics} we show the distribution in the absolute values of the dimensionless drag force exerted by the fluid on each individual sphere, $F_{t,i}$, extracted from the numerical simulations prior to ensemble averaging. Fig.~\ref{fig:statistics}(a) presents the frequency distribution of the drag force exerted by the Newtonian fluid on each individual sphere, $F_{t,i}$, for one representative configuration at a volume fraction $\phi=0.12$. The distribution obtained is approximately Gaussian, with a mean value of $\langle F_{t,i} \rangle = 112.7$ and standard deviation of $\sigma_{F_{t,i}}=27.7$. Fig.~\ref{fig:statistics}(b) shows the effect of the Weissenberg number, $Wi$, on the mean and standard deviation of the frequency distribution of the drag force on the ensemble of spheres, for $\phi=0.20$. Notice that the error bars represent the standard deviation $\pm \sigma_{F_{t,i}}$. The results obtained show that the distribution of the results (as measured by the standard deviation values) is similar for all Weissenberg numbers employed in the calculations, and that the mean value of the drag force first decreases for $Wi<1$ and then increases as elasticity starts to play a progressively more important role. Fig.~\ref{fig:statistics}(c) shows the effect of varying the location of the spheres on the average drag force $\langle F_{t,i} \rangle$, for a fixed solid volume fraction $\phi=0.12$, and $Wi=0$ and $2$, through changes in the variable $L$ (which represents the total length of the square-cross section duct in which the particles are confined, see Fig.~\ref{fig:DNSChannel}). From the results obtained it can be concluded that for $L/a \geq 30$ the average drag force distribution has converged. As before, the standard deviation values for the average drag force distribution represented by the error bars, $\sigma_{F_{t,i}}$, obtained for different $L/a$ lengths are similar. Notice that the non-monotonic behavior of the $\langle F_{t,i} \rangle$ values for the different values of $L/a$ can be attributed to the fact that here only one configuration ($n_c=1$) is used to estimate the average drag force of each particle in the suspension. Finally, in Fig.~\ref{fig:statistics}(d) we analyze the effect of the number of configurations, $n_c$, used in each kinematic condition, on the distribution of the dimensionless average drag force $\sigma_{\langle F \rangle}$, for $\phi=0.12$, and $Wi=0$ and $1$. We can conclude that for $n_c \geq 3$, the standard deviation $\sigma_{\langle F \rangle}$ has converged to a constant value.

\begin{figure}[H]
\captionsetup[subfigure]{justification=justified,singlelinecheck=false}
    \centering
    {\renewcommand{\arraystretch}{0}
    \begin{tabular}{c@{}c}
    \begin{subfigure}[b]{0.7\columnwidth}
        \centering
        \caption{{}}
        \includegraphics[width=0.7\columnwidth]{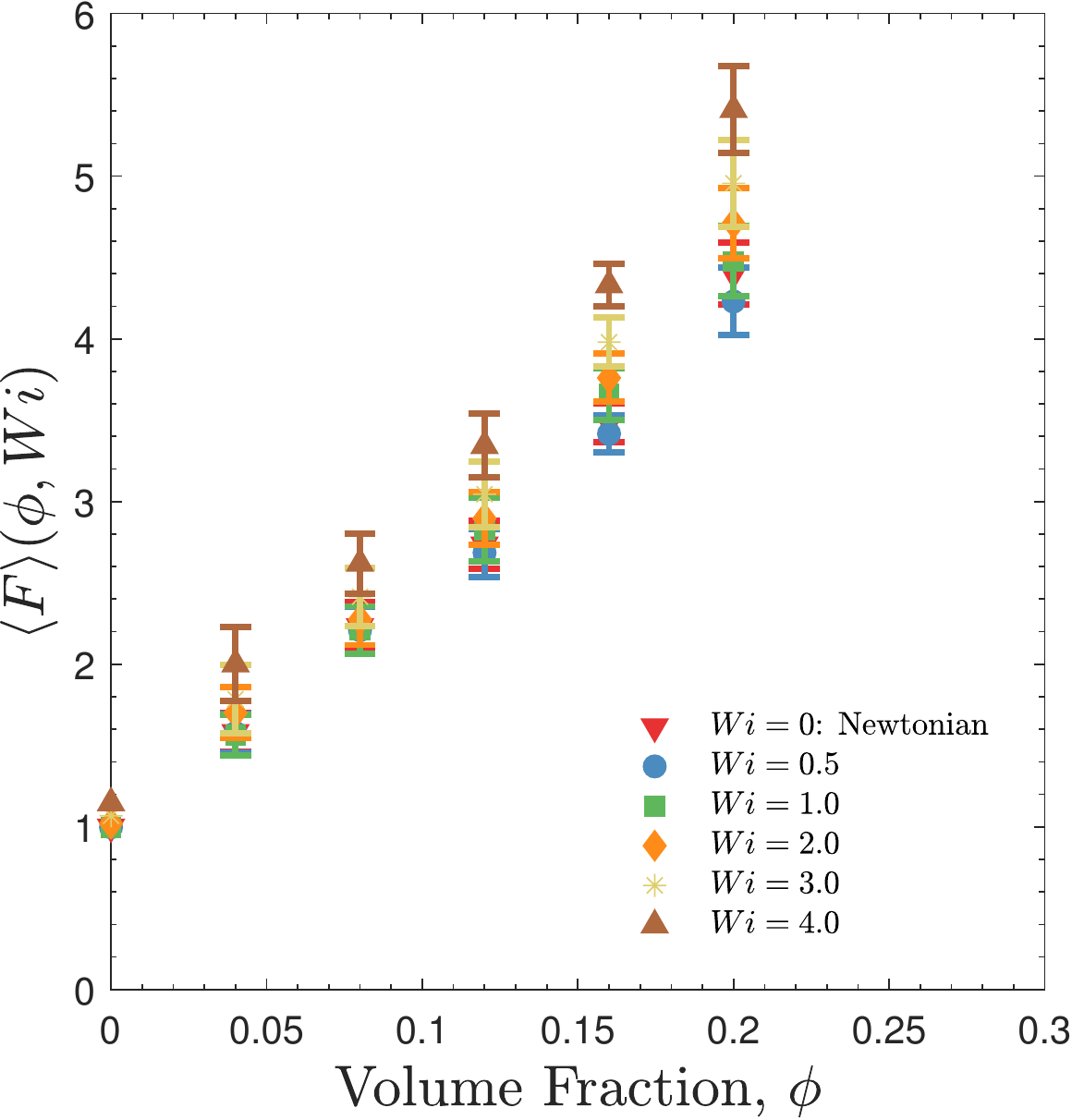}%
        \label{}
    \end{subfigure}\\
    \begin{subfigure}[b]{0.7\columnwidth}
	\centering    
	\caption{{}}
		\begin{tikzpicture}
			\node[anchor=south west,inner sep=0] (image) at (0,0)
			{\includegraphics[width=0.7\textwidth]{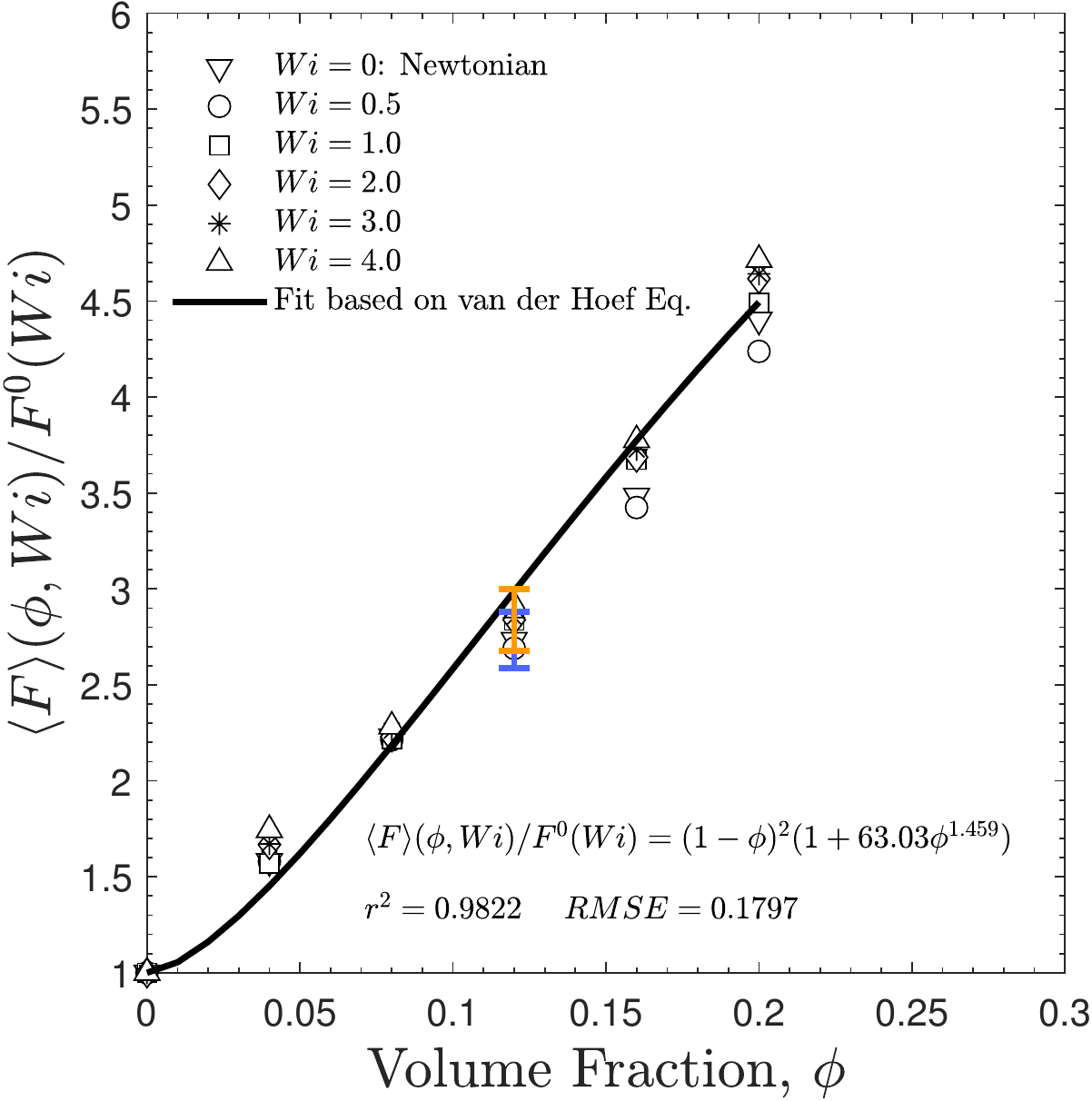}};
				\begin{scope}[x={(image.south east)},y={(image.north west)}]\node[anchor=south west,inner sep=0] (image) at (0.625,0.28)							{\includegraphics[width=0.23\textwidth]{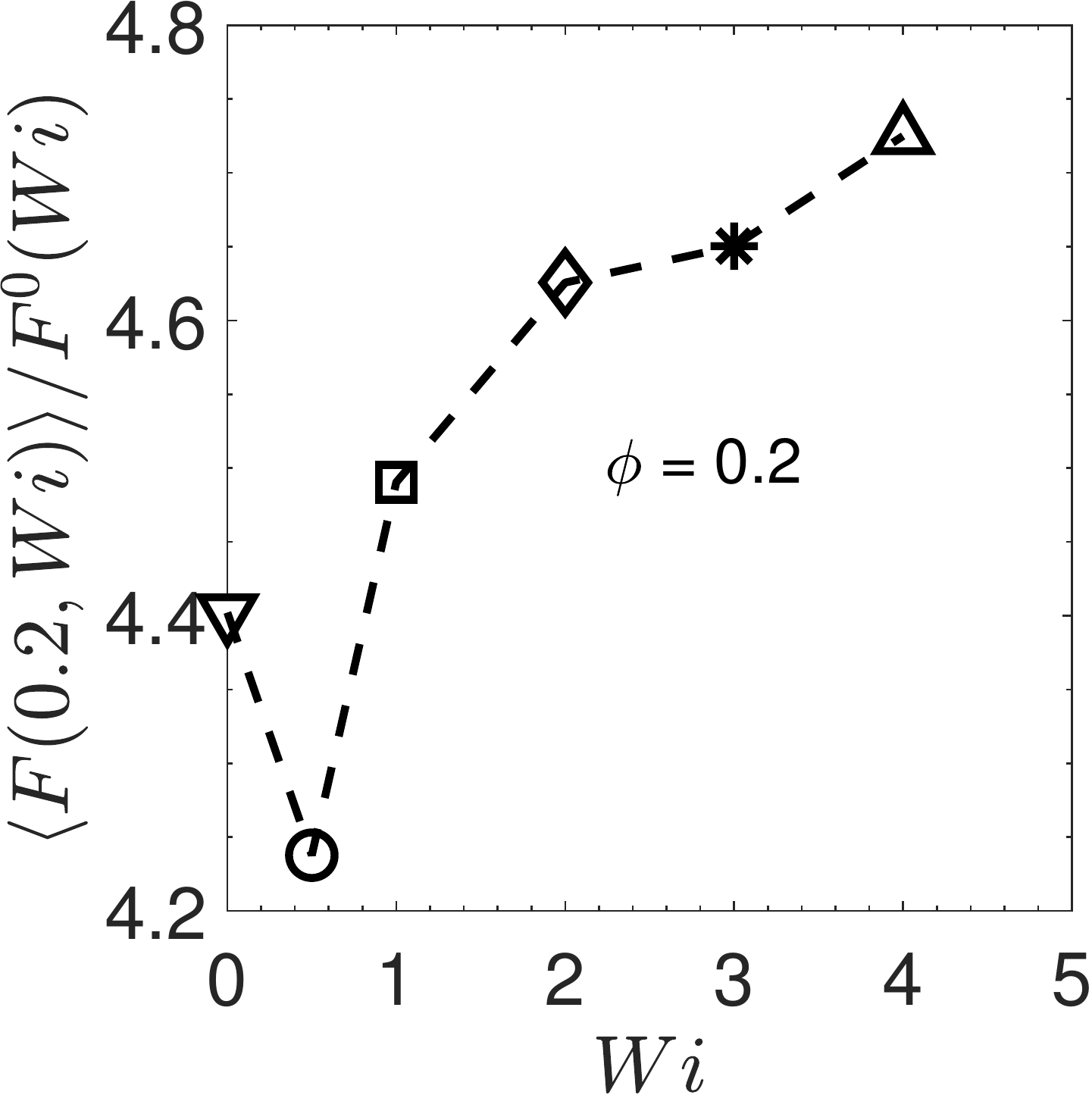}};
	     		\end{scope}
		\end{tikzpicture}
   		 \label{}
    \end{subfigure}\\
    \end{tabular}}
    \caption[]
    {Variation of (a) dimensionless average drag force $\langle F(\phi,Wi)\rangle$ and (b) normalized drag force $\langle F(\phi,Wi)\rangle/F^0(Wi)$ with Weissenberg number for random arrays of fixed particles with solid volume fractions $0<\phi\leq 0.2$ within an Oldroyd-B viscoelastic matrix-based fluid. Here $F^0(Wi)$ represents the drag force exerted by the Oldroyd-B fluid on a single particle, as described in \citet{Salah2019} and Eq.~(\ref{eq:F0}).} 
    \label{fig:dragViscoelastic}
\end{figure}

With the goal of finding a closure model for the drag force exerted by an Oldroyd-B fluid on random arrays of particles at creeping flow conditions and retardation ratio $\beta=0.5$, in Fig.~\ref{fig:dragViscoelastic}(a) we show the dimensionless average drag force, $\langle F(\phi,Wi)\rangle$, exerted on the particles for the solid volume fractions and Weissenberg numbers presented in Table~\ref{tab:oldroydB}, along with the standard deviation errors. The results obtained allow us to conclude that the variations with $\phi$ are much larger than the variations with $Wi$. Additionally, in Fig.~\ref{fig:dragViscoelastic}(b) we show the values of the normalized drag force, $\langle F(\phi,Wi)\rangle/F^0(Wi)$, exerted on the particles for the solid volume fractions and Weissenberg numbers referred above. Notice that here we have normalized the dimensionless average drag force $\langle F(\phi,Wi)\rangle$ by $F^0(Wi)$, which is the drag coefficient of a single sphere translating through an unbounded Oldroyd-B fluid under creeping flow conditions and given by the closure model presented in \citet{Salah2019} (see Eqs.~(19a) and (19b) therein, using $\zeta=1-\beta=0.5$). For the sake of completeness we write here the expression we employed for $F^0(Wi)$ with $\zeta=1-\beta=0.5$,
\begin{eqnarray}
F^0(Wi) = \begin{cases} 
	\displaystyle 1-\frac{0.0015955Wi^2+0.0295475Wi^4-0.017345Wi^6}{0.0534+3.2325Wi^2+Wi^4} & \text{if } Wi\leq 1\\
	\\
	\displaystyle 1+\frac{-0.0123176Wi^4+0.0078197Wi^6+0.000142825Wi^8}{0.2444225Wi^2+Wi^4} & \text{if } Wi > 1.
	\end{cases} 
\label{eq:F0}
\end{eqnarray}
The numerical results presented in Figure~\ref{fig:dragViscoelastic}(b) show that this rescaled drag force can be considered independent from any statistically-significant trend with Weissenberg number for the range of solid volume fractions computed in this work; i.e, when we normalize the dimensionless average drag force we compute in our simulations by $F^0(Wi)$ it helps to collapse $\langle F(\phi,Wi)\rangle$ at any given value of $\phi\leq 0.20$. 

In the last section we obtained a good agreement between the numerical results for the average drag force exerted on an ensemble of particles immersed on a Newtonian fluid with the model given by \citet{Hoef2005}, we thus propose fitting an equation of the same form to the computational results obtained in a viscoelastic fluid, i.e., 
\begin{eqnarray}
\langle F(\phi,Wi) \rangle/F^0(Wi) = (1-\phi)^2\left(1 + k_1\phi^{k_2}\right),
\label{eq:fit}
\end{eqnarray}
where $F^0(Wi)$ is the infinitely dilute ($\phi\to 0$) result for the drag force presented in \citet{Salah2019} for $\zeta=0.5$ (cf. Eq.~(\ref{eq:F0})). Notice that we could also have used a similar expression for the correction to the drag coefficient as that given by \citet{faroughi2015unifying} in Eq.~(\ref{eq:salah}) to fit our viscoelastic results, however, due to its  additional simplicity we have here chosen the functional form of the \citet{Hoef2005} expression. The correlation constants obtained using a suitable fitting procedure (conducted by solution of a nonlinear least-squares problem using the iterative Levenberg-Marquardt algorithm \cite{Levenberg2005}) led to $k_1=63.03$ and $k_2=1.459$. The resulting model accounts for 98.22\% of the variance of the numerical data, a root mean square error (RMSE) of 0.1797, and an average error computed between the proposed model and the numerical data of 5.7\%. The inset in Figure~\ref{fig:dragViscoelastic}(b) shows the evolution of the normalized drag force, $\langle F(\phi,Wi) \rangle/F^0$, with Weissenberg number, for $\phi=0.2$. There is a residual systematic trend in the rescaled values with $Wi$ indicating that the average drag force is not perfectly factorizable into functions of $\phi$ and $Wi$ respectively. However, the variation shown is less than the RMSE error between simulations and we neglect this henceforth. 

\begin{figure}[H]
\centering
\includegraphics[scale=0.35]{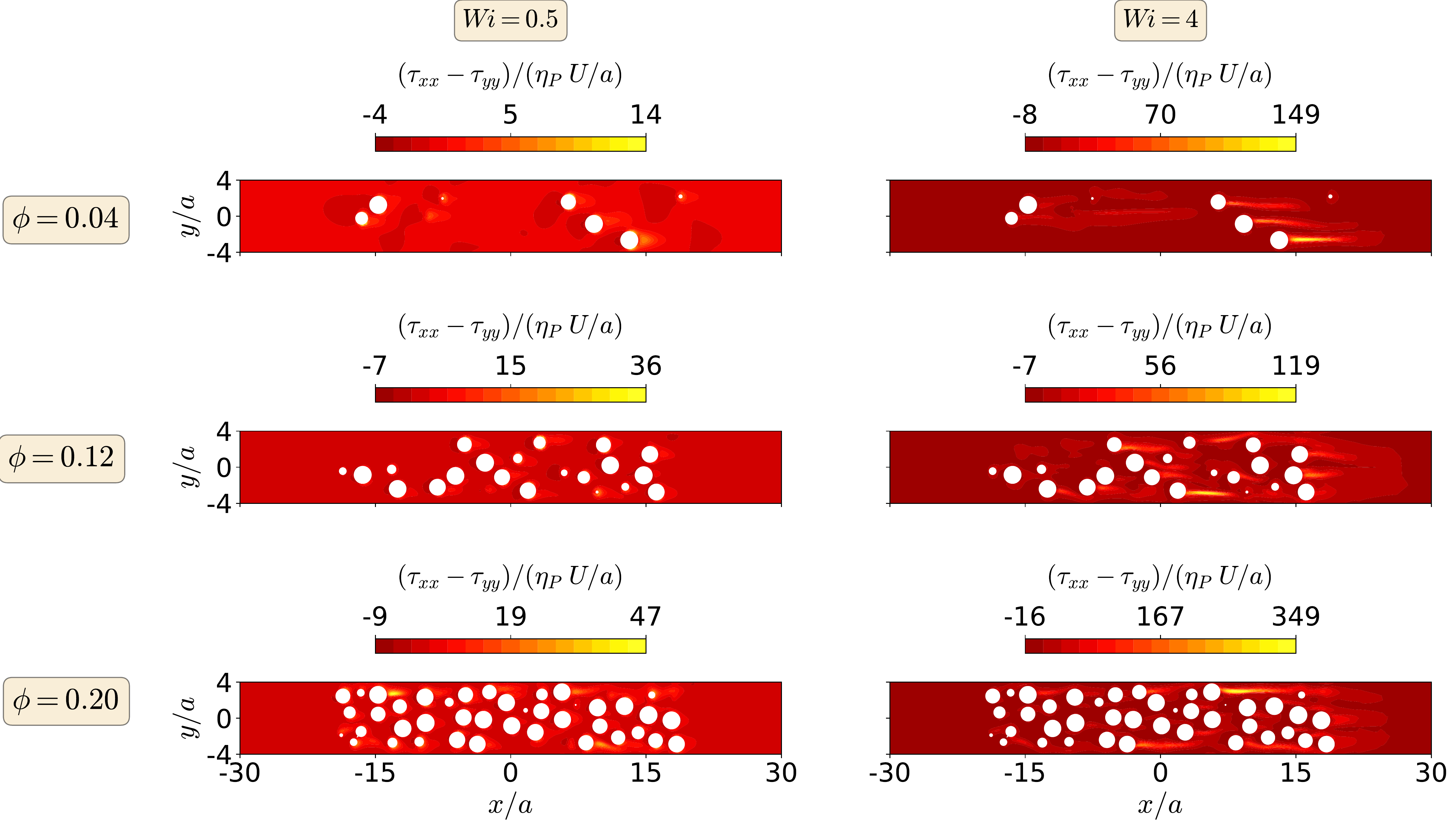}
\caption{Dimensionless first normal stress difference distribution around the random particle arrays in a channel filled with an Oldroyd-B viscoelastic matrix-based fluid.} 
\label{fig:N1}
\end{figure}

\begin{figure}[H]
\captionsetup[subfigure]{justification=justified,singlelinecheck=false}
    \centering
    {\renewcommand{\arraystretch}{0}
    \begin{tabular}{c@{}c}
    \begin{subfigure}[b]{.5\columnwidth}
        \centering
        \caption{{}}
        \includegraphics[width=\columnwidth]{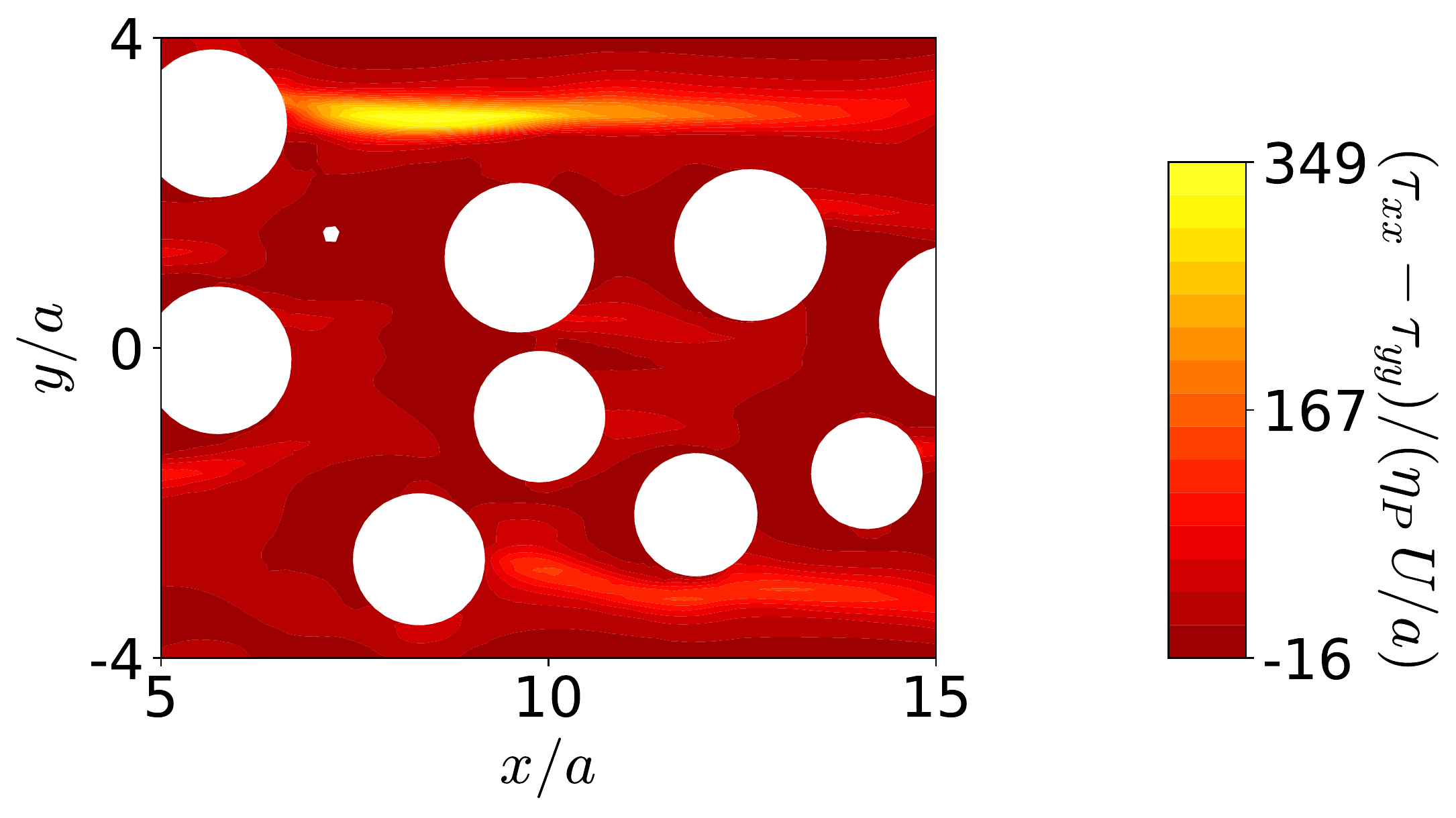}%
        \label{}
    \end{subfigure}&
    \begin{subfigure}[b]{.5\columnwidth}  
        \centering
        \caption{{}}
        \includegraphics[width=0.85\columnwidth]{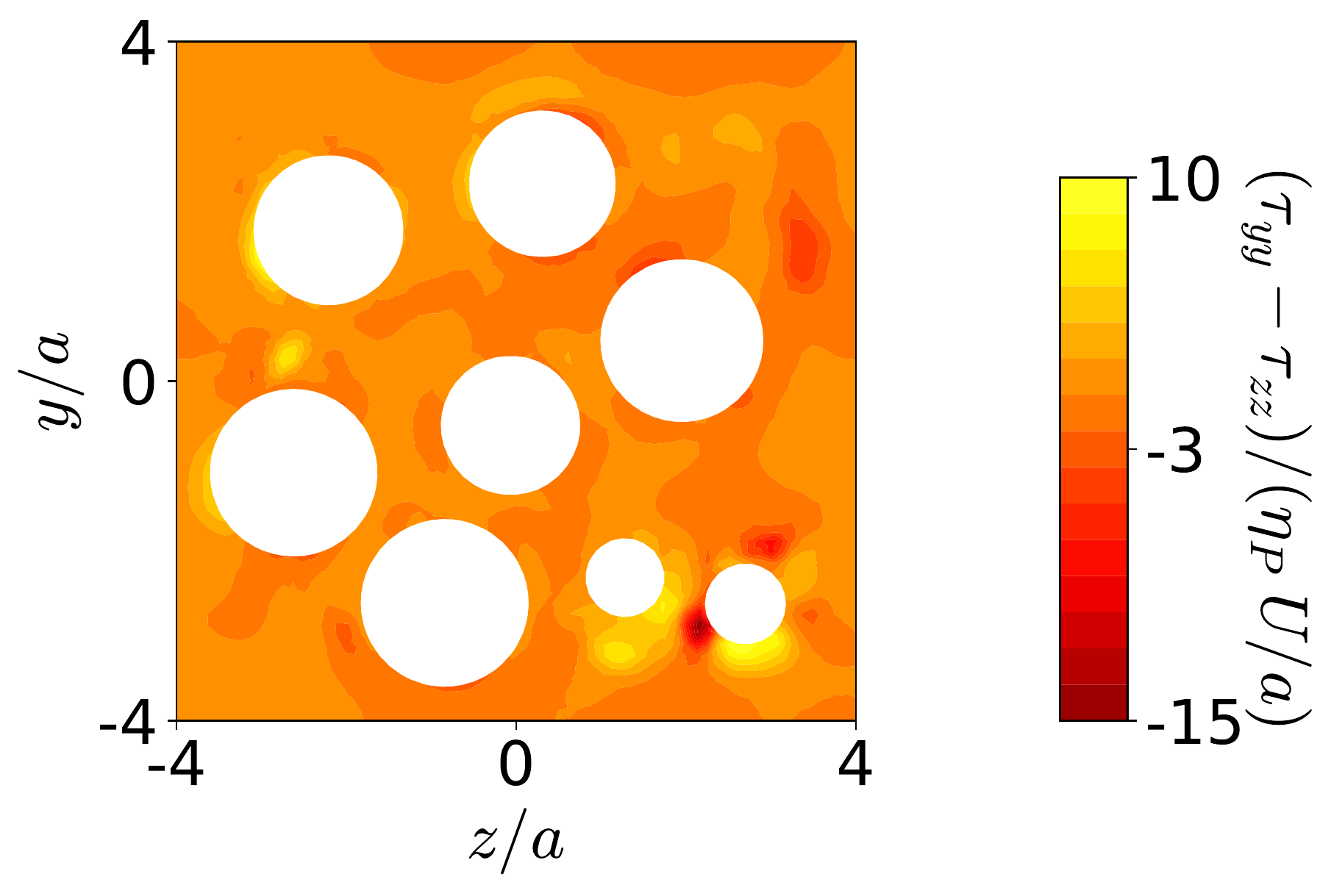}%
        \label{}
    \end{subfigure}\\
    \begin{subfigure}[t]{.5\columnwidth}   
        \centering 
        \includegraphics[width=\columnwidth]{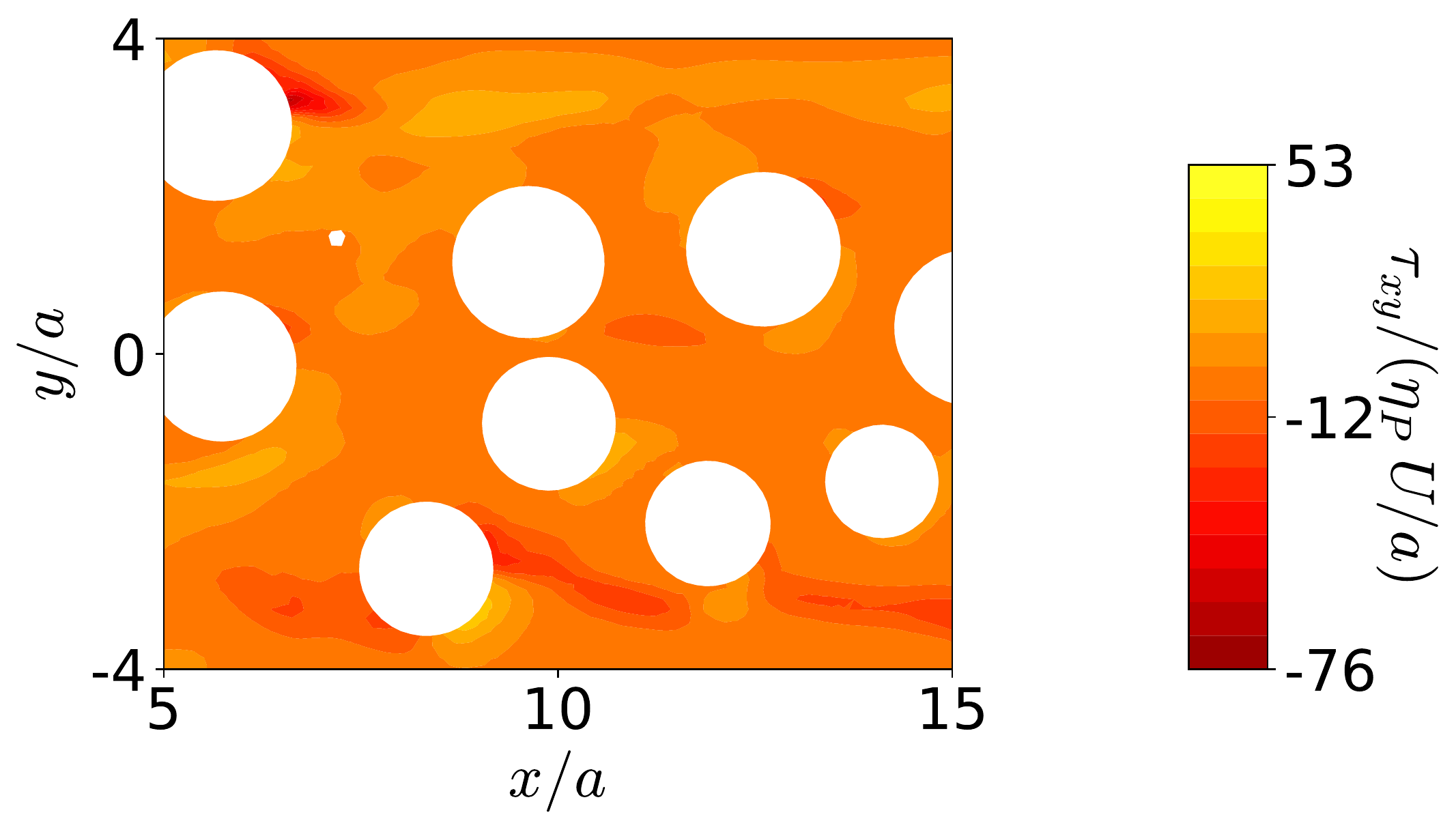}%
        \label{}
    \end{subfigure}&
    \begin{subfigure}[t]{.5\columnwidth}   
        \centering 
        \includegraphics[width=0.85\columnwidth]{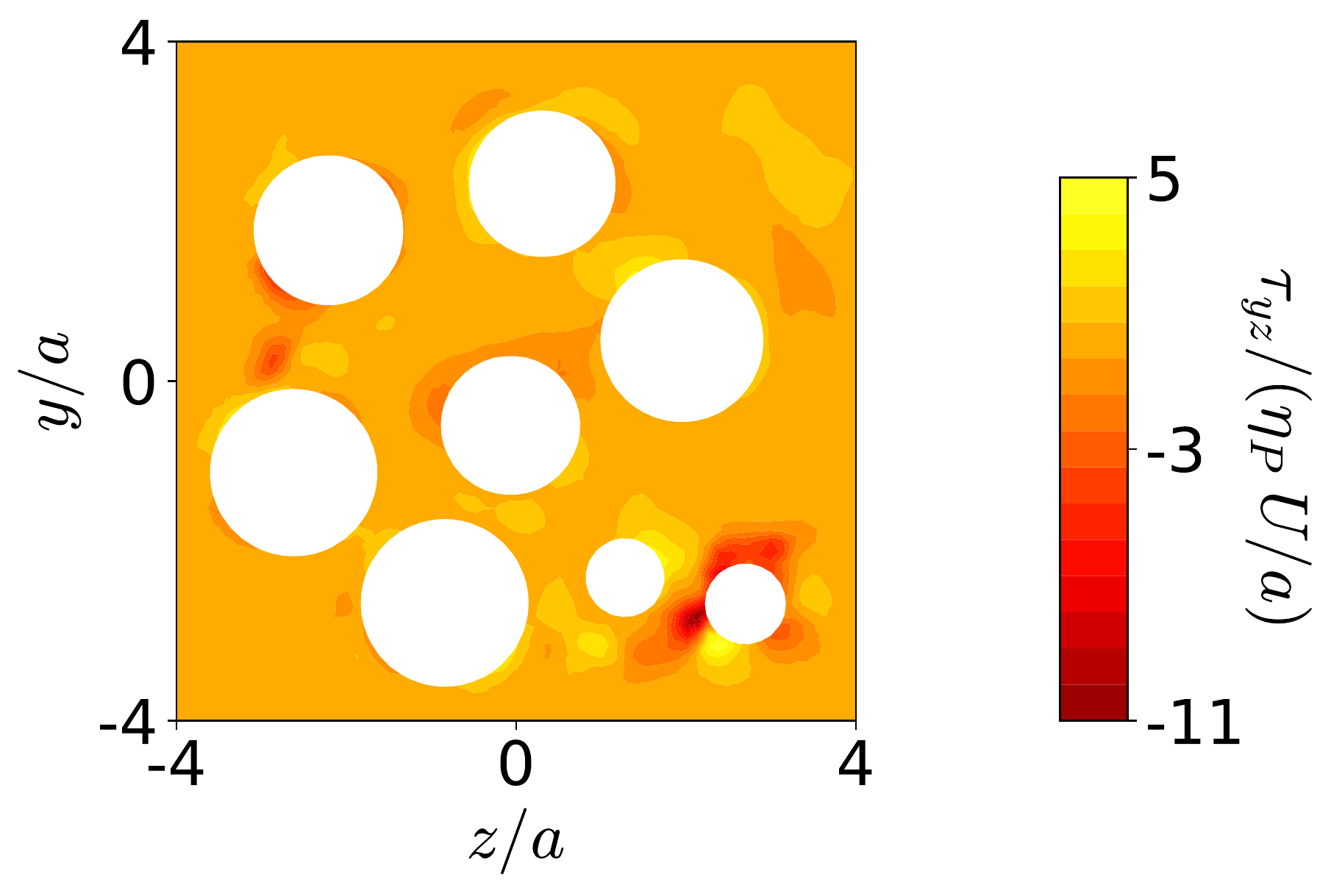}%
        \label{}
    \end{subfigure}
    \end{tabular}}
    \caption[]
    {Zoomed-in region in a plane at (a) fixed $z/a=0$ showing the dimensionless $\tau_{xx}-\tau_{yy}$ and $\tau_{xy}$ distributions, and at (b) fixed $x/a=-10$ showing the dimensionless $\tau_{yy}-\tau_{zz}$ and $\tau_{yz}$ distributions, for $Wi = 4$ and $\phi=0.2$.} 
    \label{fig:zoomStress}
\end{figure}

In Fig.~\ref{fig:N1} and Fig.~\ref{fig:zoomStress} we show contours of the dimensionless first normal stress difference, defined as $(\tau_{xx}-\tau_{yy})/(\eta_P U/a)$, obtained from the numerical simulations. The first normal stress difference is mainly generated near the no-slip surfaces and in the wake of each of the spheres. Notice that increasing the fluid elasticity (increasing $Wi$ number), i.e. from the left to the right panels in Fig.~\ref{fig:N1}, promotes a strong elastic wake and an increase in the magnitude of the first normal stress difference. Only by use of the log-conformation approach for computing the polymeric extra-stress tensor components were we able to stabilize the numerical algorithm. Additionally, increasing the particle volume fraction, i.e., from top to bottom panels, increases the magnitude of the first normal stress difference that is generated on the front stagnation point of the particles. Finally, from Fig.~\ref{fig:zoomStress} we see that the magnitude of the first normal stress difference near the rear stagnation point is much larger than the one developed upstream near the front stagnation point, and this difference becomes progressively larger for higher $Wi$.  
A zoomed in region at $z/a=0$ and $x/a=-10$, of the normal stress distributions $\tau_{xx}-\tau_{yy}$ and $\tau_{yy}-\tau_{zz}$, as well as the shear stress distributions $\tau_{xy}$ and $\tau_{yz}$ are shown in Fig.~\ref{fig:zoomStress} for $Wi=4$ and $\phi=0.2$. In the plane $z/a=0$ (i.e. the $xy$-plane, corresponding to a longitudinal section in the flow direction, Fig.~\ref{fig:zoomStress}(a)), the contour plot showing the first normal stress distribution reveals the formation of large elastic wakes at the rear stagnation point of the spheres, and the shear stress $\tau_{xy}$ indicates the formation of quadrupolar structures at $\pm 45\degree$ angles in the quadrants around each of the spheres. The magnitudes of the first normal stress difference $\tau_{xx}-\tau_{yy}$ and $\tau_{xy}$ are much greater than the magnitudes of the second normal stress difference $\tau_{yy}-\tau_{zz}$ and $\tau_{yz}$ shown in the plane $x/a=-10$ (i.e. the $zy$-plane, corresponding to a section perpendicular to the flow direction, Fig.~\ref{fig:zoomStress}(b)).

\section{Proppant transport during the hydraulic-fracture process}

In this section we develop a computational framework, based on the Eulerian-Lagrangian formulation \cite{Celio2018}, capable of numerically describing, as a proof-of-concept, proppant transport in a viscoelastic matrix-based fluid that can be characterized by the Oldroyd-B constitutive model. The newly-developed algorithm takes into account the effect of the particle volume fraction (the Lagrangian phase) on the viscoelastic fluid phase (the Eulerian phase). For this purpose, we extend the formulation presented in \citet{Celio2018} (and references therein) to be able to take into account the viscoelastic behavior of the fluid.

Consider the motion of an incompressible viscoelastic fluid phase in the presence of a secondary particulate phase, which is governed by the volume-averaged continuity equation
\begin{equation}
\begin{aligned}
\frac{\partial \epsilon_f}{\partial t}+\nabla \cdot (\epsilon_f \textbf{U}^f) = 0,
\end{aligned}
\label{eq:mass_equation}
\end{equation}
and Cauchy momentum equation
\begin{equation}
\begin{aligned}
\frac{\partial(\epsilon_f \textbf{U}^f)}{\partial t}+\nabla \cdot (\epsilon_f \textbf{U}^f \textbf{U}^f) = - \nabla P - S_p + \nabla \cdot (\epsilon_f \boldsymbol\tau_f) + \epsilon_f\textbf{g},
\end{aligned}
\label{eq:momentum_equation}
\end{equation}
where $\epsilon_f$ is the fluid porosity field satisfying $\epsilon_f = 1-\phi$, $\textbf{U}^f$ is the fluid velocity, $P$ is the modified pressure ($p/\rho_f$, with $p$ being the dynamic pressure and $\rho_f$ the fluid density), $\textbf{g}$ is the gravity acceleration vector and the fluid-phase stress tensor $\boldsymbol\tau_f$ is given by:
\begin{equation}
\begin{aligned}
\boldsymbol\tau_f= \frac{1}{\rho_f}\left[\eta_S\left((\nabla \textbf{U}^f)+(\nabla \textbf{U}^f)^T\right)+\boldsymbol\uptau_P\right],
\end{aligned}
\label{eq:viscous_stress_equation}
\end{equation}
where $\boldsymbol\uptau_P$ is the polymeric extra-stress tensor computed using the Oldroyd-B viscoelastic matrix-based constitutive model given by Eq.~(\ref{eqn:oldroydBeqdimensionless}).

The two-way coupling between the fluid phase and particles is enforced via the source term $S_p$ in the momentum balance equations, Eq.~(\ref{eq:momentum_equation}), of the fluid-phase. Because the fluid drag force, $\textbf{F}_{d,i}$, acting on each particle $i$ is known (see Section~\ref{sec:viscoelasticDrag}), then according to Newton's third law of motion the source term is computed as a volumetric fluid-particle interaction force given by:
\begin{equation}
\begin{aligned}
S_p = \frac{\displaystyle\sum_{i=1}^{N_p}{\textbf{F}_{d,i}}}{\rho_f V_{cell}},
\end{aligned}
\label{eq:source_term_equation}
\end{equation}
where $V_{cell}$ is the volume of a computational cell, and $N_p$ is the number of particles located in that cell.

In this work, we consider two different formulations to describe the contact between particles, the spring-dashpot model \cite{Cundall197947} and a Multi-Phase Particle In Cell (MPPIC) model \cite{Rourke2009}. The former allows us to explicitly handle each contact between two particles and, therefore, is very computationally intensive. The latter can be used to represent the particle collisions on average without resolving particle-particle interactions individually. In the MPPIC method the particle-particle interactions are computed by models which utilize mean values calculated on the Eulerian mesh \cite{MPPICOpenFOAM}. For that purpose, in the present work we have employed a collision damping model to represent the mean loss in kinetic energy which occurs as particles collide, and helps to produce physically realistic scattering behaviour \cite{MPPICOpenFOAM}. Finally, a collision isotropy model is also employed to spread the particles uniformly across cells \cite{MPPICOpenFOAM}.  

Two case studies were performed to validate the newly-developed $DPMviscoelastic$ solver. In the first case, we study proppant transport and sedimentation in a long conduit of rectangular cross section, a typical geometry to study flow in hydraulic fracturing \cite{Steven2020}. In the second case, we study the segregation phenomena which occurs in cement casing for horizontal wells. Both case studies were performed using Newtonian and viscoelastic carrier fluids. For the first case, particle collisions are modelled using the MPPIC model in order to handle $O(10^6)$ particles, and for the latter, the Hertzian spring-dashpot model \cite{Tsuji1992239} is employed with a total of $125,000$ particles. The following sections present comparisons between the numerical results obtained for the aforementioned case studies and results found in the scientific literature.

\subsection{Rectangular channel flow}

Despite many advances in hydrocarbon reservoir modeling and technologies \cite{dahi2011numerical,faroughi2013prompt,bordbar2018pseudo,han2016numerical,bakhshi2020numerical}  especially for unconventional resource development,  the efficiency of hydrocarbon recovery in shale reservoirs is still very low \cite{seales2017recovery}.  One of the leading issues causing this inefficiency is the lack of proper proppant placement in the fracture networks.   Proppant emplacement within  fractures directly impacts productivity because it controls both short- and long-term conductivities of the fractured wells \cite{gomaa2014viscoelastic}.  Proppant particles must be carried over large distances to ensure successful  placement, which requires a spatially homogeneous distribution of particles \cite{faroughi2018rheological}.  However, flows of non-Brownian particles, such as proppant, often result in non-homogeneous patterns, in which particle sedimentation is commonly observed \cite{Steven2020}. The proppant particles that will be modeled in this section are glass microspheres of diameter 73 $\upmu\mathrm{m}$ and of density 2.54 g/cm$^3$, in order to simulate the experiments conducted by \citet{Steven2020}. The carrier fluid is a glycerol/water mixture of 85:15 w/w$\%$ with a viscosity and density of 0.1 Pa$\mathpunct{.}$s and 1.22 g/cm$^3$ at ambient temperature, respectively.   

\begin{figure}[H]
\centering
\includegraphics[scale=0.5]{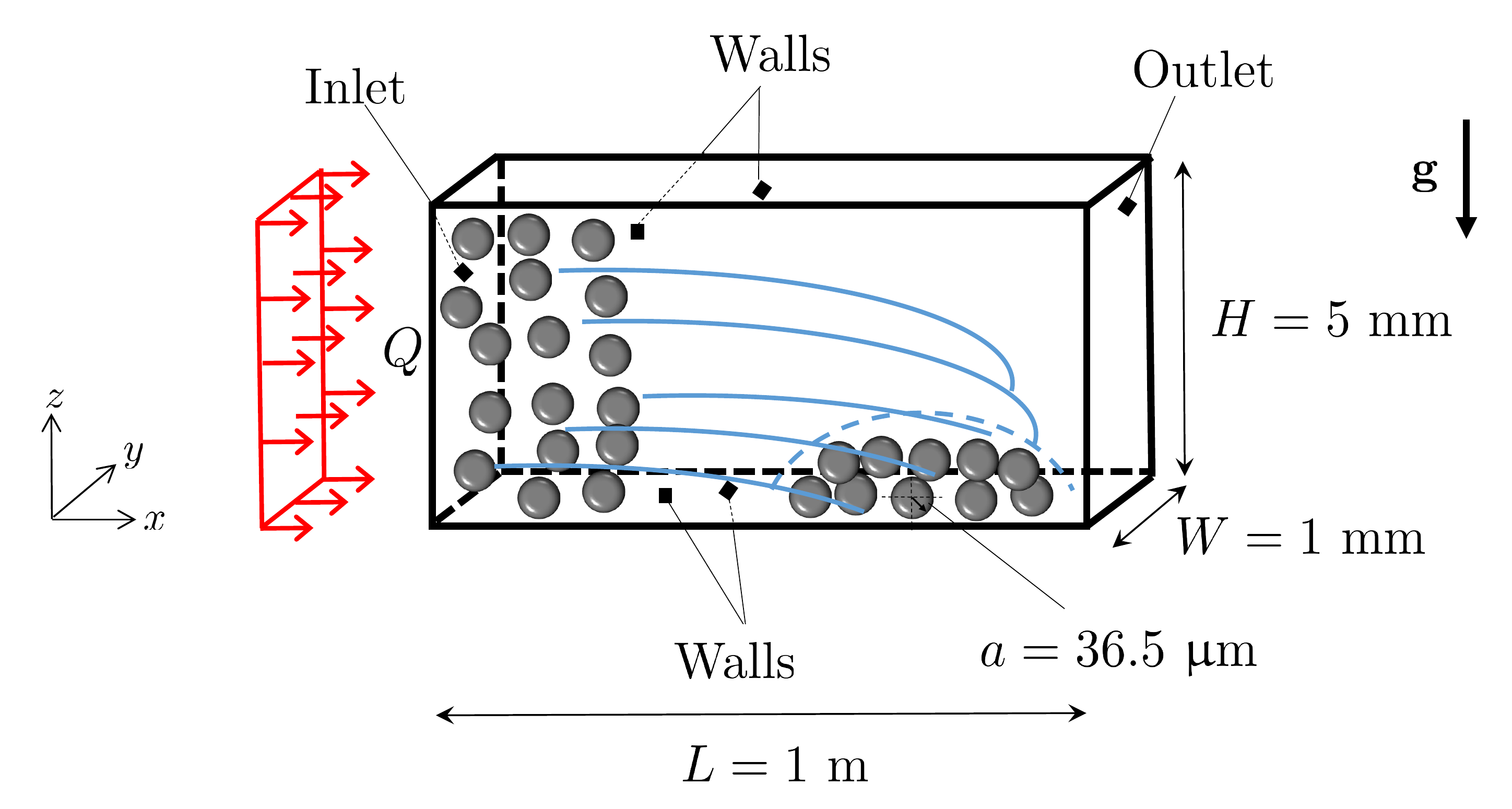}
\caption{Schematic of the channel cross-section used for simulating suspensions of particles ($\phi= 0.025,~0.038$ and $0.05$) settling in a confined channel ($Re_W = 0.06832$ and $0 < Wi \le 2.1$), which mimic a hydraulic fracture vertical fissure.} 
\label{fig:rectangularChannel}
\end{figure}

The computational setup used to study the suspension transport is shown schematically in Fig.~\ref{fig:rectangularChannel}. First the pure carrier matrix fluid is homogeneously injected to prefill the channel at the start of the simulation. After that, the proppant suspension is injected at a flow rate $Q$ and a uniform initial volume fraction $0 < \phi_i \leq 0.05$. The channel interior, which is used to mimic a vertical fissure, has a rectangular cross section of height $H=5~$mm and width $W=1~$mm, the latter corresponds to a maximum of 14 particles across the width. The channel is $L=1~$m long. The channel exit is left open to atmospheric pressure.

To verify the numerical algorithm for 3D calculations of the carrier fluid alone we conduct a set of numerical experiments to evaluate the pressure difference $\Delta P$ as a function of flow rate $Q$. A mesh refinement sensitivity analysis is performed by using three different levels of mesh refinement on $L\times H\times W$: Mesh 1 (M1), $2500\times 13\times 3$ cells, Mesh 2 (M2), $5000\times 26\times 6$ cells and Mesh 3 (M3), $10000\times 52\times 12$ cells. Table~\ref{tab:vortexsize} gives the pressure drop results corresponding to each of the mesh refinement levels employed at the prescribed inlet flow rate $Q$. Additionally, Table~\ref{tab:vortexsize} compares the numerical results obtained and the analytical values of the pressure difference for the Hagen-Poiseuille flow of a Newtonian fluid \cite{Mortensen2005}, which for an aspect ratio of $H/W=5$ is given by $\Delta P/L = 13.7\mu Q/H W^3$. The numerical results obtained in the most refined mesh employed in the calculations (M3) are within 0.39$\%$ of the analytical values for all the flow rates tested.

\begin{table}[H]
  \begin{threeparttable}
    \caption{Pressure difference $\Delta P$ (mbar) as a function of flow rate $Q$ (cm$^3$/h) and mesh level refinement for the Hagen-Poiseuille flow of a Newtonian fluid with viscosity $\mu=0.1~$Pa$\mathpunct{.}$s and geometry parameters given in the text. The relative error (\%) between the calculated numerical result and the analytical value \cite{Mortensen2005} is $0.39\%$ for the most refined mesh M3.}  
    \vspace{0.1cm}
	\centering
	\renewcommand{\arraystretch}{1.2}
    \begin{tabular}{l r r r r}
    \toprule

$Q$ & \multicolumn{4}{c}{$\Delta P$} \\

\cmidrule{1-5}

 & M1 & M2 & M3 & Analytical  \\ 

10 & 6.20 & 7.25 & 7.58 & 7.61\\
20 & 12.41 & 14.50 & 15.16 & 15.22\\
50 & 31.01 & 36.25 & 37.89 & 38.06\\
100 & 62.03 & 72.50 & 75.79 & 76.11\\
150 & 93.04 & 108.75 & 113.68 & 114.17\\
200 & 124.06 & 145.00 & 151.57 & 152.22\\
300 & 186.09 & 217.50 & 227.36 & 228.33\\
\% error$^a$ & 18.50   & 4.74    & 0.39   & \\ 
\bottomrule
  \end{tabular}
  \label{tab:vortexsize}
      \vspace{0.1cm}
  \begin{tablenotes}
  \scriptsize
      \item[$^a$] Calculated between numerical results obtained with M1, M2 and M3, and analytical values.
  \end{tablenotes}
  \end{threeparttable}
  \end{table}%

\begin{figure}[H]
\centering
\includegraphics[scale=0.5]{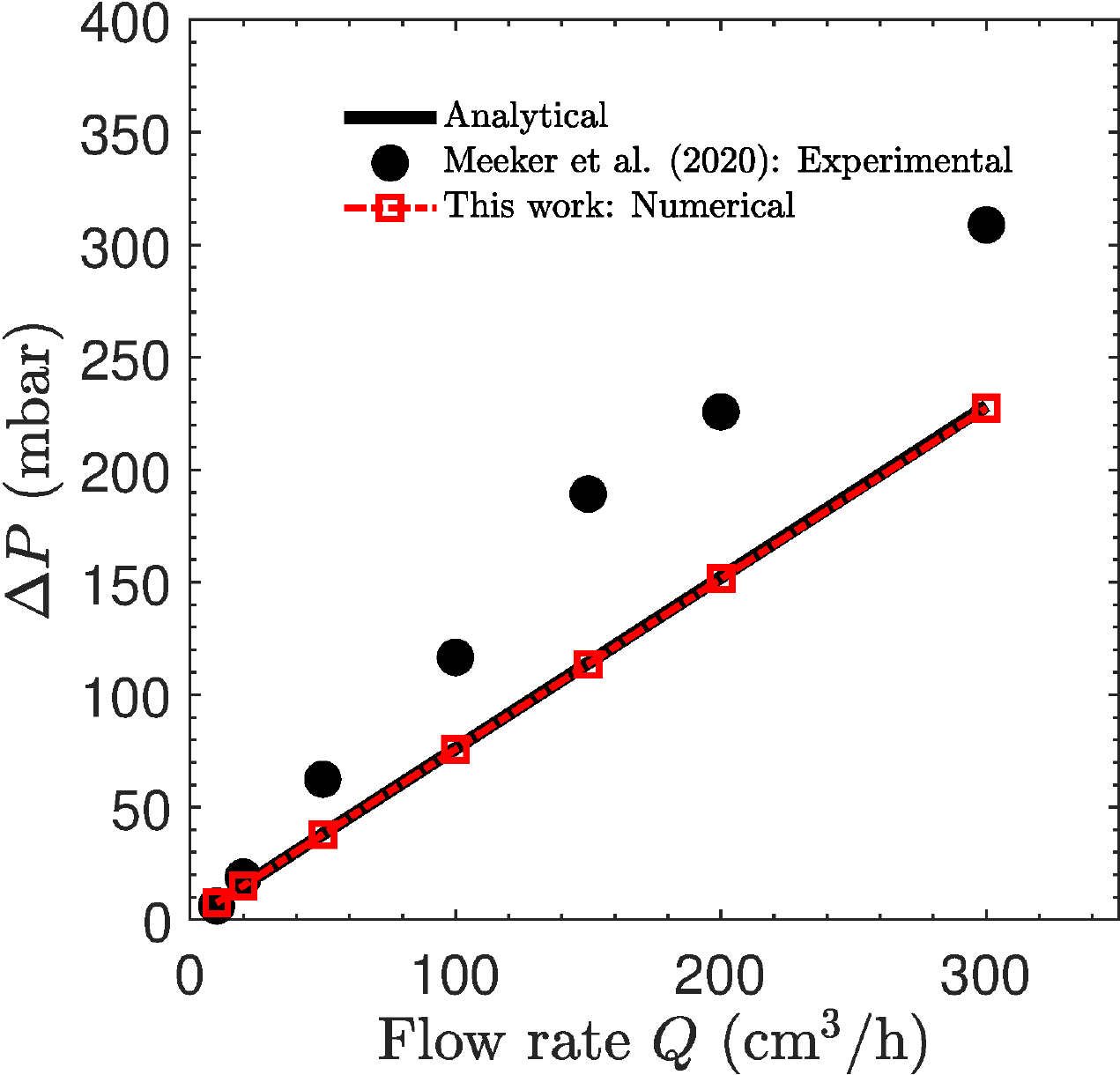}
\caption{Pressure difference, $\Delta P$ (mbar), as a function of the imposed flow rate, $Q$ (cm$^3$/h). Numerical results obtained in this work using mesh M3 are compared against analytical values for the Hagen-Poiseuille flow of a Newtonian fluid (black solid line) and to the experimental data of \citet{Steven2020} (black solid circles).} 
\label{fig:pressureDifference}
\end{figure}

Fig.~\ref{fig:pressureDifference} shows the comparison between experimental pressure difference measurements, $\Delta P$, as a function of flow rate $Q$, obtained by \citet{Steven2020} using the 0.1 Pa$\mathpunct{.}$s carrier fluid, and the computed numerical results in this work. From the experimental data, it is clear that $\Delta P$ increases linearly with $Q$ up to $\Delta P\approx 0.2$ bar, corresponding to a flow rate $Q=150~$cm$^3$/h and a Reynolds number of $Re_W=\rho(Q/WH)W/\mu = 0.10248$ for the glycerol/water mixture fluid. As noted by \citet{Steven2020}, the onset of nonlinear behavior is most likely to be caused by some deformation of the poly dimethylsiloxane elastomer at higher pressures. Therefore, the flow rate of our numerical tests is set at $Q=100~$cm$^3$/h in the following studies with suspensions. This results in an average fluid velocity $U=Q/HW=5.6\times 10^{-3}~$m/s, which corresponds to $Re_W = 0.06832$, confirming that we are in the creeping flow regime. We also note that $U$ is much greater than the average particle sedimentation velocity $U_{Stokes}=3\times 10^{-5}~$m/s, and thus, our simulations are performed under favorable transport conditions with minimal settling at the entrance. In fact, the slope of the trajectory of a sedimenting particle being transported at this average speed, $U_{Stokes}/U\approx 5\times 10^{-3}$, is similar to the ratio of channel height to length $H/L$, meaning that most of the initially suspended particles entering the channel should settle as they approach the channel exit.

\begin{figure}[H]
\centering
\includegraphics[scale=0.5]{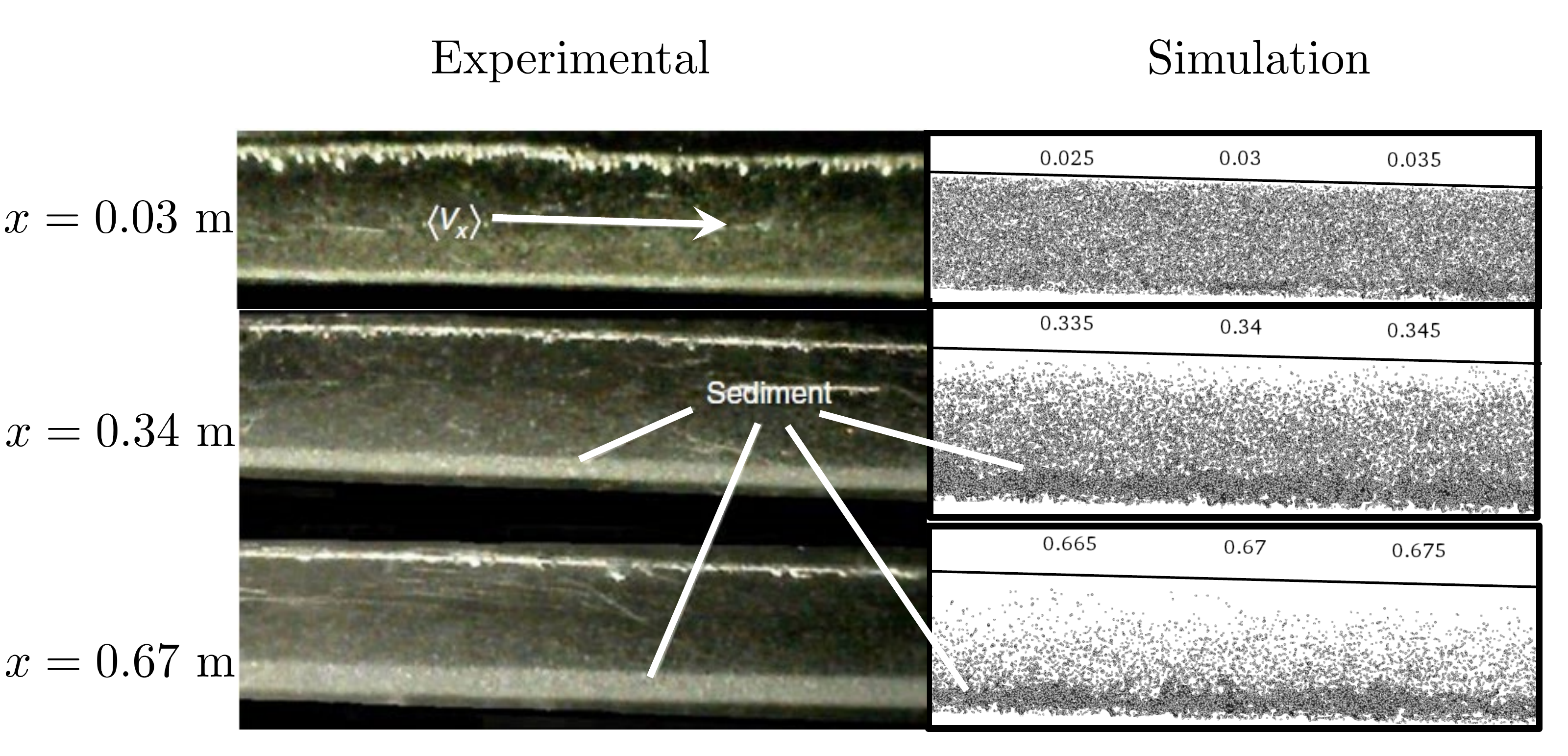}
\caption{Comparison between the experimental (left) \cite{Steven2020} and numerical (right) steady-state sedimentation heights in three different portions of the channel (corresponding to $x/H=6,~6.8$ and $13.4$) during flow of a suspension with a Newtonian matrix fluid with initial volume fraction $\phi_i = 0.05$ at a flow rate of $Q=100~$cm$^3$/h ($Re_W = 0.06832$).} 
\label{fig:expQ100Phi5}
\end{figure}

\begin{figure}[H]
\captionsetup[subfigure]{justification=justified,singlelinecheck=false}
    \centering
    {\renewcommand{\arraystretch}{0}
    \begin{tabular}{c@{}c}
    \begin{subfigure}[b]{.5\columnwidth}
        \centering
        \caption{{Suspended sediments under flow}}
        \includegraphics[width=\columnwidth]{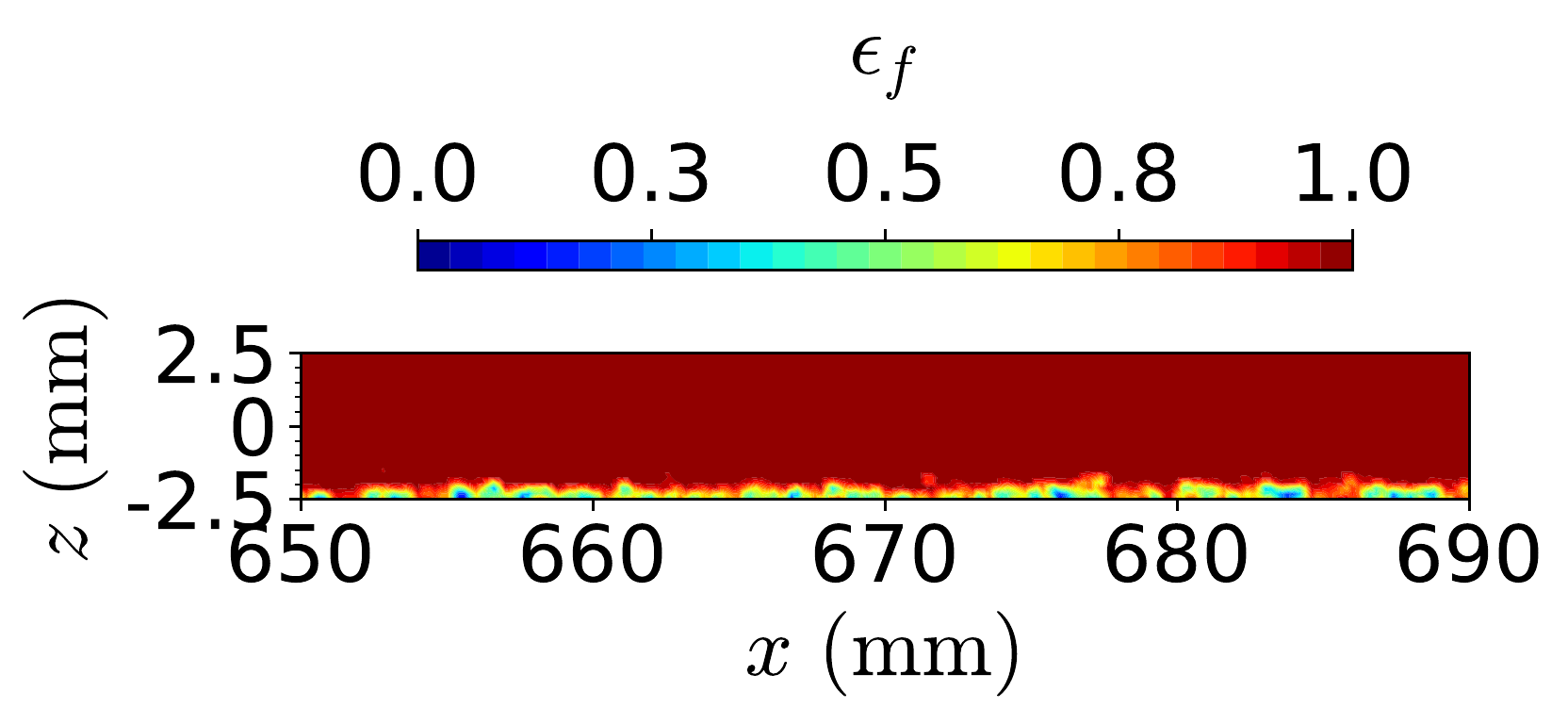}%
        \label{}
    \end{subfigure}&
    \begin{subfigure}[b]{.5\columnwidth}  
        \centering
        \caption{{Suspended sediments when the flow has ceased}}
        \includegraphics[width=\columnwidth]{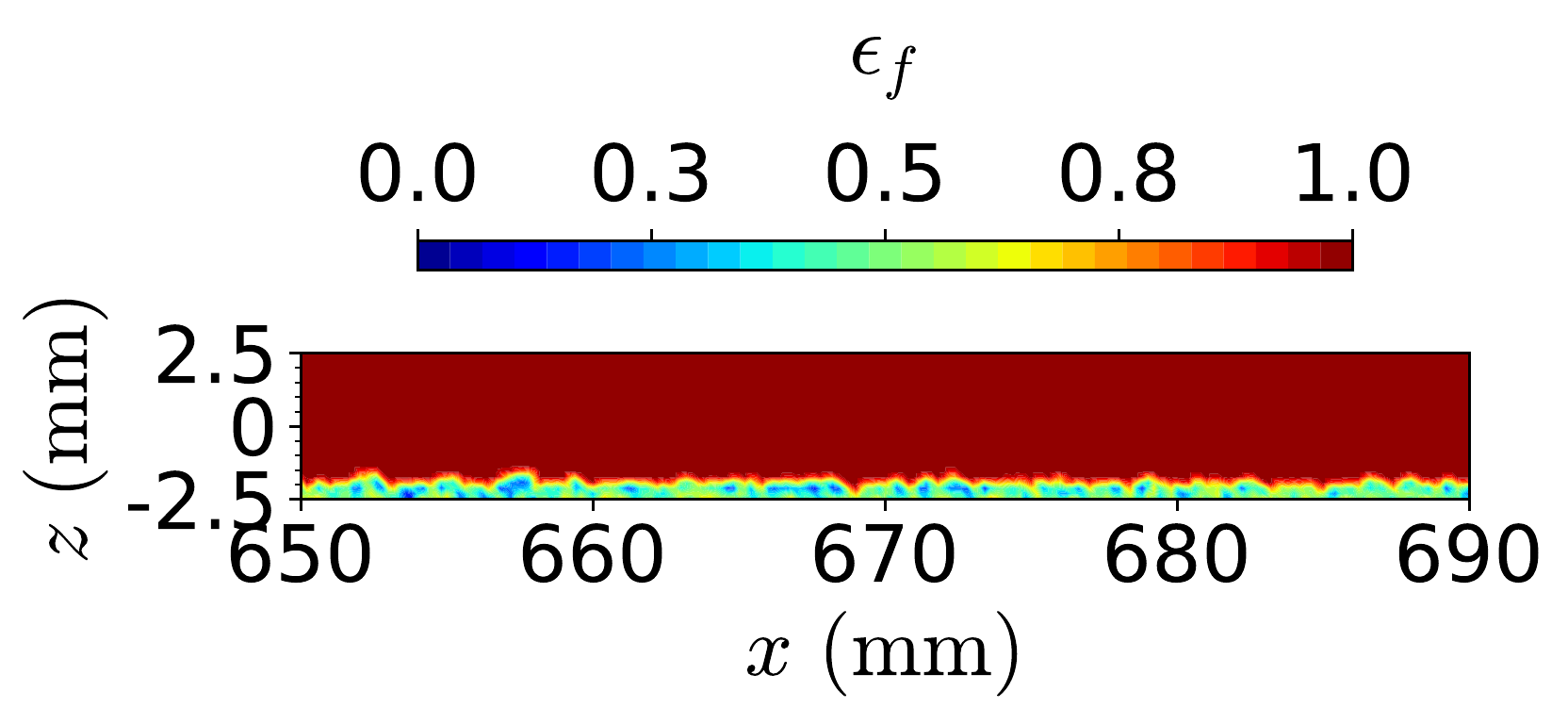}%
        \label{}
    \end{subfigure}\\
    \begin{subfigure}[t]{.5\columnwidth}   
        \centering 
        \vspace{1.5cm}
        \includegraphics[width=\columnwidth]{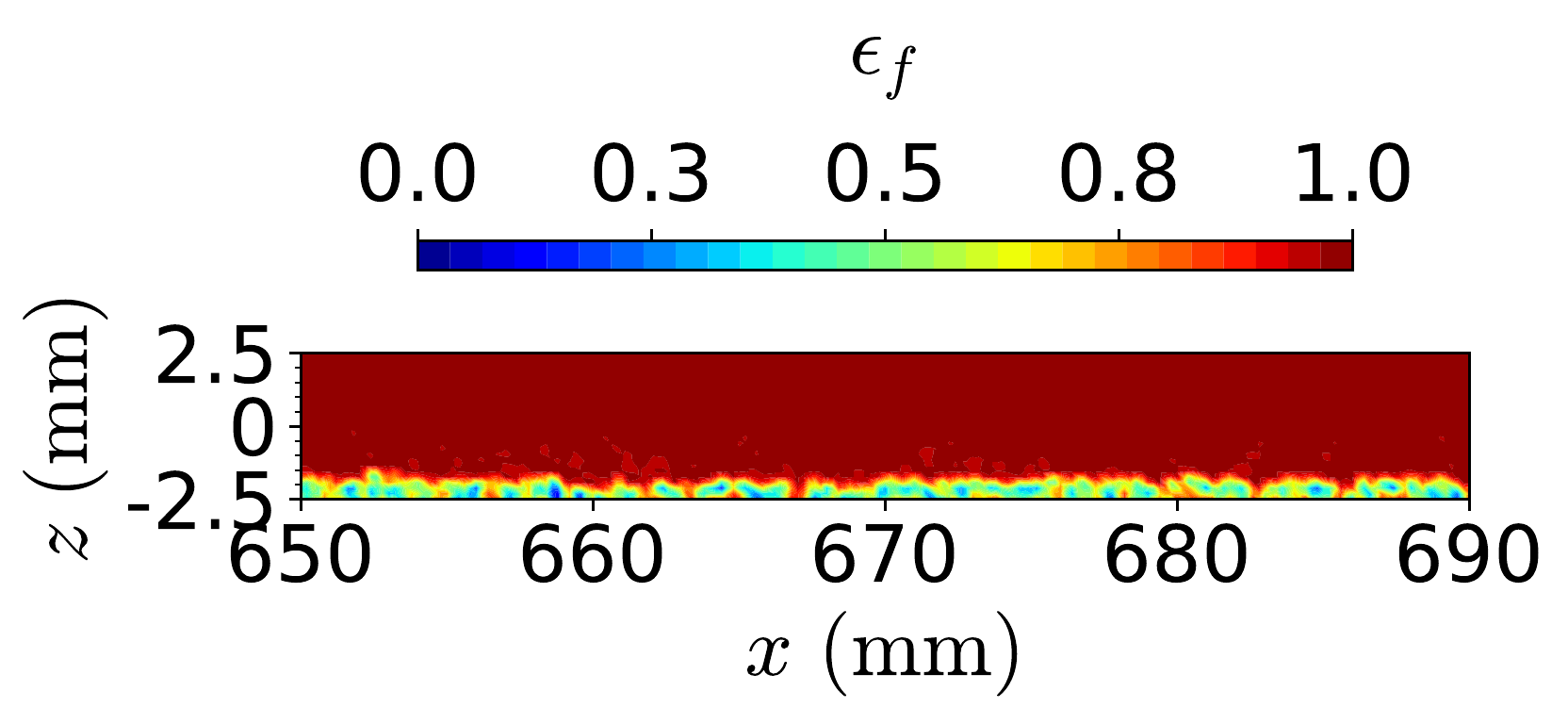}%
        \label{}
    \end{subfigure}&
    \begin{subfigure}[t]{.5\columnwidth}   
        \centering 
        \vspace{1.5cm}
        \includegraphics[width=\columnwidth]{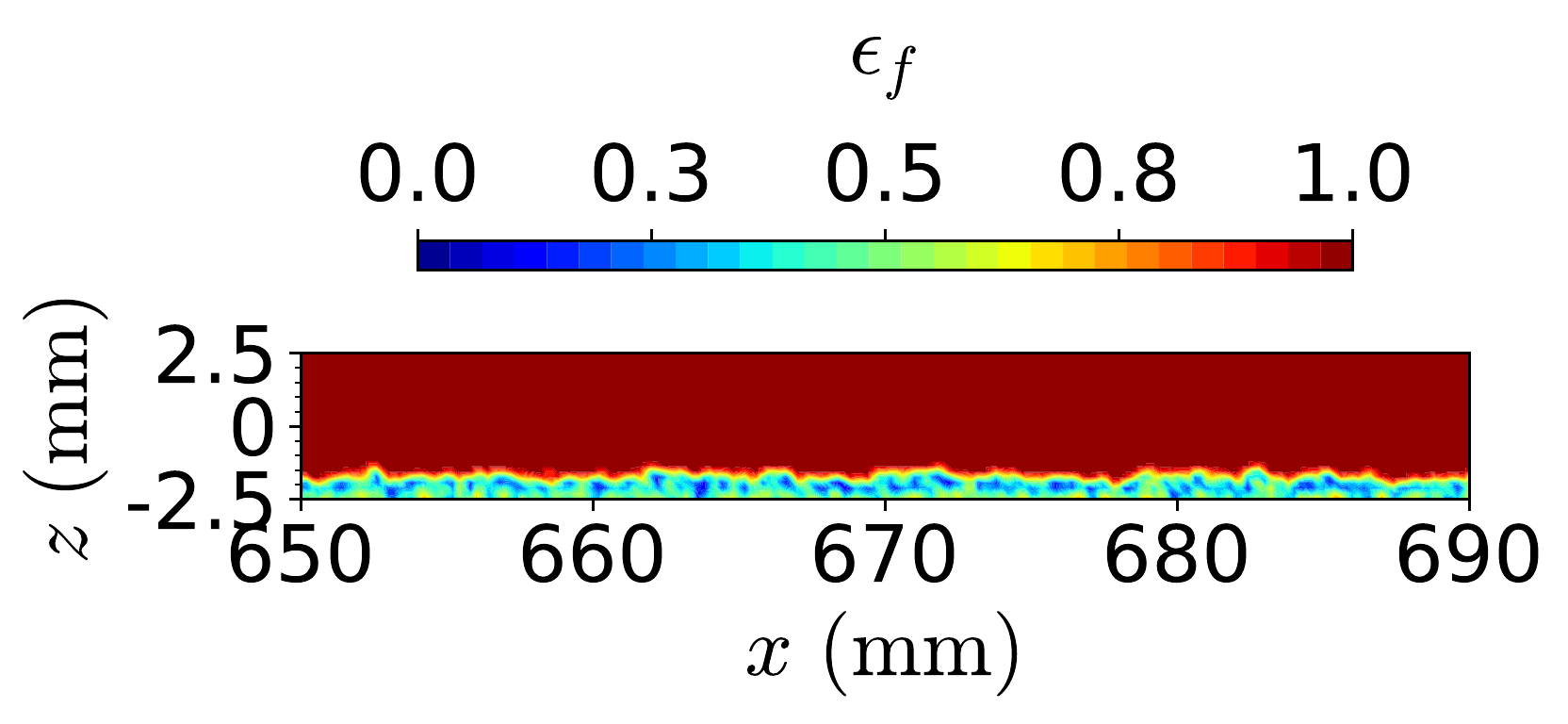}%
        \label{}
    \end{subfigure}\\
        \begin{subfigure}[t]{.5\columnwidth}   
        \centering 
        \vspace{1.5cm}
        \includegraphics[width=\columnwidth]{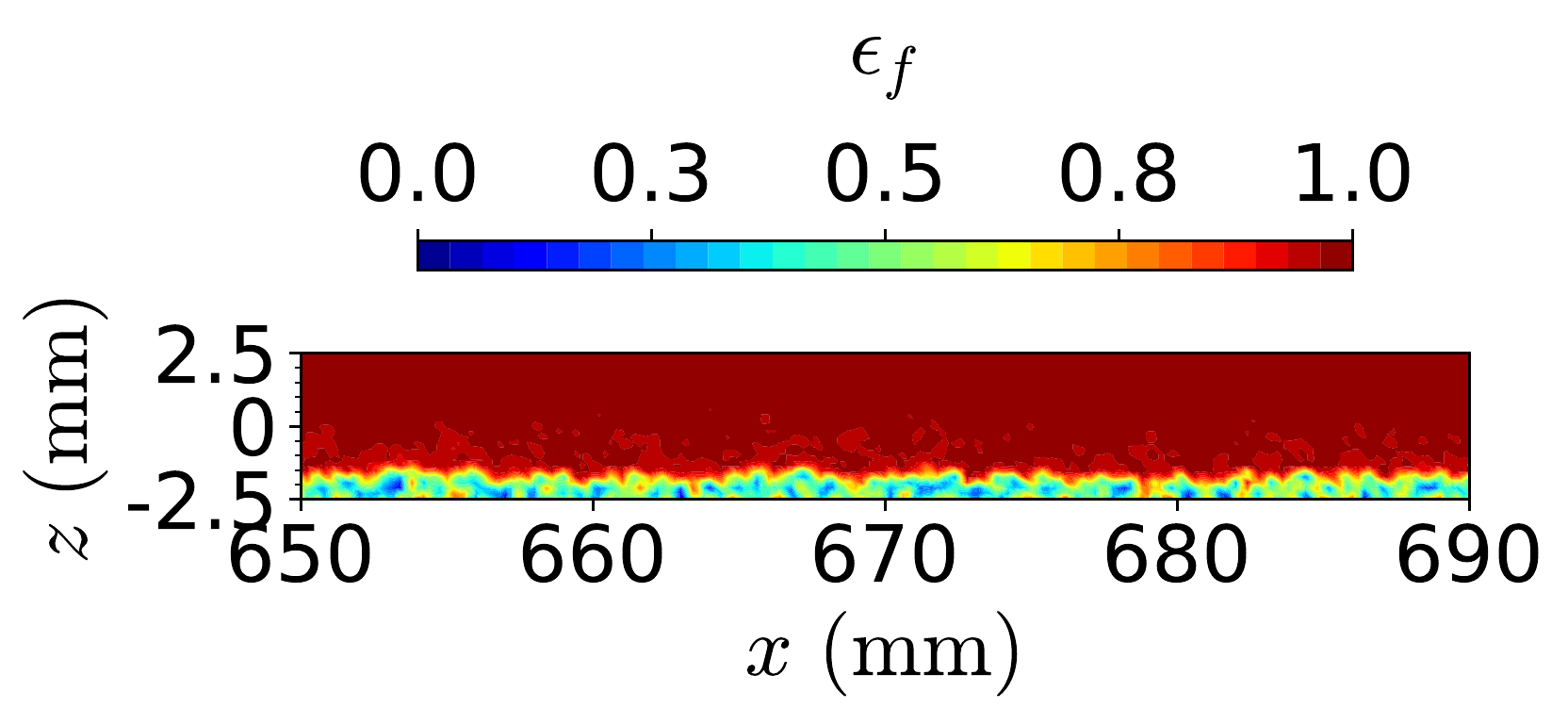}%
        \label{}
    \end{subfigure}&
    \begin{subfigure}[t]{.5\columnwidth}   
        \centering 
        \vspace{1.5cm}
        \includegraphics[width=\columnwidth]{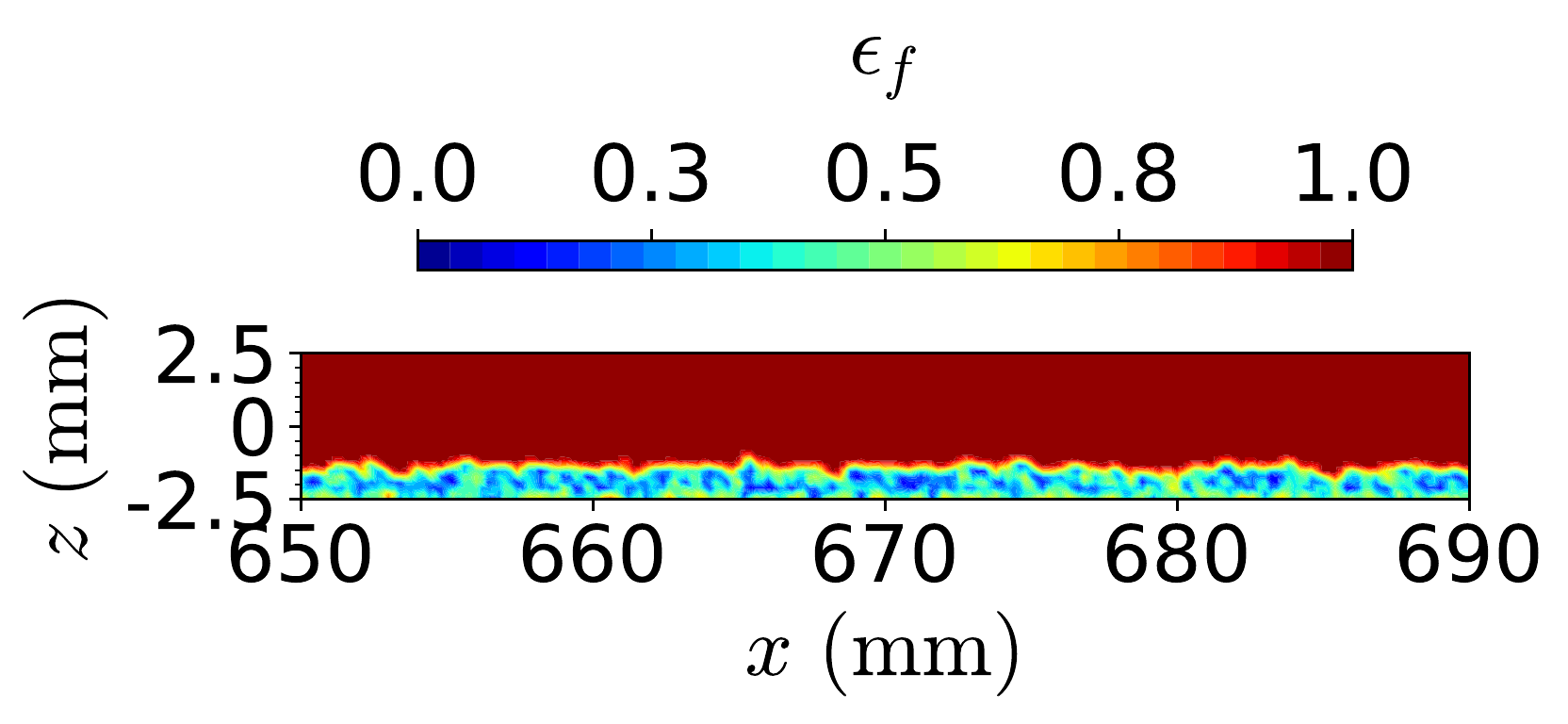}%
        \label{}
    \end{subfigure}
    \end{tabular}}
    \caption[]
    {Contours of the fluid porosity field $\epsilon_f$ as a function of the lateral position near $x\approx 67~$cm ($x/H=13.4$) and channel height $z$ for (a) steady-state measurements of suspended sediments under flow; and (b) static sediment heights, when the suspension flow has ceased, for initial suspension volume fraction $\phi_i=0.025~(\textrm{top});~0.038~(\textrm{middle})~\textrm{and}~0.05~(\textrm{bottom})$, conducted at a flow rate $Q=100~$cm$^3$/h, corresponding to $Re_W = 0.06832$ and $El=0$.} 
    \label{fig:contoursEpsilon}
\end{figure}

Fig.~\ref{fig:expQ100Phi5} shows a visual comparison of experimental and numerical simulation results for steady-state sedimentation heights in three different portions of the channel length ($x=0.03;~0.34;~\textrm{and}~0.67$~m), using an initial suspension volume fraction of $\phi_i=0.05$ and a flow rate $Q=100~$cm$^3$/h. The numerical results for the axial sediment distribution follow a similar trend as those obtained experimentally \cite{Steven2020}, i.e., after the detection of the suspension in each channel section (as indicated by the slight turbidity compared with the pure carrier fluid) the progressive buildup of an opaque particle sediment along the channel floor is observed. This is particularly noticeable in the second and third channel observation sections, in which the sediment height increases markedly. Then, the sediment height ceases to grow further quite abruptly, although the particle suspension continues to flow through the channel. This steady-state behavior persists for the duration of the flow and represents a balance between sedimentation and shear-driven resuspension. Upon cessation of flow we immediately observe a collapse in the dense phase height, i.e. the particle phase settles further to form a more compact final sedimented state. Figure~\ref{fig:contoursEpsilon} shows the contours of the fluid porosity distribution $\epsilon_f(x,z)$ under steady flow conditions, as well as the change in the sediment height after flow has ceased, for a lateral position $x\approx 67$~cm and initial suspension volume fractions $\phi_i=0.025;~0.038~\textrm{and}~0.05$ conducted at flow rate $Q=100$~cm$^3$/h ($Re_W = 0.06832$). A pure fluid phase with no particles corresponds to $\epsilon_f\to 1.0$. For random close packed spheres we expect the viscosity to diverge when $\phi\to\phi_m=0.637$ \cite{faroughi2015generalized}.  This value changes significantly based on the shape of  particles \cite{faroughi2017self} and the size ratio between particles in the pack \cite{faroughi2014crowding,faroughi2016theoretical}. Contours of $\epsilon_f \lesssim 0.4$ thus correspond to effectively solid deposits of particles. Again we can conclude that the steady-state sediment height during suspension flow increases with the initial suspension volume fraction, and that following the cessation of flow the sediment bed compactifies and reduces in height.

\begin{figure}[H]
\captionsetup[subfigure]{justification=justified,singlelinecheck=false}
    \centering
    {\renewcommand{\arraystretch}{0}
    \begin{tabular}{c@{}c@{}}
    \begin{subfigure}[b]{.6\columnwidth}
        \centering
        \caption{{}}
        \includegraphics[width=1\columnwidth]{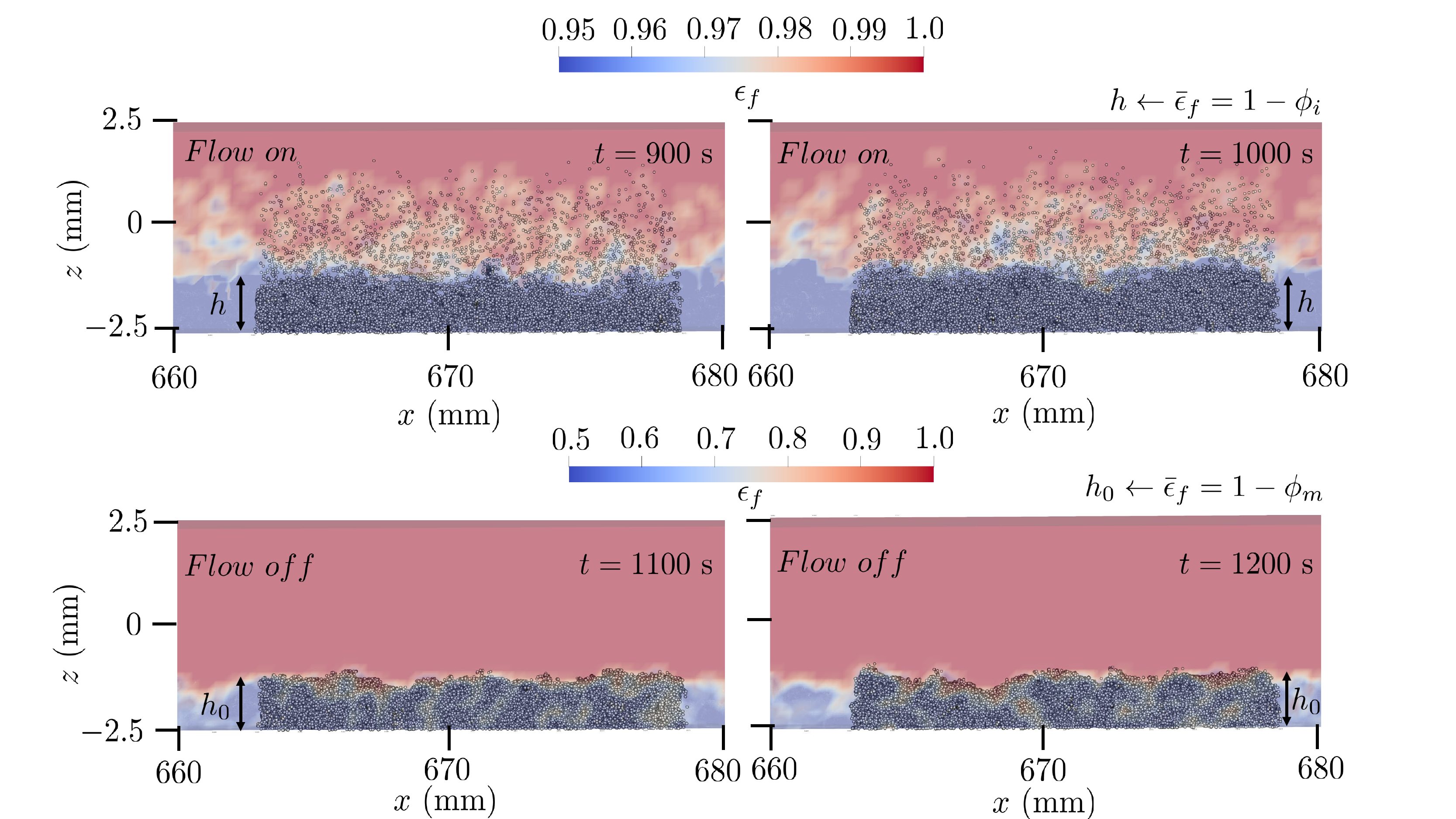}%
        \label{}
    \end{subfigure}&
    \begin{subfigure}[b]{.4\columnwidth}  
        \centering
        \caption{{}}
        \includegraphics[width=0.9\columnwidth]{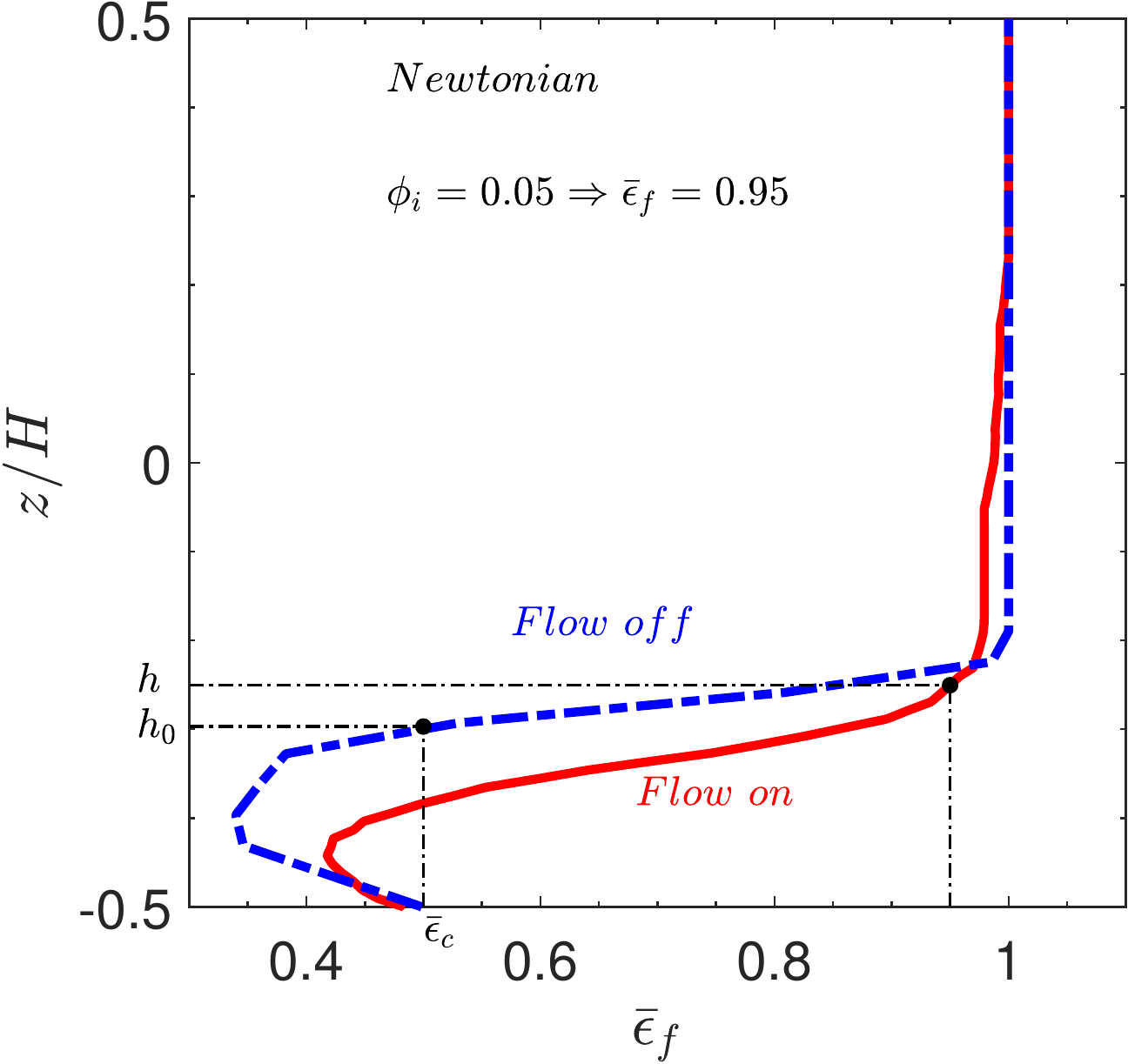}%
        \label{}
    \end{subfigure}\\
    \begin{subfigure}[b]{.6\columnwidth}   
        \centering 
        \caption{{}}
        \includegraphics[width=0.6\columnwidth]{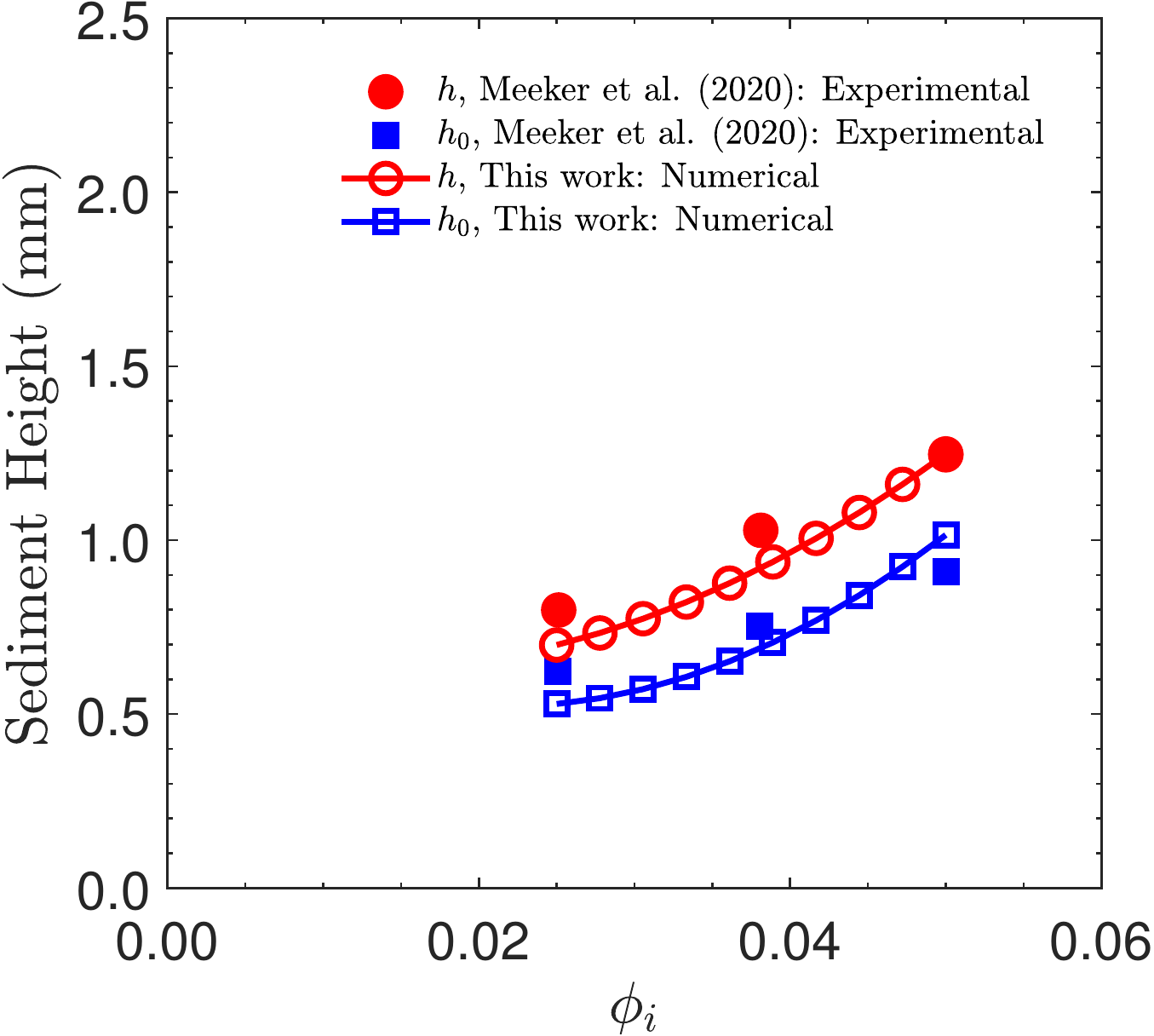}%
        \label{}
    \end{subfigure}&
    \begin{subfigure}[b]{.4\columnwidth}   
        \centering 
        \caption{{}}
        \includegraphics[width=0.9\columnwidth]{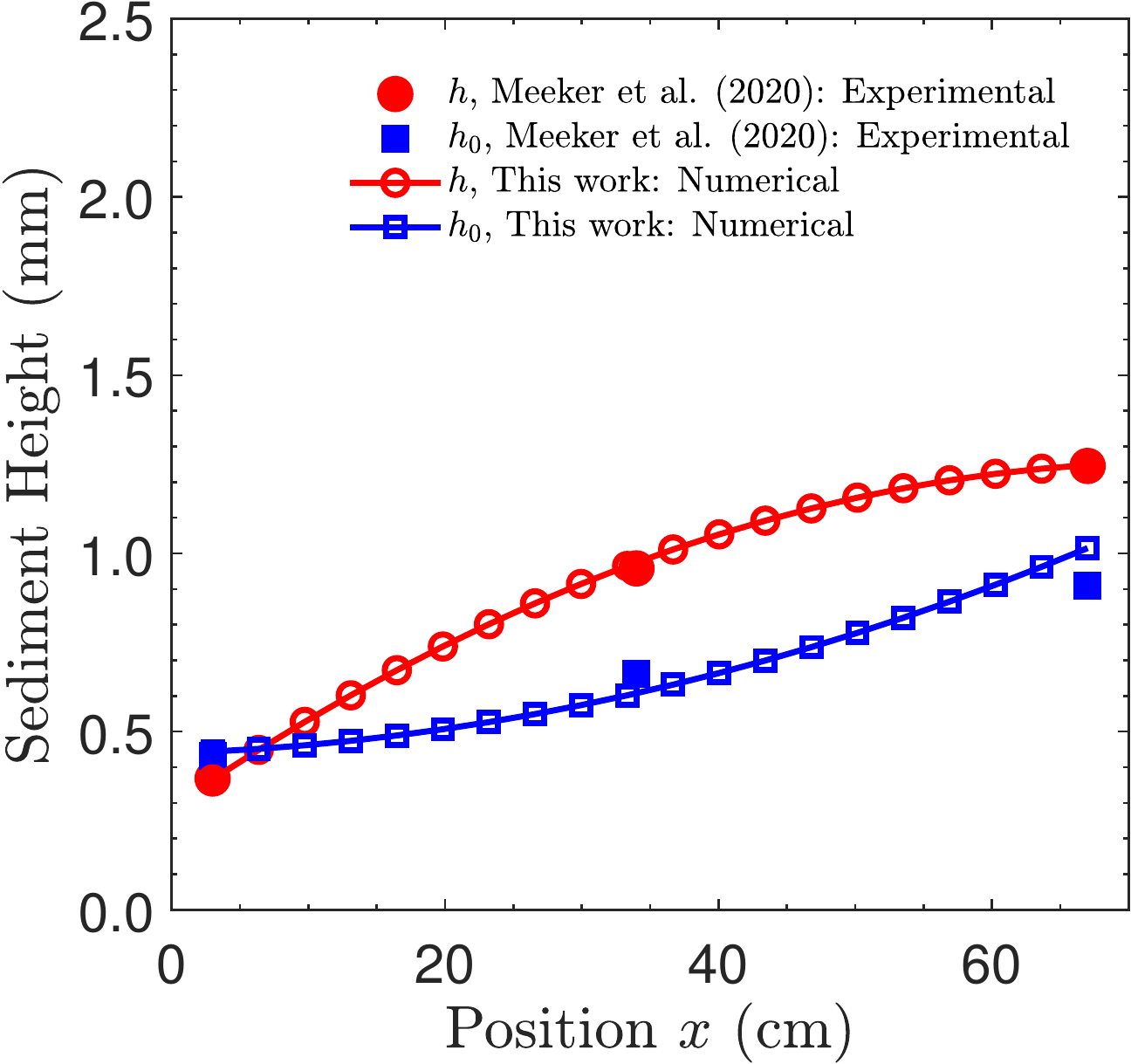}%
        \label{}
    \end{subfigure}\\
    \end{tabular}}
    \caption[]
    {Sediment height measurements under steady flow conditions with height $h$; and static sediment heights, $h_0$, when the suspension flow has ceased. (a) contours of fluid porosity $\epsilon_f$ and particle distribution for initial suspension volume fraction $\phi_i=0.05$ in the last observable channel section $x\approx 67~$cm ($x/H=13.4$), (b) average fluid porosity distribution $\bar{\epsilon}_f$ along channel height direction $z$ at $x\approx 67~$cm, (c) sediment heights for initial suspension volume fraction $\phi_i=0.025,~0.038~\textrm{and}~0.05$ in the last observable channel section $x\approx 67~$cm ($x/H=13.4$) and (d) evolution of sediment heights along the channel direction $x$ for $\phi_i=0.05$, conducted at a flow rate $Q=100~$cm$^3$/h, corresponding to $Re_W = 0.06832$ and $El=0$.} 
    \label{fig:expvsOFQ100}
\end{figure}

To quantitatively define the steady-state sediment height, $h$, under flow and the static sediment height, $h_0$, once the flow of the suspension has ceased, we compute the average fluid porosity $\bar{\epsilon}_f$ over a local section as
\begin{equation}
\begin{aligned}
\bar{\epsilon}_f = \frac{1}{H}\int_{x-H/2}^{x+H/2}\epsilon_f(x,z)dx,
\end{aligned}
\label{eq:fluid_porosity}
\end{equation}
and define appropriate characteristic values for $\bar{\epsilon}_f$ to quantify $h$ and $h_0$. For particular flow conditions of a non-Brownian suspension flowing at $Q=100$~cm$^3$/h, \citet{Steven2020} observed the buildup of a dense but flowing sediment that rapidly reaches a steady-state height $h$. \textit{The existence of this steady-state flowing sediment implies that the proppant flux leaving the channel equals that entering the channel, and thus, an ``efficient" proppant transport occurs.} Knowing this fact, we define the criteria to compute $h$ as $\bar{\epsilon}_f=1-\phi_i$ (see Fig.~\ref{fig:expvsOFQ100}(a)). Because the flow is at a low Reynolds number ($Re_W = 0.06832$), the relevant mechanism of sediment transport must be viscous resuspension (flow of an ``expanded" sediment at an equilibrium height, and  its subsequent ``collapse" once the flow ceases \cite{Leighton1986,Acrivos1993}). To quantify $h_0$, we quote the work of \citet{Steven2020} stating that \textit{for quiescent conditions the packing volume fraction when water is the suspending fluid is} $\phi_p\approx 0.58$,\textit{ which is close to} $\phi_m$. \textit{However, when the 85:15 w/w}$\%$ \textit{glycerol/water mixture (viscosity} $\approx 0.1$~Pa.s) \textit{was employed as the suspending fluid, the packing fraction decreased to} $\phi_p\approx 0.5$. Following this, we consider the criterion to compute $h_0$ as $\bar{\epsilon}_c=0.5$ to represent a dense/compact suspension bed (see Fig.~\ref{fig:expvsOFQ100}(a)). 

An example of these computations is shown in Fig.~\ref{fig:expvsOFQ100}(b), which depicts the evolution of the average fluid porosity distribution, $\bar{\epsilon}_f$, along the channel height, $z$. This is shown for the last observable channel section $x\approx 67~$cm ($x/H=13.4$) for initial suspension volume fraction $\phi_i=0.05$. Notice that the sharply inflected ``elbow" shape of the average fluid porosity distribution very close to the bottom wall of the rectangular channel is due to the random-packing which occurs in this region, where we may have either void spaces (fluid) or particles close to the contact plane at the wall. In Fig.~\ref{fig:expvsOFQ100}(c), we compare the evolution of $h$ (circular symbols) and $h_0$ (square symbols) against the experimental results obtained by \citet{Steven2020}, for initial suspension volume fractions $\phi_i=0.025;~0.038~\textrm{and}~0.05$ at $x\approx 67$~cm ($x/H=13.4$). Both the local bed height under flow ($h$) and the compact bed at rest ($h_0$) follow the same trend of the experimental data presented in \citet{Steven2020}, where monotonic increases of the sediment heights are observed with the increase of the initial suspension volume fraction. Finally, in Fig.~\ref{fig:expvsOFQ100}(d) we compare the evolution of $h$ (circular symbols) and $h_0$ (square symbols) against the experimental results obtained by \citet{Steven2020}, for initial suspension volume fraction $\phi_i=0.05$ along the channel length direction $x$. Again our numerical results follow the experimental data of \citet{Steven2020}, showing an increase of the sedimentation heights along the channel length. These results verify the robustness of our coupled Eulerian-Lagrangian technique to model the sedimentation phenomena which occur in many applications, e.g., during hydraulic fracture processes.

\begin{figure*}
\captionsetup[subfigure]{justification=justified,singlelinecheck=false}
    \centering
    {\renewcommand{\arraystretch}{0}
    \begin{tabular}{c@{}c}
    \begin{subfigure}[b]{.5\columnwidth}
        \centering
        \caption{{$El=0$}}
        \includegraphics[width=\columnwidth]{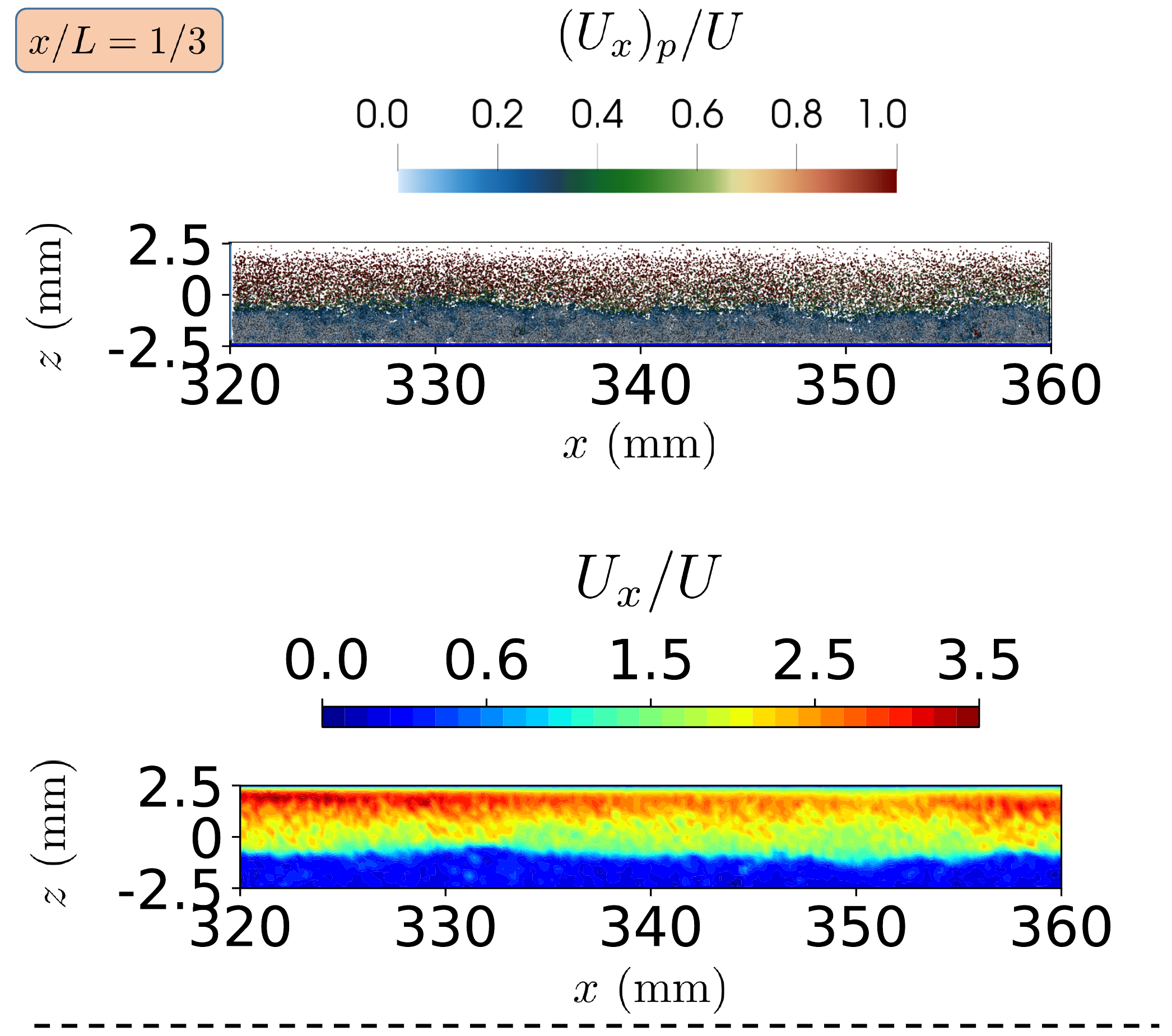}%
        \label{}
    \end{subfigure}&
    \begin{subfigure}[b]{.5\columnwidth}  
        \centering
        \caption{{$El=30$}}
        \includegraphics[width=\columnwidth]{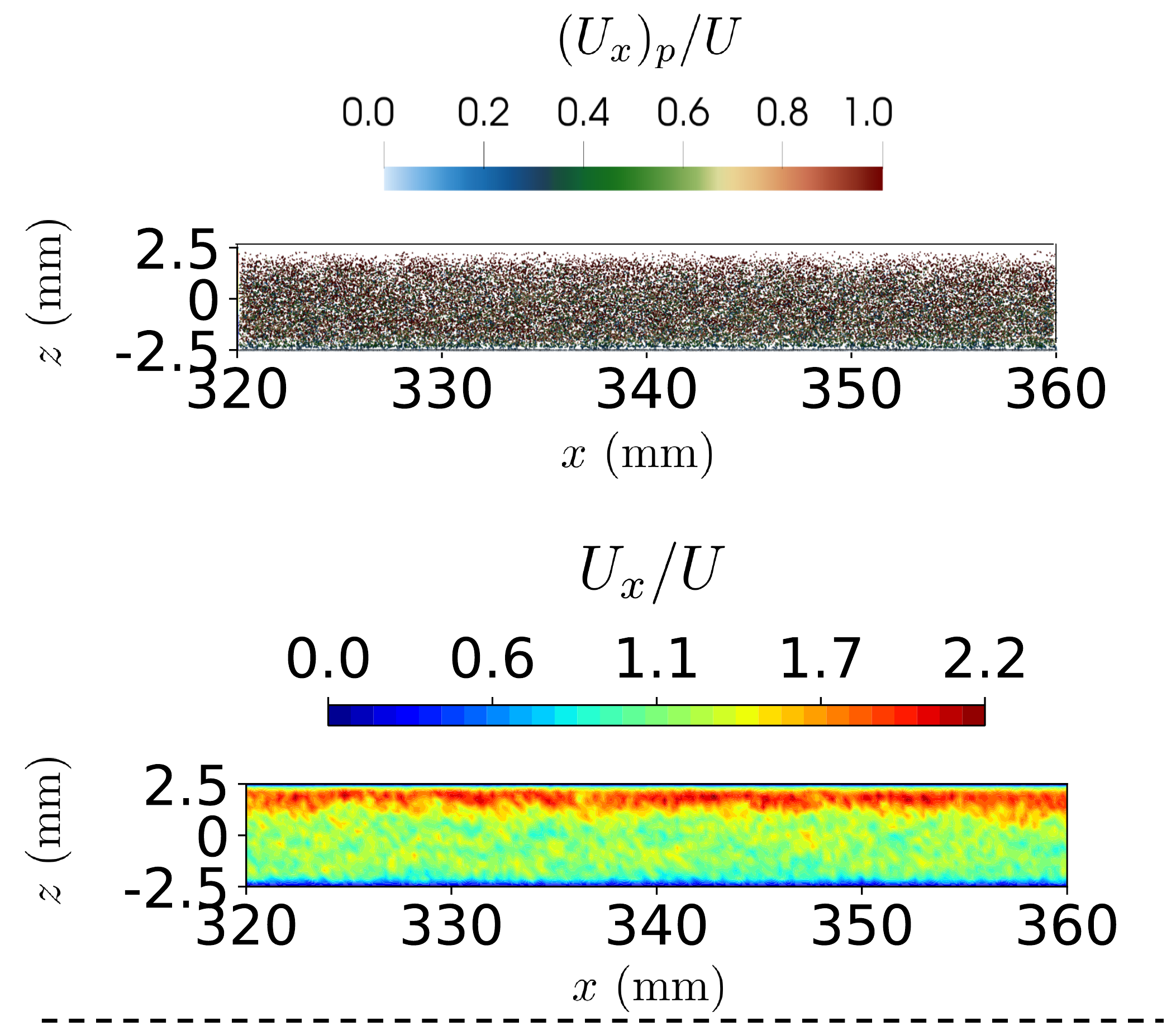}%
        \label{}
    \end{subfigure}\\
    \begin{subfigure}[t]{.5\columnwidth}   
        \centering 
        \vspace{1.5cm}
        \includegraphics[width=\columnwidth]{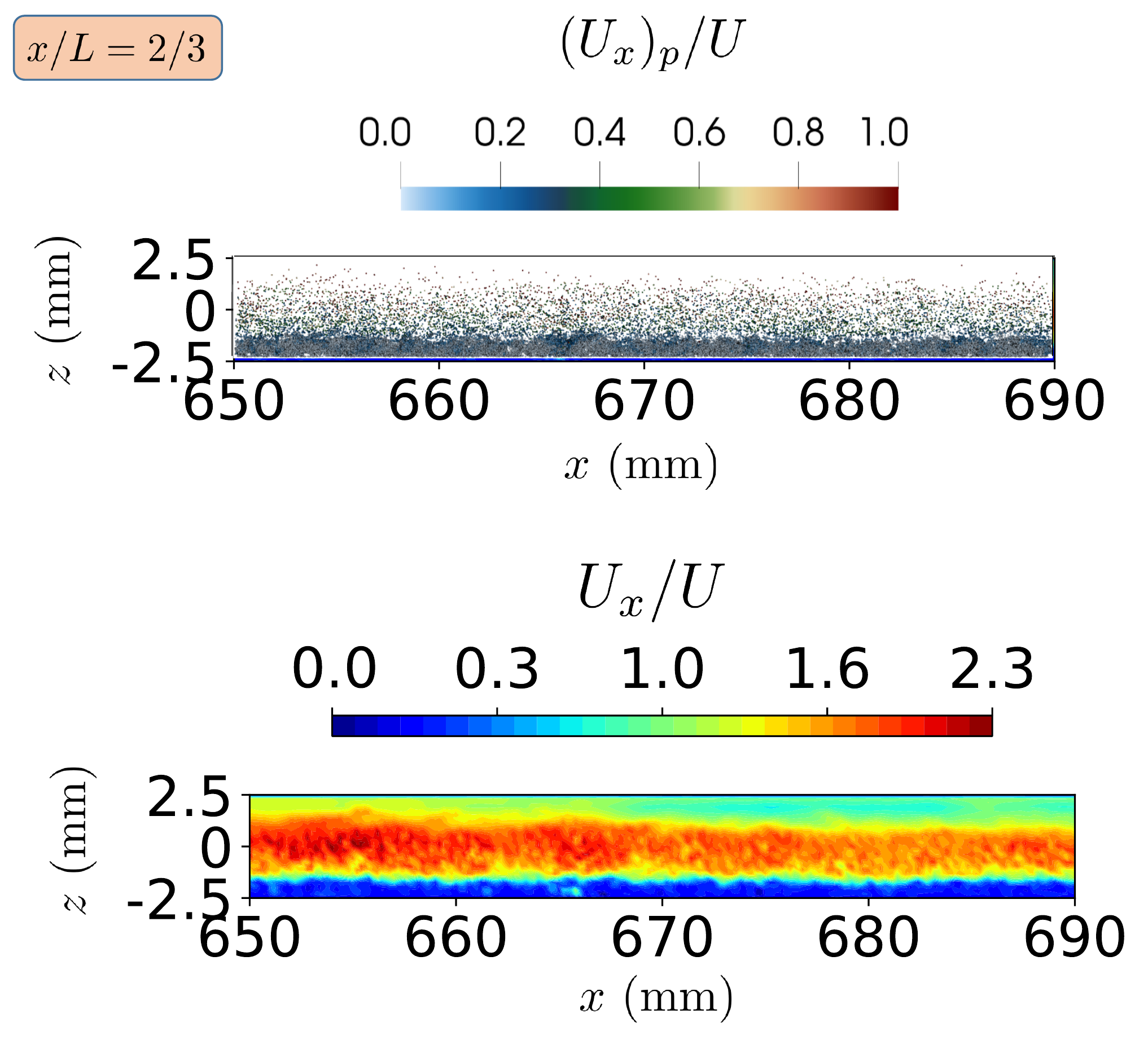}%
        \label{}
    \end{subfigure}&
    \begin{subfigure}[t]{.5\columnwidth}   
        \centering 
        \vspace{1.5cm}
        \includegraphics[width=\columnwidth]{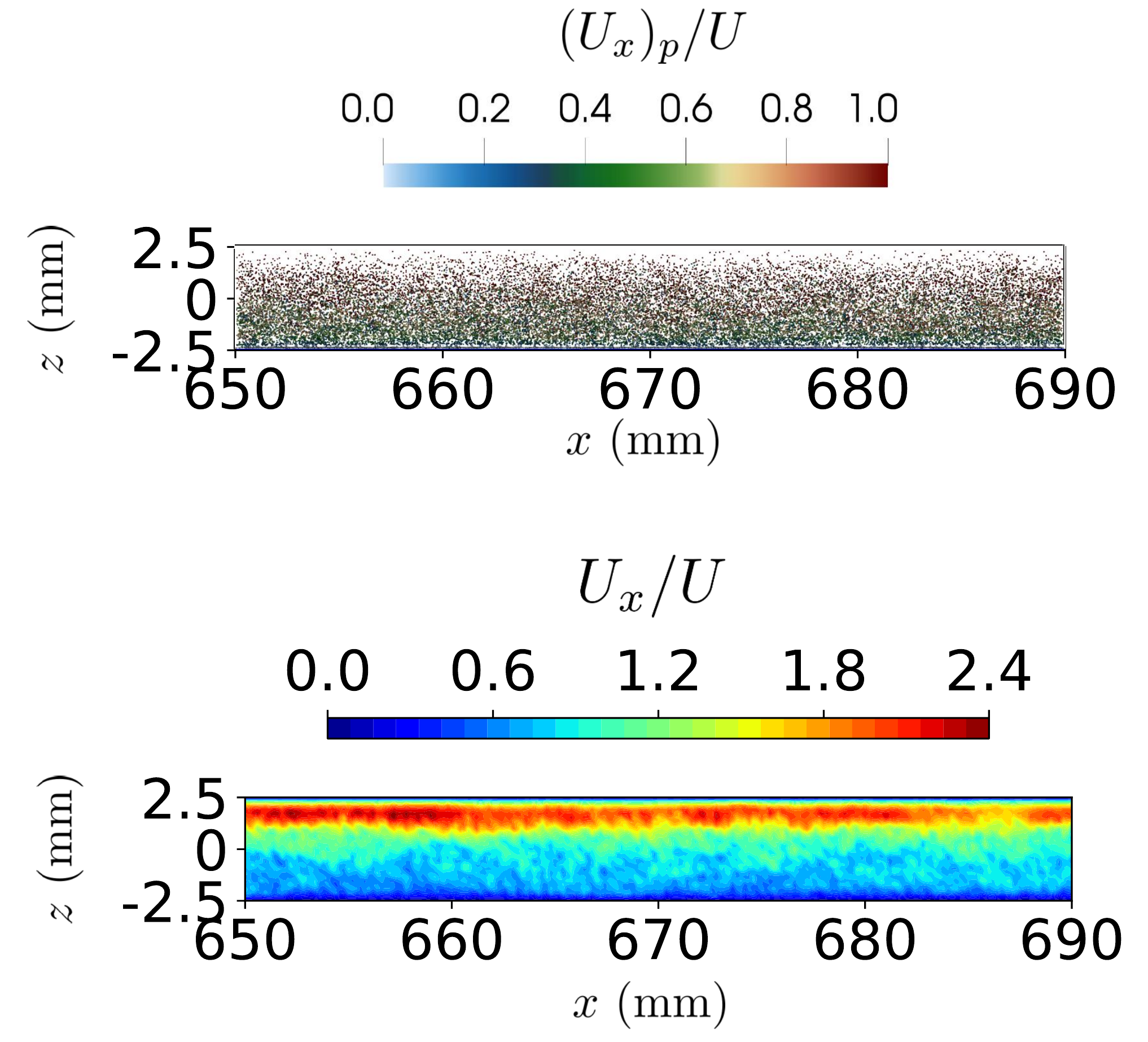}%
        \label{}
    \end{subfigure}\\
    \end{tabular}}
    \caption[]
    {Velocities of the settling particles (first and third rows of images) and the matrix fluid (second and fourth rows of images) at $1/3$ (top) and $2/3$ (bottom) of the channel length $L$ for (a) $El = 0$ (left column) and (b) $El = 30$ (right column), with $Re_W=0.06832$ and $\phi_i=0.05$.} 
    \label{fig:phiXElYSF}
\end{figure*}

Finally, we simulate the sedimentation of particle-laden Oldroyd-B viscoelastic fluids using the newly-developed $DPMviscoelastic$ solver and fluid drag model (see Eq.~(\ref{eq:fit}) in Section~\ref{sec:viscoelasticDrag}). Figure~\ref{fig:phiXElYSF} shows contours of the particle and fluid velocity fields for $El = Wi/Re = 0$ (Newtonian flow) and $El = 30$ at $1/3$ and $2/3$ of the channel length $L$ for initial suspension volume fraction $\phi_i=0.05$, conducted at flow rate $Q=100$~cm$^3$/h ($Re_W = 0.06832$).
For the Newtonian fluid (Fig.~\ref{fig:phiXElYSF}(a)) at $El=0$, when analyzing the particle distribution at $x=L/3$ (top left images), we notice that there is a significant sedimentation layer where the particle velocity is zero. In the middle and top zones of the channel, both the matrix fluid and particles flow smoothly and axially along the channel. The particles continue to slowly sediment and eventually join the deposited layer with $(U_x)_p\to 0$.
For the quasi-linear Oldroyd-B viscoelastic fluid (Fig.~\ref{fig:phiXElYSF}(b)) at $El=30$, when analyzing the particle distribution at $x=L/3$ (top right images), we notice that the distribution of particle velocities is almost uniform along the channel height, and only a thin layer of sedimentation is observed. This is in contrast to the Newtonian fluid behavior. At $x/L=2/3$, the behavior of the particles and fluid constituents is not substantially changed from that at $x/L=1/3$, and, therefore, there is no significant particle settling zone along the channel floor for an elastic fluid with $El=30$. 

\begin{figure}[H]
\centering
\includegraphics[scale=0.5]{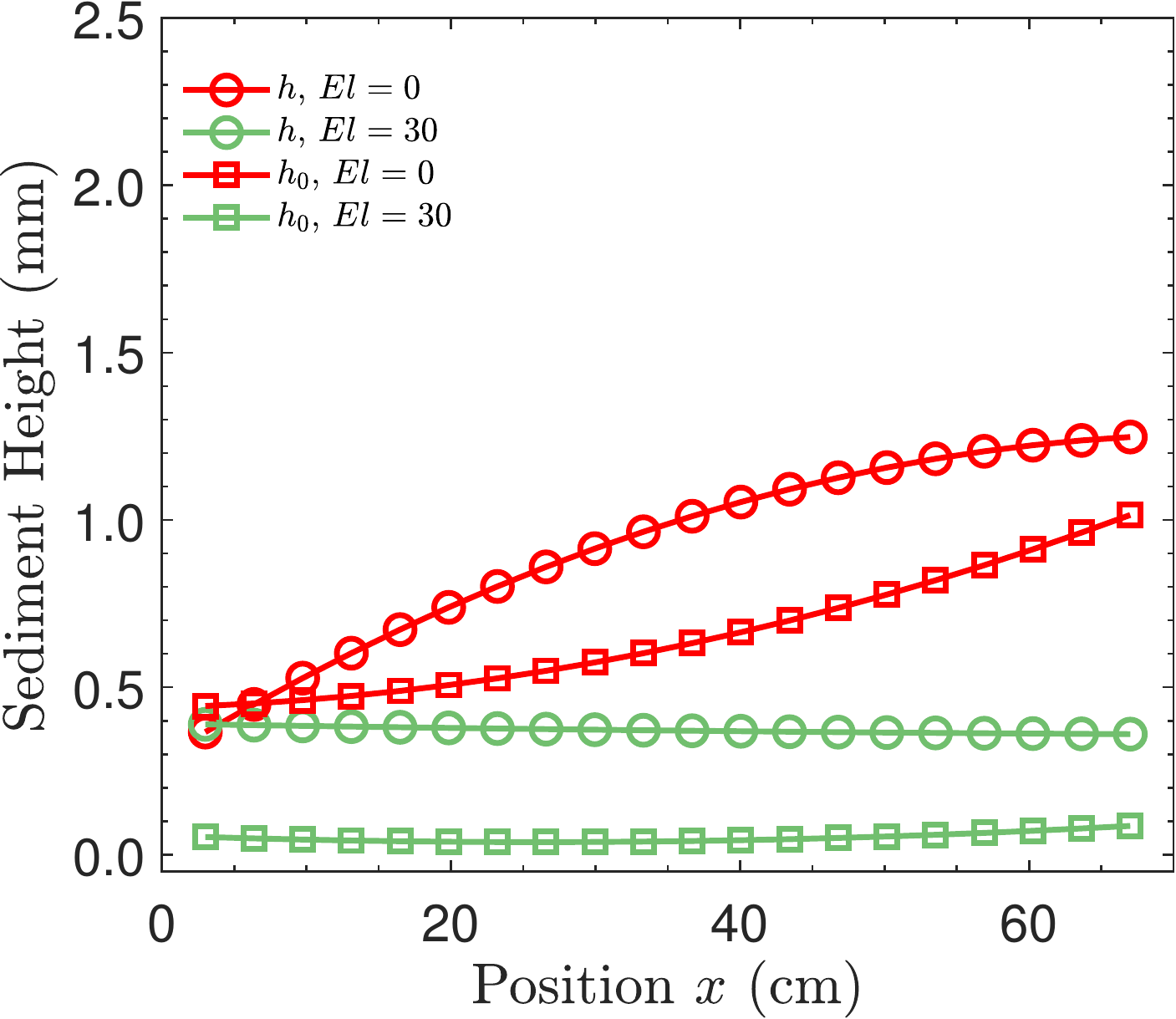}
\caption{Evolution in sediment height along the channel length direction for steady-state measurements of sediments under flow, height $h$; and when the suspension flow has ceased, static sediment heights $h_0$ are compared at $El=0$ and $El=30$, with $Re_W=0.06832$ and $\phi_i=0.05$.} 
\label{fig:NewtVsVisc}
\end{figure}

Figure~\ref{fig:NewtVsVisc} shows the evolution of the sedimentation heights, $h$ and $h_0$, along the channel length for both $El=0$ and $El=30$. As before, we compute $h$ and $h_0$ using the definition of $\bar{\epsilon}_f$ in Eq.~(\ref{eq:fluid_porosity}). The sedimentation heights are nearly constant along the channel length for the viscoelastic case ($El=30$), contrarily to the Newtonian case ($El=0$) where a progressive increase of the sedimentation height $h(x)$ is observed. This indicates that the addition of fluid viscoelasticity hinders the settling of particles through the viscoelastic enhancement of the drag coefficient. This can be extremely helpful to identify fluid formulations that can improve proppant transport in hydraulic fracturing operations in long rectangular channels/cracks. In the future, we will use this numerical framework to further explore other types of fluid rheologies, specifically with shear-thinning rheology and non-zero second normal stress differences as captured by the Giesekus fluid model. The resulting numerical framework can also be applied to other migration and settling problems in the industrial and life science fields, e.g. in the circulatory system. 

\subsection{Annular pipe flow}
Particle segregation in pumped concrete is one of the big challenges encountered when creating casing for horizontal drilled wells \cite{Faroughi2017,robisson2020}. In this case, particulate solids tend to segregate axially across the pipe due to differences in the size, density, shape and other properties of the constituent phases.  The corresponding increase in the percentage of cementitious particles in the bottom part of the casing increases the chance of shrinkage and formation of cracks in the upper portion of the cemented casing. These cracks, often large in size, can easily transport hydrocarbons and other toxic chemicals into the formation which is a concern. Tuning the rheology of the conveying fluids systematically by considering the hindrance effect (i.e., reduction in the relative settling velocity of a particle due to the presence of other particles) can help minimize this issue. 

\begin{figure}[H]
\centering
\includegraphics[scale=0.5]{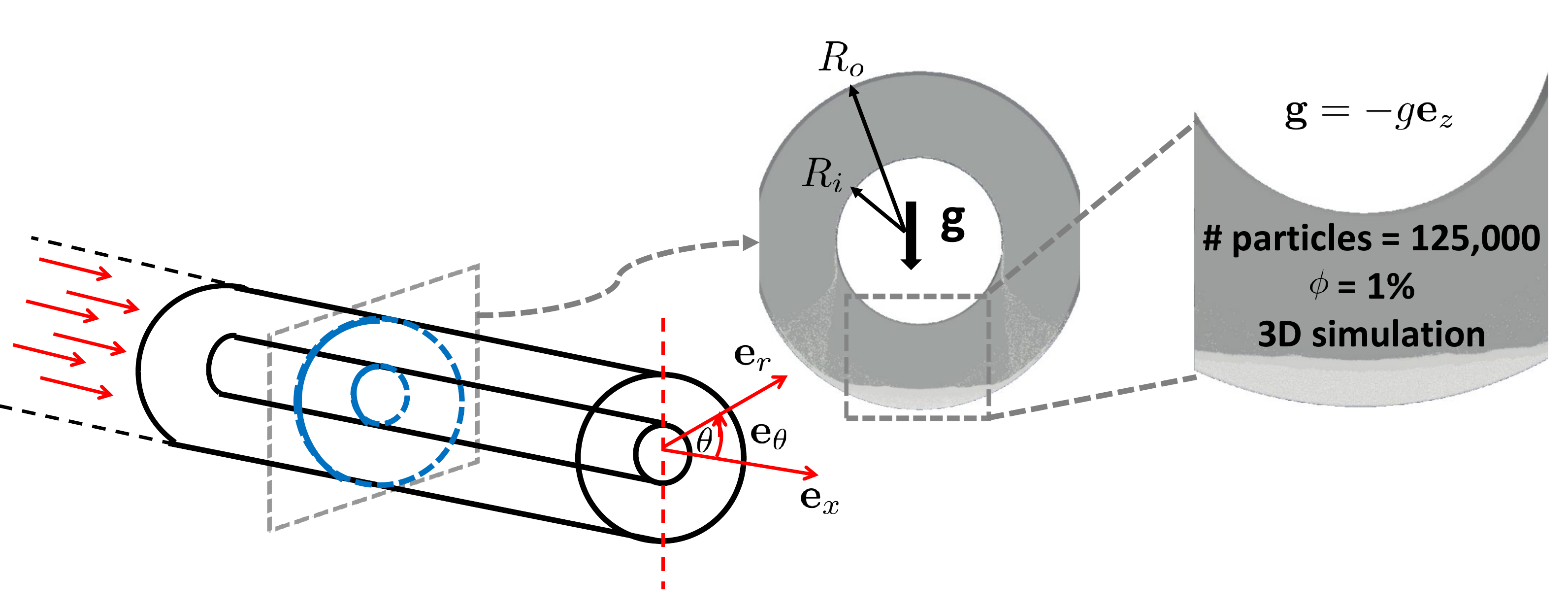}
\caption{Schematic of the annular pipe channel cross-section used for simulating suspensions of particles in horizontal wells. The axis of the pipe is denoted by the $x-$direction for consistency with Fig.~\ref{fig:DNSChannel} and gravity is aligned in the $-\textbf{e}_z$ direction.} 
\label{fig:eccentricflowGeometry}
\end{figure}

Here we study numerically the particle segregation in a simplified annular pipe geometry. The setup used to study the particle segregation is shown schematically in Fig.~\ref{fig:eccentricflowGeometry}. The channel interior, which is used to mimic an horizontal well, has an annular cross section with inner ($R_i$) and outer ($R_o$) radius of $25~$mm and $50~$mm, respectively, and a depth of $10~$mm. The particles have a diameter of 200 $\upmu\mathrm{m}$ and density of 5 g/cm$^3$. The carrier fluid is a Newtonian silicone oil with a constant viscosity of 0.01 Pa$\mathpunct{.}$s and density of 1 g/cm$^3$.  

\begin{figure*}
\captionsetup[subfigure]{justification=justified,singlelinecheck=false}
    \centering
    {\renewcommand{\arraystretch}{0}
    \begin{tabular}{c@{} c c@{}}
    \begin{subfigure}[b]{.33\columnwidth}
        \centering
        \caption{{}}
        \includegraphics[width=0.7\columnwidth]{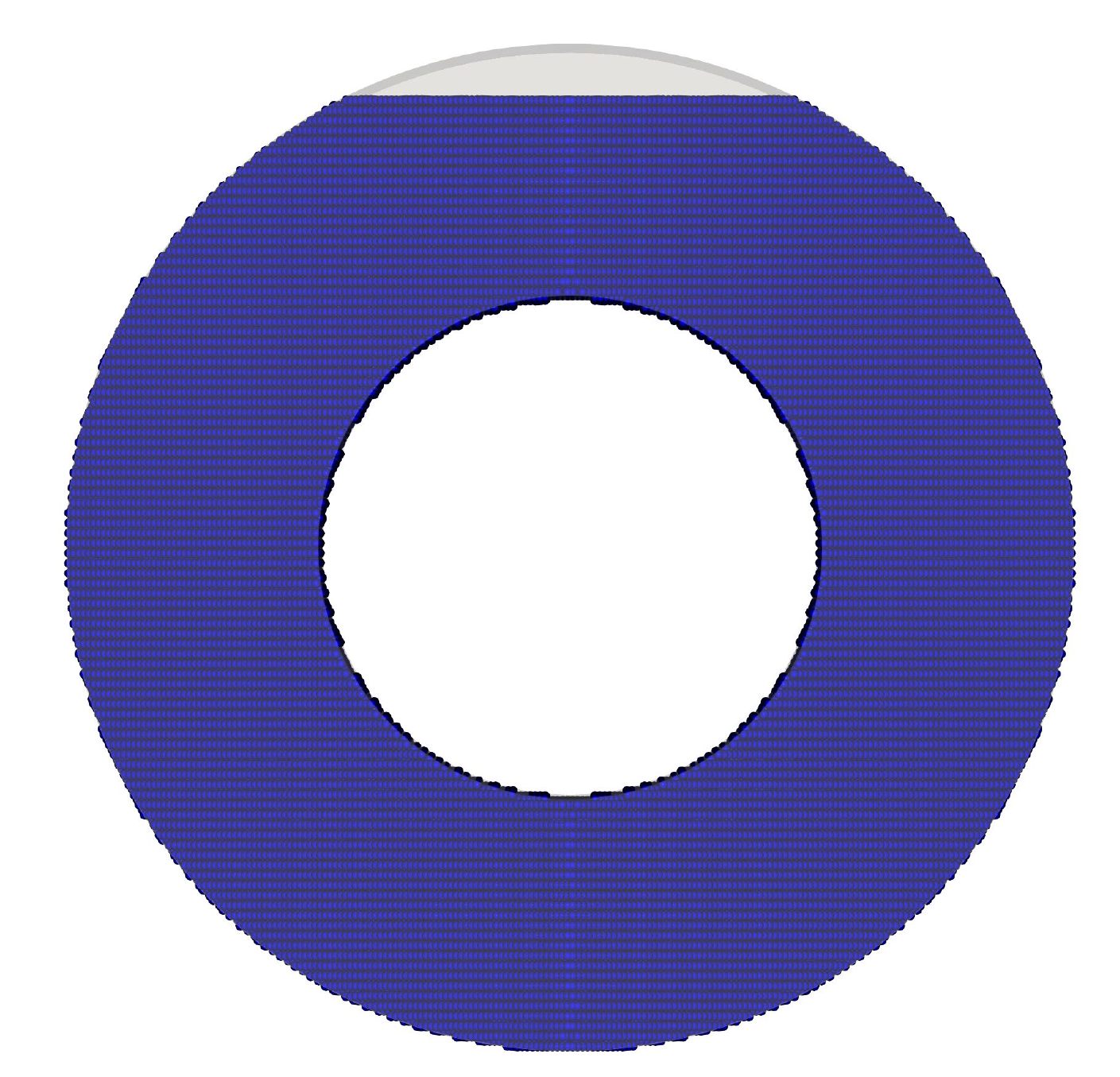}%
        \label{}
    \end{subfigure}&
    \begin{subfigure}[b]{.33\columnwidth}  
        \centering
        \caption{{}}
        \includegraphics[width=1.1\columnwidth]{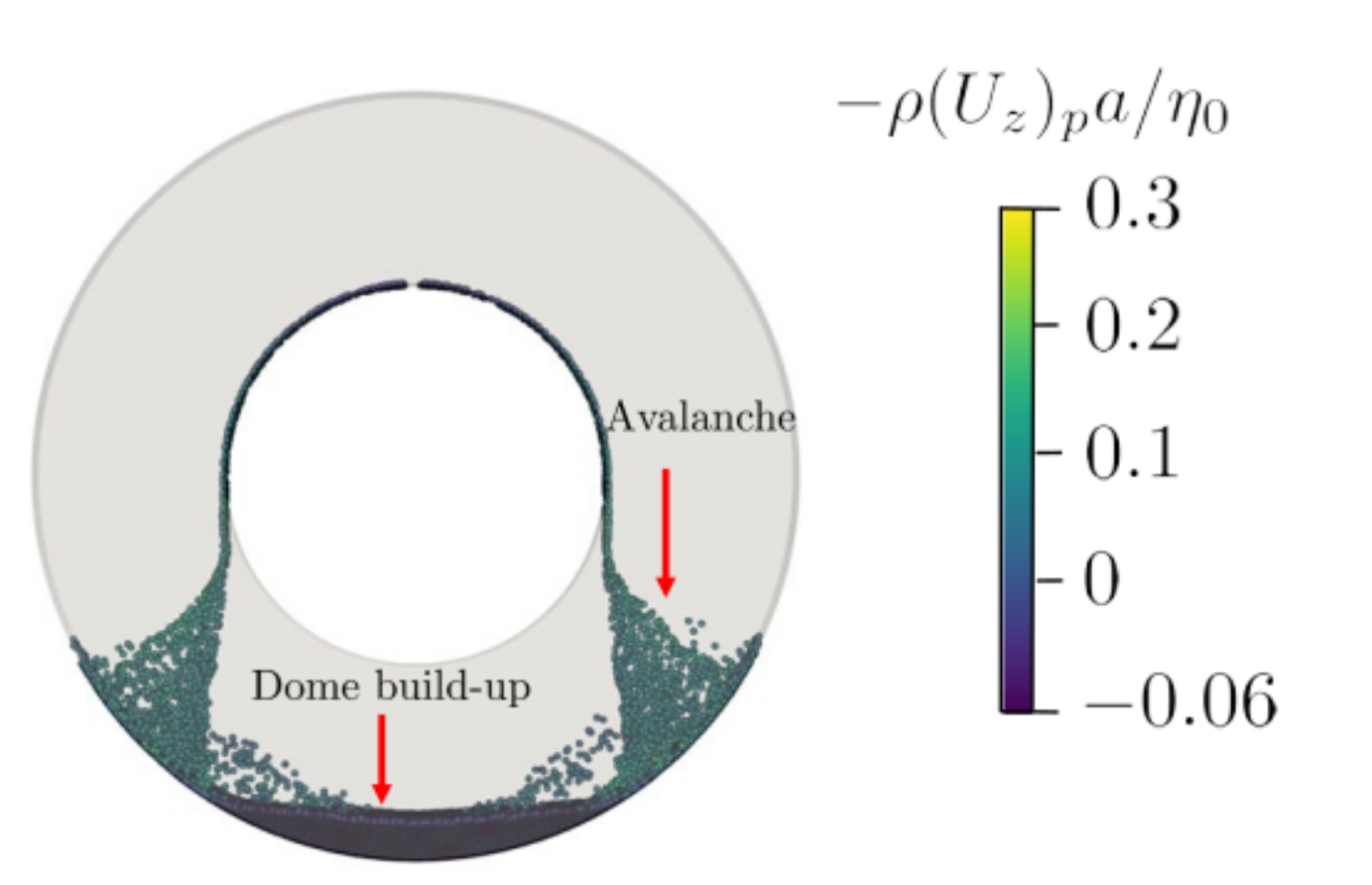}%
        \label{}
    \end{subfigure}&
    \begin{subfigure}[b]{.33\columnwidth}   
        \centering 
        \caption{{}}
        \includegraphics[width=1.0\columnwidth]{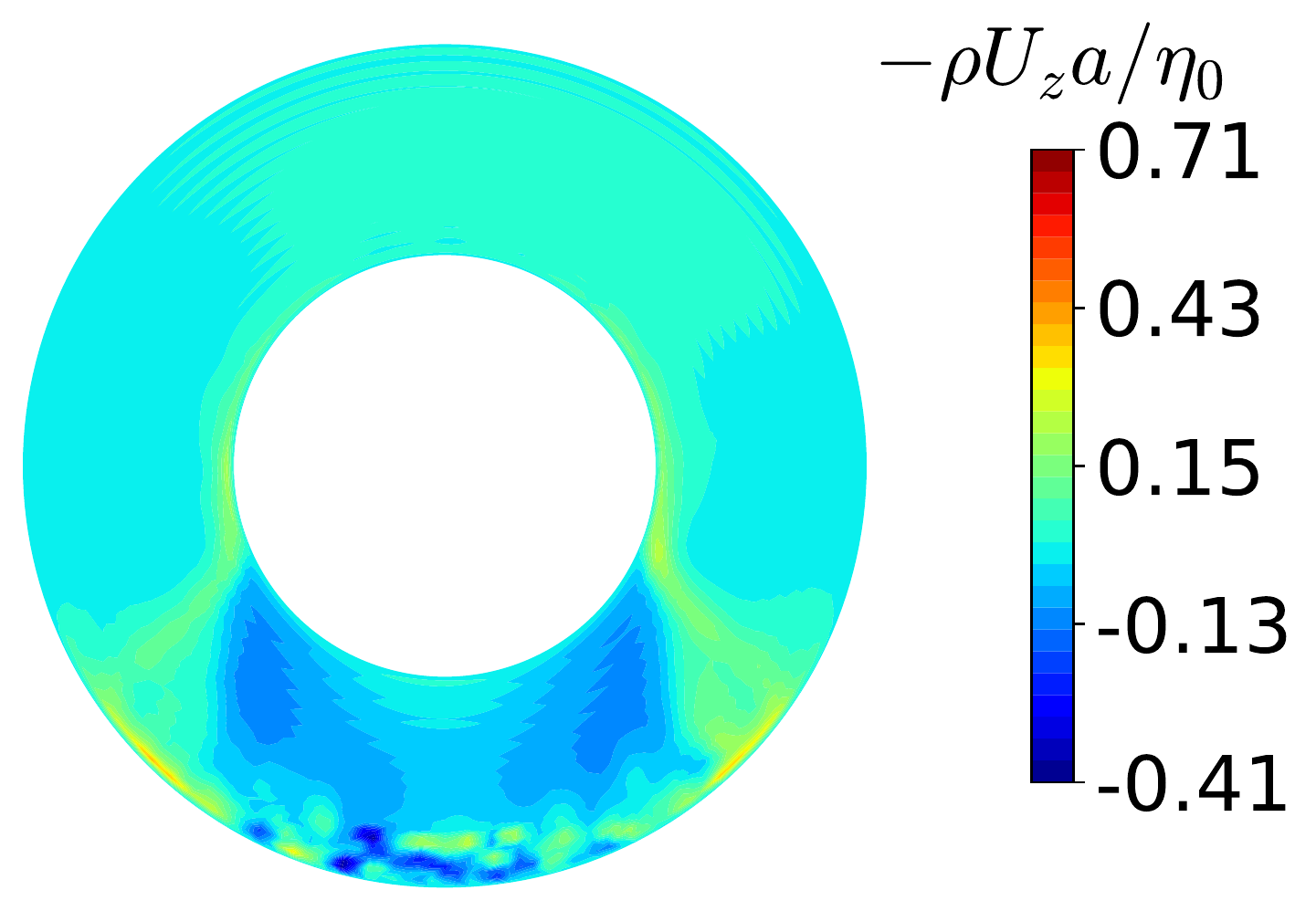}%
        \label{}
    \end{subfigure}\\
    \end{tabular}}
    \caption[]
    {Panel (a) shows the computational annular pipe setup ($R_i=25~$mm and $R_o=50~$mm) and the initial homogeneous distribution of the particles at $\phi_i=1\%$ when $t/t_c=0$, where $t_c=R_o/U_{Stokes}$. Panel (b) shows the numerical result for a Newtonian fluid at $t/t_c \approx 1.2$ that reveals the development of strong spatially inhomogeneous particle distributions as avalanches along the rigid wall of the annulus and a static deposited dome builds up. Panel (c) shows the fluid velocity distribution. The results are shown in a slice through the midplane of the computational domain. Velocities in red, orange, yellow and green are aligned with gravity (which points in the $-\textbf{e}_z$ direction) and velocities in blue indicate backflow.} 
    \label{fig:eccentricflow}
\end{figure*}

The initial setup for this computational study is shown schematically in Fig.~\ref{fig:eccentricflow}(a). A total of 125,000 particles, representing $1\%$ of the annular cavity volume, is used in this case study. The particles are distributed evenly throughout the stagnant fluid at time zero (see Fig.~\ref{fig:eccentricflow}(a)). The particle positions at time $t=0$ are generated using a nearest neighbor algorithm. Gravity is applied vertically across the thin annular geometry ($\textbf{g}=-g\textbf{e}_z$), breaking the azimuthal symmetry and mimicking the onset of concrete particle settlement right after injection is stopped. The goal here is to capture the settling dynamics and test our numerical code to reproduce those dynamics. The code can be then used to analyze different rheological tuning mechanisms to minimize settling over the required time-scale for the concrete to harden. Figures \ref{fig:eccentricflow}(b) and (c) illustrate the numerical result obtained for a Newtonian fluid. It can be readily seen that our code captures local azimuthal avalanches along the rigid walls of the inner pipe, and ultimately static dome build-up effects on longer time-scales of order ($t_c$) where the characteristic settling time is defined as $t_c=R_o/U_{Stokes}$ \cite{Faroughi2017,robisson2020} (see also the movie in the supplementary material). This confirms the accuracy of our 4-way coupling model in which the continuous fluid matrix affects particle motion, local densification effects of the particle lead to enhanced gravitational body forces (per unit volume) driving the sedimenting flow, and local compaction as $\phi\to\phi_m$ leads ultimately to flow arrest. This coupled model is used below to study how these effects change with elastic contributions to the stress field provided by polymeric fluid additives.

\begin{figure*}
\captionsetup[subfigure]{justification=justified,singlelinecheck=false}
    \centering
    {\renewcommand{\arraystretch}{0}
    \begin{tabular}{c@{}c}
    \begin{subfigure}[b]{.5\columnwidth}
        \centering
        \caption{{}}
        \includegraphics[width=0.86\columnwidth]{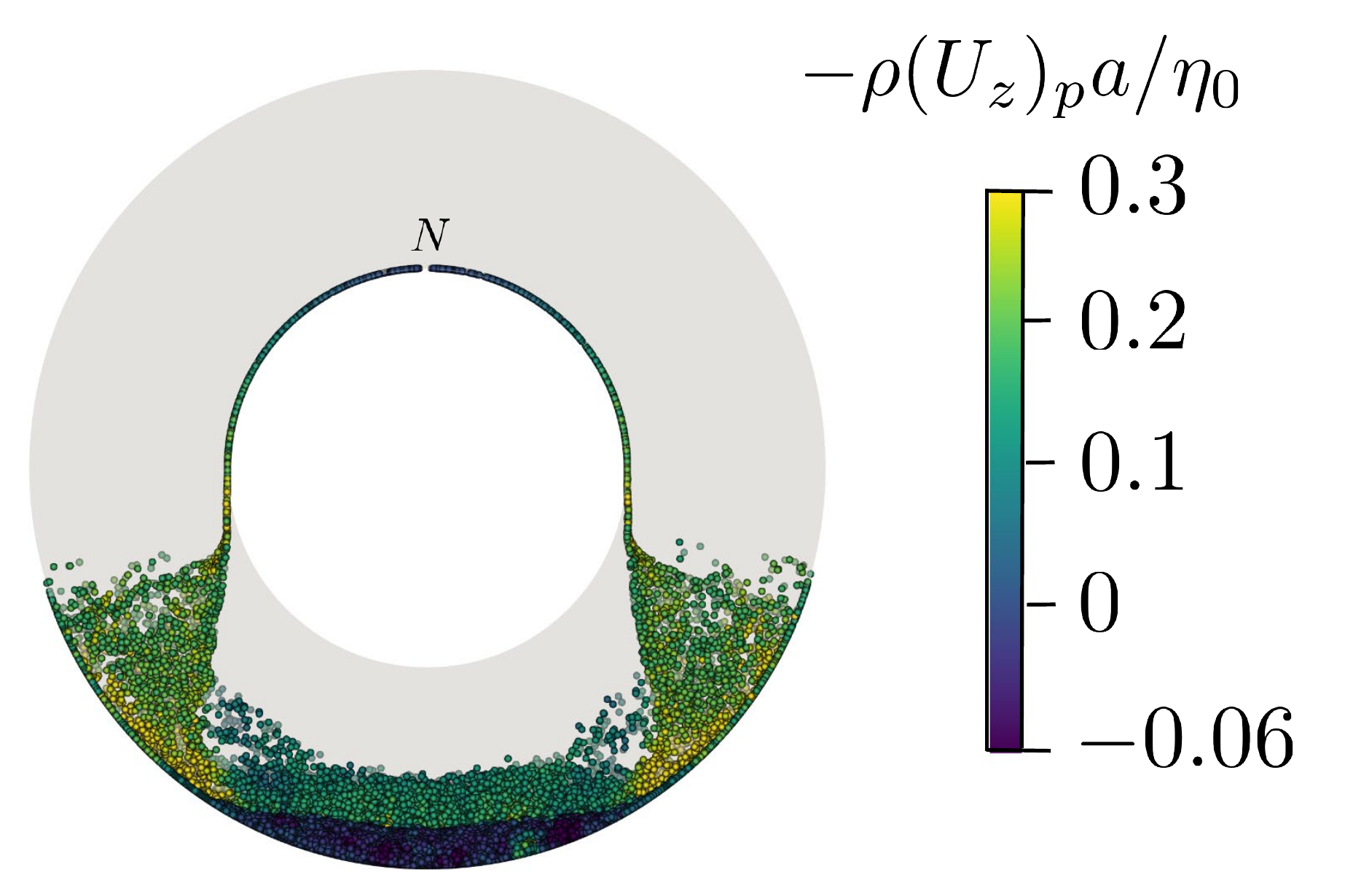}%
        \label{}
    \end{subfigure}&
    \begin{subfigure}[b]{.5\columnwidth}  
        \centering
        \caption{{}}
        \includegraphics[width=0.9\columnwidth]{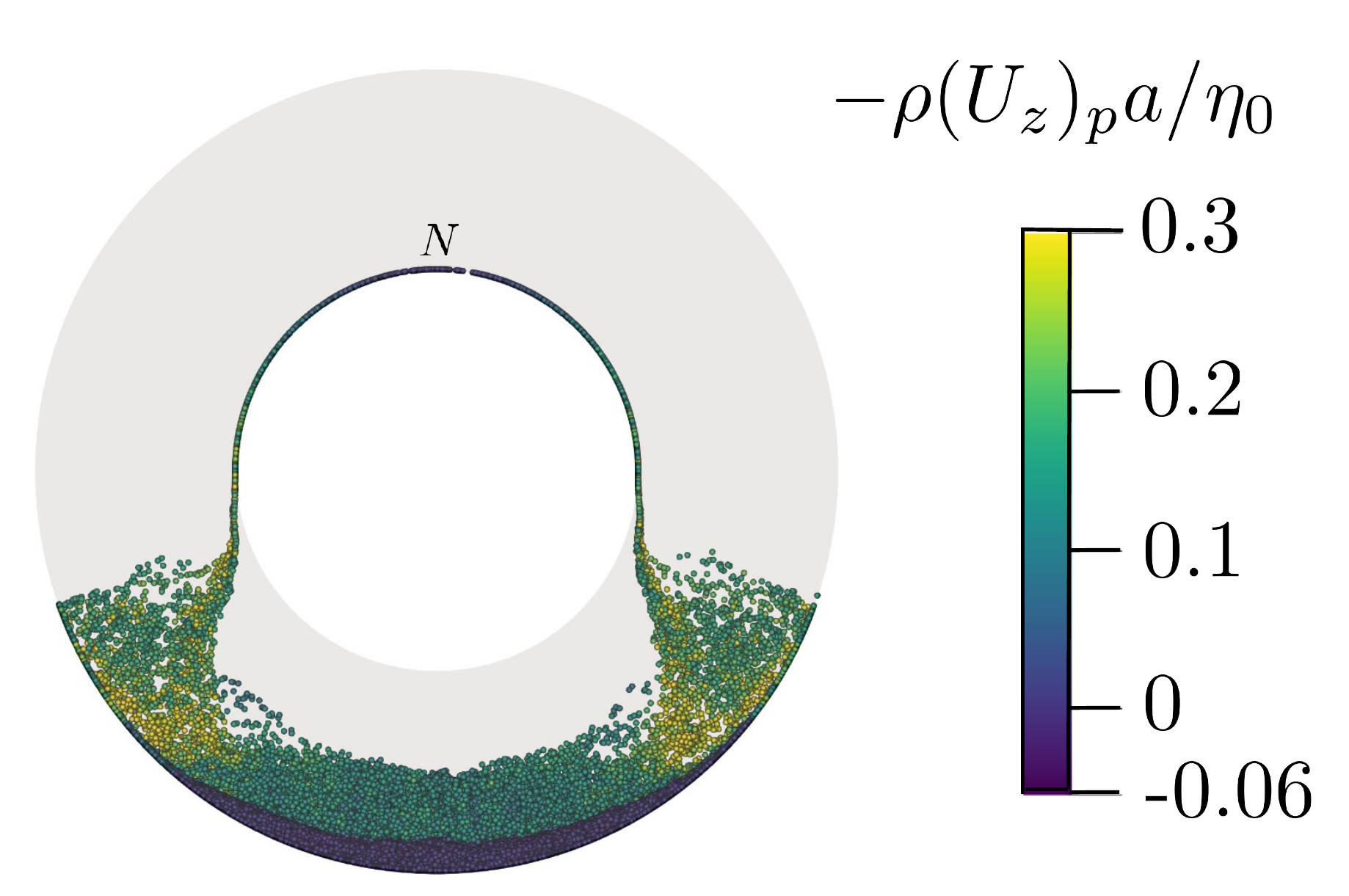}%
        \label{}
    \end{subfigure}\\
    \begin{subfigure}[t]{.5\columnwidth}   
        \centering 
        \includegraphics[width=0.9\columnwidth]{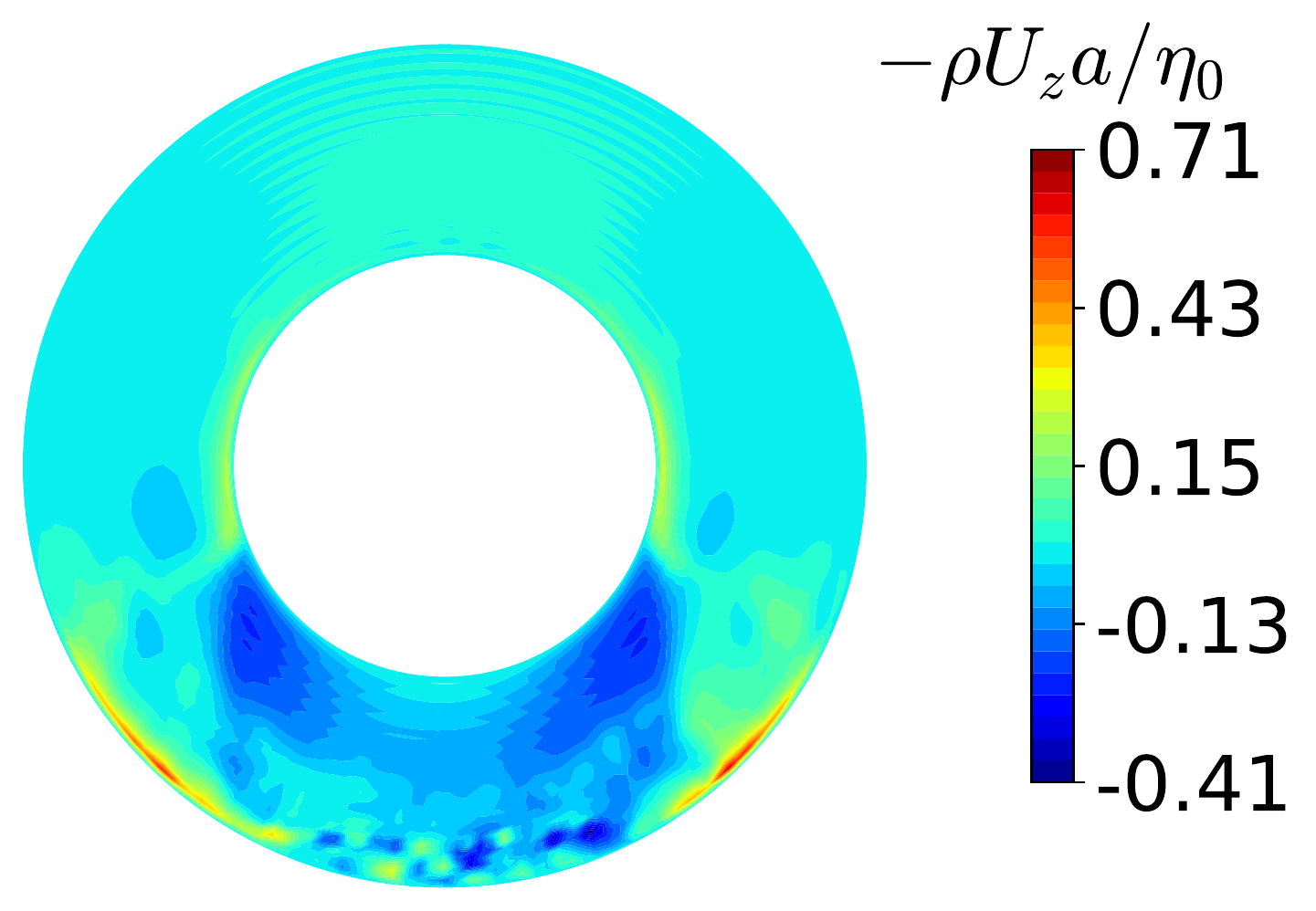}%
        \label{}
    \end{subfigure}&
    \begin{subfigure}[t]{.5\columnwidth}   
        \centering 
        \includegraphics[width=0.9\columnwidth]{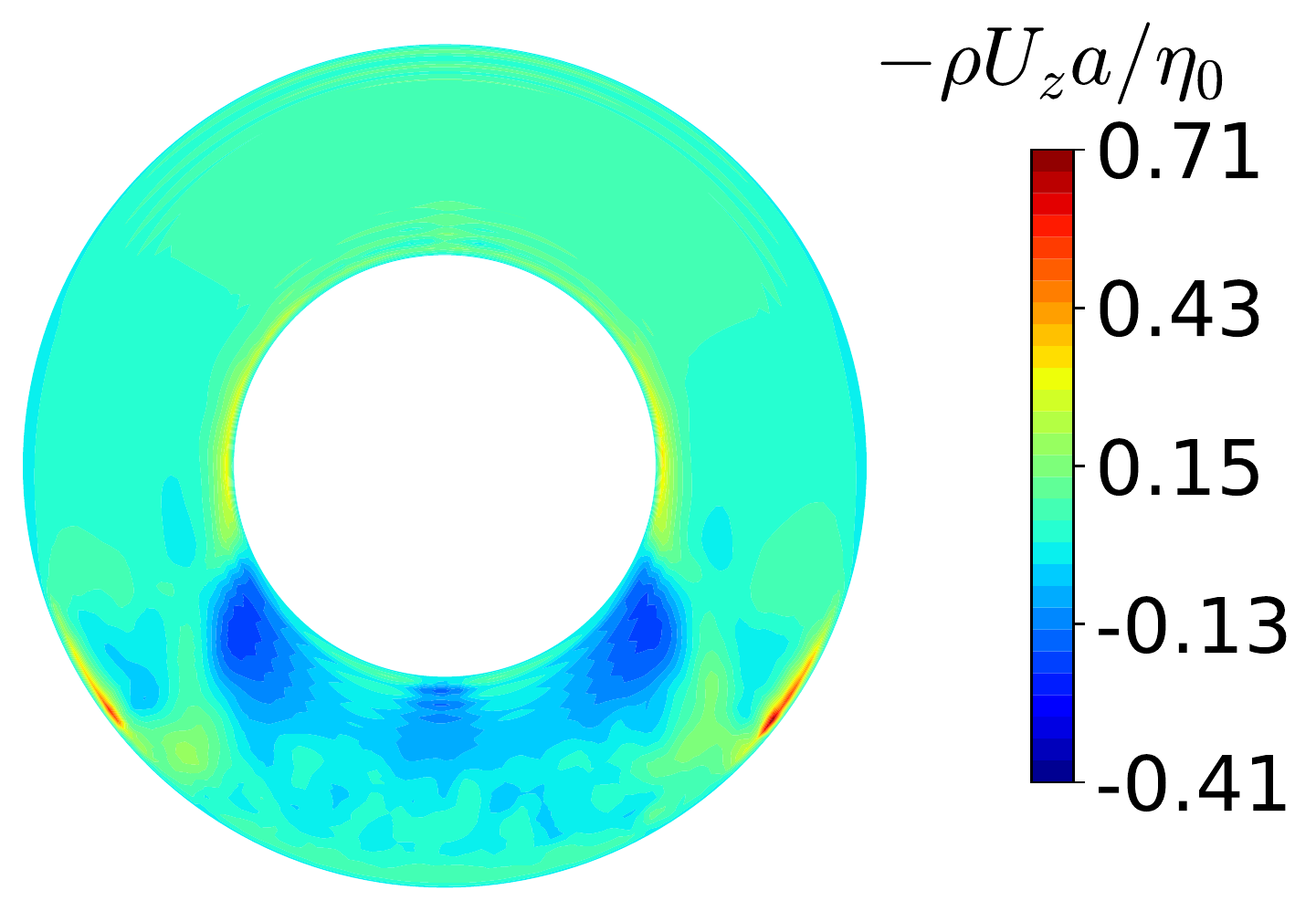}%
        \label{}
    \end{subfigure}
    \end{tabular}}
    \caption[]
    {Velocities of the settling particles (top) and fluid (bottom) in an Oldroyd-B fluid with $\zeta=0.5$ for (a) $El = 0.1$ and (b) $El = 5$ at $ t/t_c \approx 1.2$. The results are shown on a slice through the midplane of the computational domain. Velocities in red, orange, yellow and green are aligned with gravity (which points in the $-\textbf{e}_z$ direction) and velocities in blue indicate backflow.} 
    \label{fig:eccentricEl}
\end{figure*}

Figure~\ref{fig:eccentricEl} shows the dimensionless $z$-component of the velocity of the particles, $-\rho(U_z)_p a/\eta_0$, and fluid, $-\rho U_z a/\eta_0$, computed numerically for a viscoelastic matrix fluid described by the Oldroyd-B constitutive model at $El=Wi/Re=0.1$ and $5$. To vary the elasticity number, we kept the Reynolds number and the settling velocity fixed and changed the Weissenberg number accordingly (i.e. effectively changing the relaxation time of the fluid).

For these two elasticity numbers the particle distributions are similar, with a settling zone and avalanche zones at the bottom and lateral walls of the annular pipe domain, respectively. Additionally, in the settling zone, a backflow of particles occurs due to fluid displaced by the sedimentation and net accumulation of particles in this region. At the north pole (point $N$ in Fig.~\ref{fig:eccentricEl}) of the inner cylinder wall the particles have a backflow velocity, which makes them bounce and slide along the inner cylinder wall. Subsequently, the particles approach the most unsteady settling zone of the annular pipe channel, where a mixture of fluid backflow and gravity-induced velocities are present. Regarding the differences in the results obtained for the two elastic fluids, $El=0.1$ and $El=5$, we can see that the fluid velocity distribution results in a larger region of positive (upwards) velocity near the dome region for the less elastic case, which indicates a migration of the particles to the avalanche zone. In fact, from the particle velocity distributions, we see that the stronger migration of the particles to the avalanche zone at $El=0.1$ causes an increase in the suspension bed height when comparing to the higher elastic case with $El=5$. In fact, the calculated final packed bed height for the cases where $El=0$ or $El=0.1$ is 4.5 mm and for $El=5$ is 3.5 mm.        

\section{Conclusions}
\label{sec:conclusions}

Direct numerical simulations (DNS) of random arrays of spherical particles immersed in Newtonian and constant-viscosity viscoelastic fluids were performed using a finite-volume method. The overall procedure solves the equations of motion coupled with the viscoelastic Oldroyd-B constitutive equation using a log-conformation approach, with a SIMPLEC (Semi-Implicit Method for Pressure-Linked Equations-Consistent) method. The drag forces on individual particles were calculated with the aim of providing an approximate closed-form model to describe numerical simulation data obtained for the unbounded flow of Newtonian and Oldroyd-B fluid past random arrays of spheres. This expression can then be integrated into a Eulerian-Lagrangian solver that enables coupled simulations of the fluid flow and particle migration over a wide range of kinematic conditions. For this purpose, the DNS consisted of a total of 150 different configurations, in which the average fluid-particle drag force is obtained for solid volume fractions $\phi$ $(0 < \phi \leq 0.2)$ and Weissenberg number $Wi$ $(0 \leq Wi \leq 4)$.

The proposed DNS methodology was first tested and verified for the creeping flow of random arrays of spheres immersed in a Newtonian fluid. It was found that the numerical results obtained agree with the Lattice-Boltzmann results of \citet{Hill2001} and can be described by the best-fit model of \citet{Hoef2005}. Statistical accuracy was achieved by averaging the DNS results at each value of $\phi$ over five random configurations, resulting in errors below $3.5\%$ of the average drag force. Subsequently, the same DNS methodology was used to perform finite volume simulations of viscoelastic creeping flows (using the Oldroyd-B constitutive equation with $\zeta=0.5$) past the same fixed random configurations of particles. A simple factorized closure model for the viscoelastic drag coefficient of random arrays of spheres (corresponding to moderately dense suspensions $\phi\leq 0.2$) translating in a quasi-linear Oldroyd-B viscoelastic fluid was proposed, by fitting the DNS results with an equation of the same form as the \citet{Hoef2005} model, combined with the viscoleastic drag force correction on a single sphere proposed by \citet{Salah2019}. The resulting regression model accounts for 98.2\% of the variance of the numerical data, with an average error of 5.7\%.

Finally, a numerical formulation for Eulerian-Lagrangian simulation of solid particles in viscoelastic fluids, $DPMviscoelastic$, was presented and implemented using a combination of the finite-volume and the discrete particle methods. The implementation was carried out by extending the solver $DPMFoam$ from the open-source $OpenFOAM$ library. The algorithm solves the motion of an incompressible viscoelastic fluid phase in the presence of a secondary particulate phase, in which the volume-averaged continuity and Navier-Stokes equations are employed together with a viscoelastic constitutive equation to describe the fluid flow, and a discrete particle method is used  to update the particle movements. This approach guarantees the coupling between the dynamics of the continuous fluid and the discrete solid phases, by imposing a two-way coupling between the two phases. The coupling is provided by momentum transfer through the drag force expression proposed here, which is exerted by the fluid on the solid particles. Additionally, we consider two different formulations to describe the contact between particles, the Hertzian spring-dashpot and Multi-Phase Particle In Cell (MPPIC) models.

As a proof-of-concept, the newly-developed algorithm was assessed for accuracy in two case studies. First, we studied the proppant transport and sedimentation during pumping (a phenomenon typical of hydraulic fracturing operations) in a long channel of rectangular cross section. For the case in which the fluid matrix is Newtonian, the resulting axial distribution of particle sedimentation profiles was compared with experimental data available in the literature for suspensions formulated with Newtonian matrix fluids and different initial particle volume fraction, and good agreement was obtained. Subsequently, the $DPMviscoelastic$ solver was tested on the same problem using an Oldroyd-B fluid. Analysis of the particle distribution and fluid velocity profiles at an elasticity number of $El=30$, showed that fluid elasticity inhibits the rate of particle settling and prevents the formation of a dense sedimented layer along the floor of the channel.

Subsequently, segregation phenomena which occurs when pumping a casing material along horizontal wells was also studied in an annular pipe domain. Numerical simulations using a Newtonian fluid were performed, and we were able to capture the avalanche and dome build-up effects observed in experimental observations of the particle distributions \cite{robisson2020}. Additionally, a viscoelastic fluid was also employed at two different elasticity numbers $El=0.1$ and $5$. The particles were found to sediment with two markedly contrasting zones, a highly disordered and unsteady region where a mixture of fluid backflow and gravity-induced settling velocities are present and a sedimented zone where particles are closely packed together and the fluid velocity is almost zero. It was found that the stronger migration of the particles to the avalanche zone at $El=0.1$ cause an increase in the suspension bed height when comparing to the higher elastic case with $El=5$.

In summary, the DNS computational methodology presented here allows us to construct a closed-form expression for the drag force exerted by an Oldroyd-B viscoelastic fluid on random arrays of particles, which can be incorporated in a newly-developed Eulerian-Lagrangian viscoelastic code, $DPMviscoelastic$, using an open-source framework. The resulting code can predict the flow patterns and particle distributions that develop in moderate volume fraction suspensions with viscoelastic matrix fluids. We hope that in the future this open-source code can be used to help understand other migration and settling phenomena in complex fluids which are commonly encountered in a range of industrial and biological applications.

\section*{Acknowledgments}
\label{section:Acknowledgments}

This work is funded by FEDER funds through the COMPETE 2020 Programme and National Funds through FCT (Portuguese Foundation for Science and Technology) under the projects UID-B/05256/2020, UID-P/05256/2020 and MIT-EXPL/TDI/0038/2019 - APROVA - Deep learning for particle-laden viscoelastic flow modelling (POCI-01-0145-FEDER-016665) under MIT Portugal program. The authors would like to acknowledge the University of Minho cluster under the project NORTE-07-0162-FEDER-000086 (URL: http://search6.di.uminho.pt), the Minho Advanced Computing Center (MACC) (URL: https://
macc.fccn.pt) under the project CPCA\_A2\_6052\_2020, the Texas Advanced Computing Center (TACC) at The University of Texas at Austin (URL: http://www.tacc.utexas.edu), the Gompute HPC Cloud Platform (URL: https://www.gompute.com), and PRACE - Partnership for Advanced Computing in Europe under the project icei-prace-2020-0009, for providing HPC resources that have contributed to the research results reported within this paper.

\section*{Appendix A}
\label{section:Appendix}

\begin{figure}[H]
\captionsetup[subfigure]{justification=justified,singlelinecheck=false}
    \centering
    {\renewcommand{\arraystretch}{0}
    \begin{tabular}{c@{}c}
    \begin{subfigure}[b]{0.5\columnwidth}
        \centering
        \caption{{}}
         	\includegraphics[width=0.9\columnwidth]{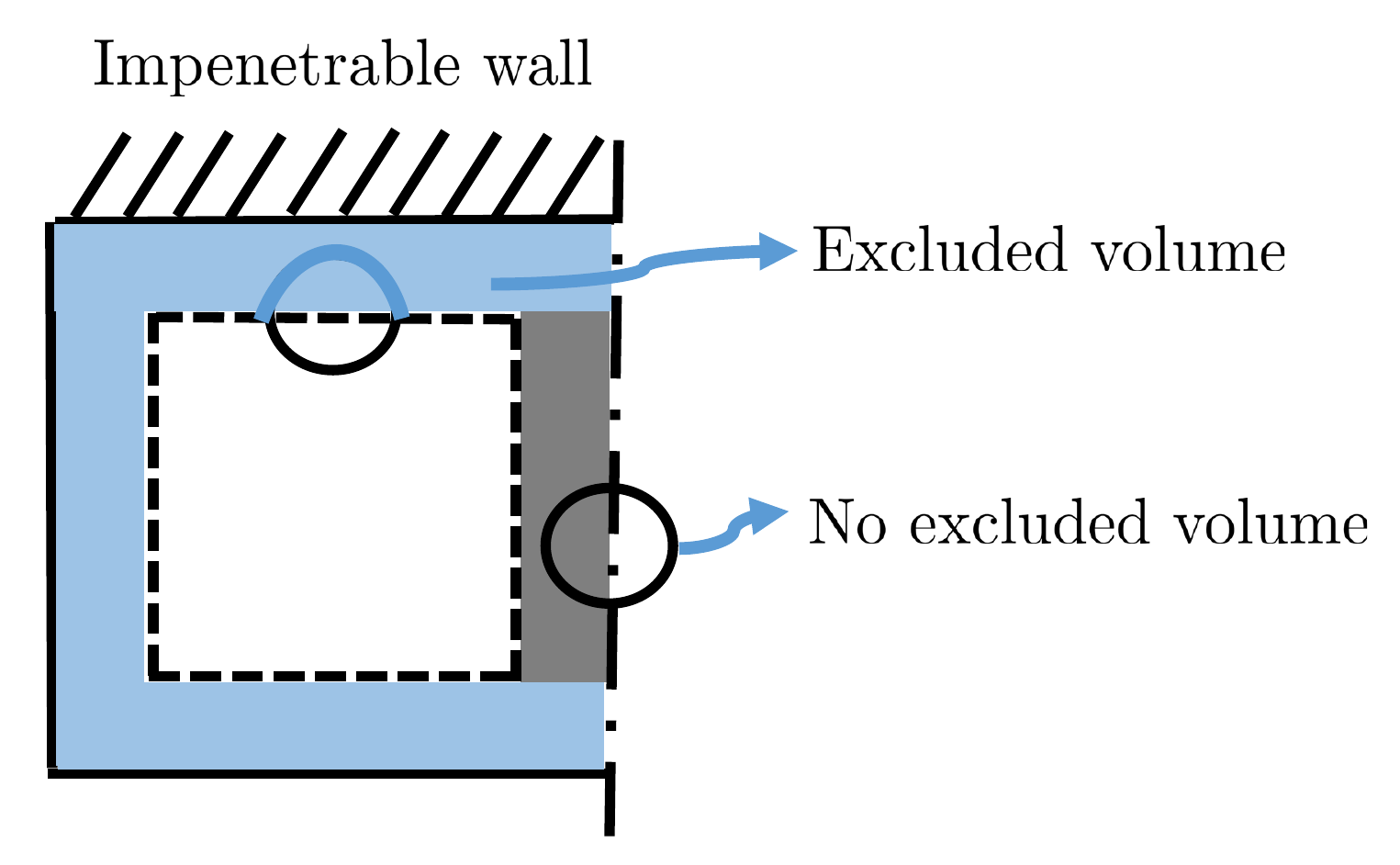}%
        \label{}
    \end{subfigure}&
    \begin{subfigure}[b]{0.4\columnwidth}
	\centering    
	\caption{{}}
\hspace{-1.5cm}        \includegraphics[width=0.7\columnwidth]{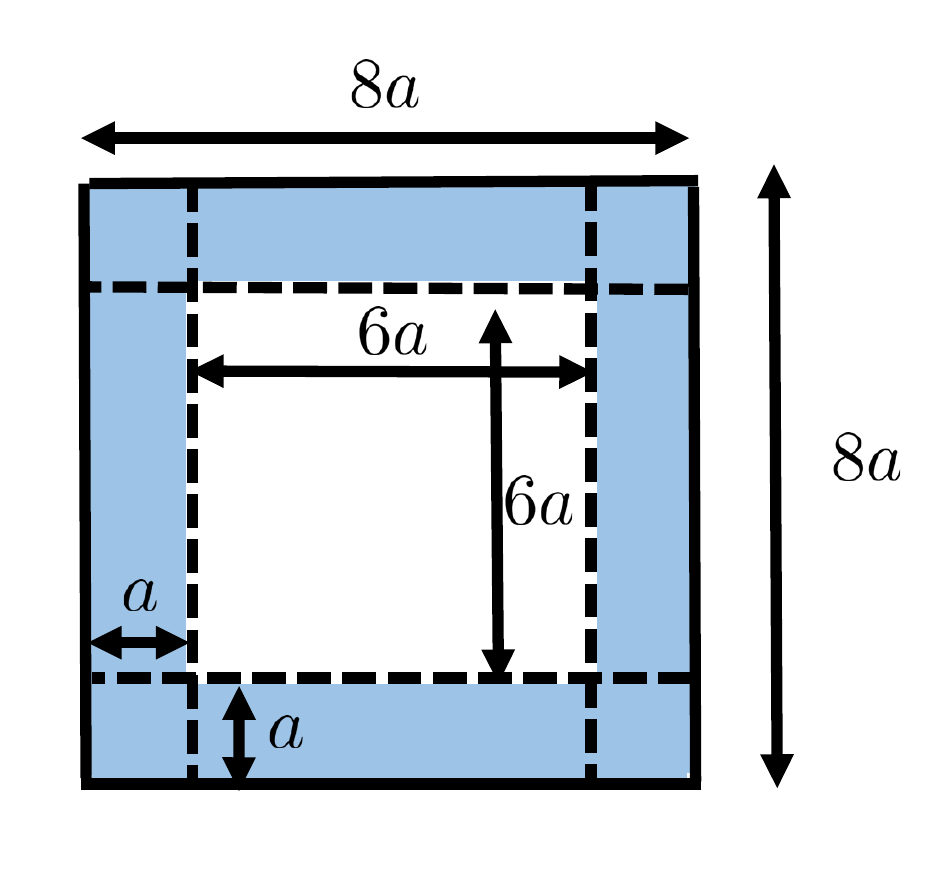}%
   		 \label{}
    \end{subfigure}\\
    \end{tabular}}
    \caption[]
    {Schematic representation of (a) excluded and no-excluded volume regions and (b) area fraction in the excluded volume region.} 
    \label{fig:draftEV}
\end{figure}

In the DNS study presented in section~\ref{sec:DNS}, we considered two different domain configurations, one with spheres having centroids in a wall region of thickness $a$ around all four lateral edges of the flow domain and another in which the sphere centroids are excluded from this wall region. We refer to these cases as the no-excluded volume and excluded volume configurations, respectively. As shown in Fig.~\ref{fig:draftEV}(a) when spheres are allowed to be located in the wall region (blue color), i.e., when their centroid is located less than one radius from the wall, then the boundary acts as a perfectly periodic wall. In the opposite case the boundary walls exclude the spheres and act like rigid stress free periodic walls. In fact, based on Fig.~\ref{fig:draftEV}(b), we can calculate the probability of a single sphere being located in the wall region. Assuming a square cross-section with a width of $8a$, the total cross-sectional area is $64a^2$. Regarding the blue annular area, i.e., the region which excludes the spheres near the wall, the area is of ($64a^2-36a^2=28a^2$). Hence, the probability of a randomly placed sphere being located in the excluded region is equal to $28a^2/64a^2=0.4375$ and the overall/fraction area of spheres (of volume fraction $\phi$) in this region can be as large as $0.4375\phi$. Therefore, as $\phi$ increases it is important that when particles are randomly distributed in the domain they should be allowed to be placed with centroids near the walls. 

\begin{figure}[H]
\captionsetup[subfigure]{justification=justified,singlelinecheck=false}
    \centering
    {\renewcommand{\arraystretch}{0}
    \begin{tabular}{c@{}c}
    \begin{subfigure}[b]{0.5\columnwidth}
        \centering
        \caption{{}}
\hspace{-2cm}		\includegraphics[width=0.8\columnwidth]{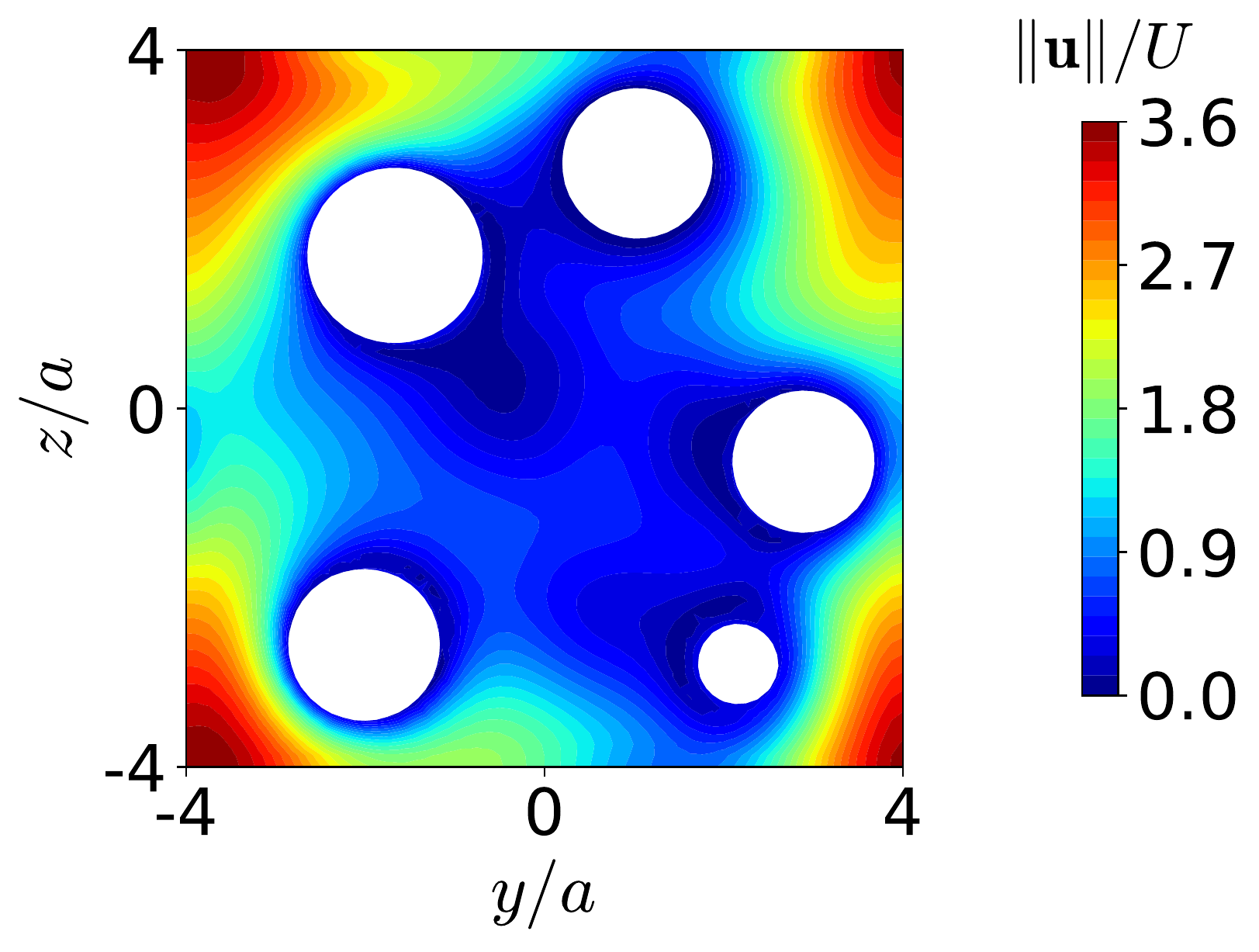}%
        \label{}
    \end{subfigure}&
    \begin{subfigure}[b]{0.5\columnwidth}
	\centering    
	\caption{{}}
\hspace{-2cm}        \includegraphics[width=0.8\columnwidth]{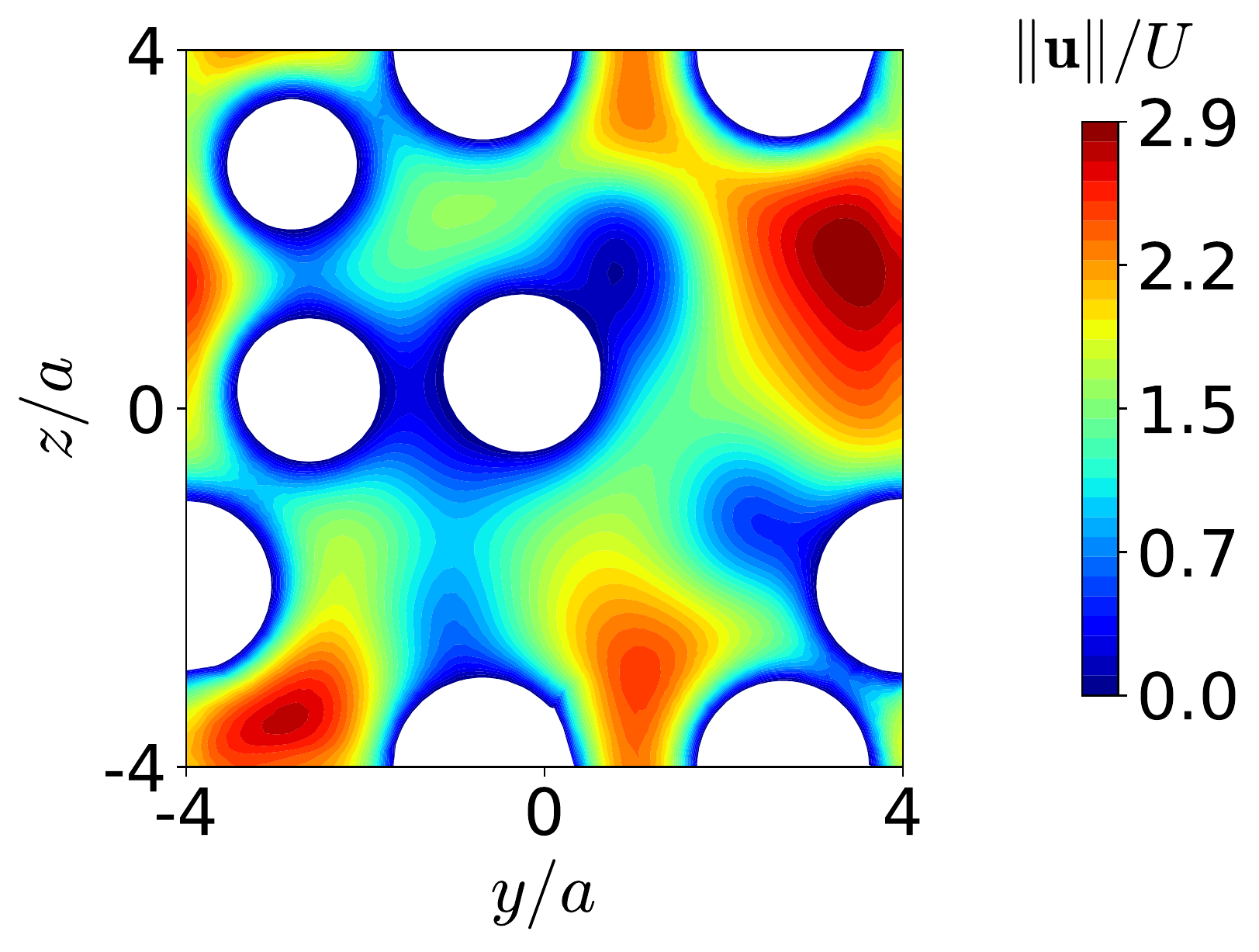}%
   		 \label{}
    \end{subfigure}\\
    \end{tabular}}
    \caption[]
    {Steady flow field around one representative random array of particles in a channel filled with Newtonian fluid. Contours of the velocity magnitude field $\|\textbf{u}\|$ (with the inflow direction pointed out of the plane of the page) are represented for particle volume fraction $\phi=0.2$, in the $y-z$ plane with (a) excluded volume near the walls and (b) a no-excluded volume configuration. In the latter case the velocity field is more evenly distributed across the entire cross-section of the domain.} 
    \label{fig:EVvsNEV}
\end{figure}

In Fig.~\ref{fig:EVvsNEV}, we show contours of the velocity magnitude in the transverse $y-z$ plane for configurations with an excluded volume and no-excluded volume region. From the distribution of velocity magnitude contours, it can be seen that in the excluded volume configuration (i.e., Fig.~\ref{fig:EVvsNEV}(a)), the larger local concentration of rigid impenetrable spheres in the middle of the square channel push the strongest fluid flow ownwards towards the walls causing a stagnant region near the channel center. On the other side, in the no-excluded volume configuration (i.e., Fig.~\ref{fig:EVvsNEV}(b)), the fluid flow is more evenly distributed across the entire channel. This affects the average drag force exerted on the spheres as shown in Table~\ref{tab:newtonian} and Fig.~\ref{fig:streamDe0}.

\label{section:References}

\def\mybibdoicolor{\color{black}}
\newcommand*{\doi}[1]{\href{\detokenize{#1}} {\raggedright\mybibdoicolor{DOI: \detokenize{#1}}}}

\bibliographystyle{unsrtnat}
\bibliography{header.bbl}

\end{document}